\documentclass[aps,showpacs,prb,reprint,longbibliography,superscriptaddress]{revtex4-2}

\usepackage[colorlinks=true,linkcolor=blue,citecolor=blue,urlcolor=blue]{hyperref}
\usepackage{setspace} 
\usepackage{graphicx}
\usepackage{amsmath}
\usepackage{color}
\usepackage{amsmath}
\usepackage{mathtools}
\usepackage{amssymb}
\usepackage{verbatim}
\usepackage{physics}
\usepackage{latexsym}
\usepackage{enumerate} 
\usepackage{bm} 
\usepackage[caption=false]{subfig}
\usepackage{multirow}
\usepackage{booktabs}
\usepackage{siunitx}
\usepackage{bm,bbm}

\usepackage{float}

\usepackage{lipsum}

\begin{document}


\title{Compressing Hamiltonians with \emph{ab initio} downfolding for simulating strongly-correlated materials on quantum computers}

\author{Antonios M. Alvertis}
\email{antoniosmarkos.alvertis@nasa.gov}
\affiliation{KBR, Inc., NASA Ames Research Center, Moffett Field, California 94035, United States}
\author{Abid Khan}
\affiliation{Department of Physics, University of Illinois Urbana-Champaign, Urbana, IL, United States 61801}
\author{Norm M. Tubman}
\email{norman.m.tubman@nasa.gov}
\affiliation{NASA Ames Research Center, Moffett Field, California 94035, United States}

\date{\today}
\begin{abstract}

The accurate first-principles description of
strongly-correlated materials is an important
and challenging problem in condensed matter physics. \emph{Ab initio}
downfolding has emerged as a way of 
deriving compressed
many-body Hamiltonians that maintain the essential physics of
strongly-correlated materials. The solution of these material-specific models is still exponentially difficult to generate on classical computers, but quantum algorithms allow for a significant speed-up in obtaining the ground states of these compressed Hamiltonians.  
Here we demonstrate that utilizing quantum algorithms for obtaining the properties of
downfolded Hamiltonians 
can indeed yield high-fidelity solutions. 
By combining \emph{ab initio} downfolding and variational quantum eigensolvers, we correctly predict  the antiferromagnetic
state of one-dimensional cuprate $\text{Ca}_2\text{CuO}_3$, the excitonic
ground state of monolayer $\text{WTe}_2$, and
the charge-ordered state of correlated metal
$\text{SrVO}_3$. Numerical simulations 
utilizing a classical tensor
network implementation of variational quantum eigensolvers allow us to simulate large
models with up to $54$ qubits and encompassing up to four bands in the correlated subspace, which is indicative
of the complexity that our framework can address.   Through these methods we demonstrate the potential of classical pre-optimization and downfolding techniques for enabling efficient materials simulation using quantum algorithms.

\end{abstract}

\maketitle

\section{Introduction}
Materials with strong electronic correlations exhibit a wealth
of interesting properties, including superconductivity~\cite{doi:10.1126/science.1071122,Li2019}, charge-ordered and spin-ordered states~\cite{doi:10.1080/0001873021000057114,Clay2019}, Mott insulating~\cite{doi:10.1126/science.aam9189,Grytsiuk2024} and excitonic insulating behavior~\cite{Jia2022,Ma2021}. This
constitutes strongly-correlated materials of great interest within
condensed matter physics, and achieving a deep understanding of
their properties central towards novel technological breakthroughs. It is therefore important
to develop computational methods which can predict the properties of strongly-correlated systems from first quantum
mechanical principles. Density functional theory (DFT) has
been the main workhorse of computational materials science
for decades~\cite{Haunschild2016}, however it generally fails to capture strong electronic correlations, at least in its most commonly employed semilocal formulation~\cite{Perdew2009}. 

The full many-body Hamiltonian describing a general material includes $M^4$ terms, where $M$ the number of orbitals. For realistic systems this can be prohibitively large.
A promising computational
method for generating a compressed representation of materials from first principles is \emph{ab initio}  downfolding~\cite{Nakamura2008,Nakamura2010,Arita2015,Zheng2018}. This technique utilizes a DFT starting point in order to derive a material-specific many-body Hamiltonian within a strongly-correlated active space of interest, which can be subsequently
solved using a range of techniques, including exact diagonalization~\cite{Yoshimi2021,Botzung2024,tubman2016deterministic,tubman2020modern}, quantum Monte Carlo~\cite{Ma2015,kim2018qmcpack,tubman2015molecular}, diagrammatic approximations~\cite{PhysRevB.100.115154}, and beyond. 
The many-body Hamiltonian resulting from downfolding and describing the physics of the strongly-correlated region is typically of the extended Hubbard form~\cite{Nakamura2021}:
\begin{align}
    \label{eq:Hamiltonian}
    H= \sum_{\sigma}\sum_{\mathbf{R}\mathbf{R}'}\sum_{ij}t_{i\mathbf{R}j\mathbf{R}'}a_{i\mathbf{R}}^{\sigma \dagger}a_{j\mathbf{R}'}^{\sigma}+\nonumber\\ \frac{1}{2}\sum_{\sigma \rho}\sum_{\mathbf{R}\mathbf{R}'}\sum_{ij}U_{i\mathbf{R}j\mathbf{R}'}a_{i\mathbf{R}}^{\sigma \dagger}a_{j\mathbf{R}'}^{\rho \dagger}a_{j\mathbf{R}'}^{\rho}a_{i\mathbf{R}}^{\sigma},
\end{align}
where $\sigma,\rho$ are spin indices, $\mathbf{R},\mathbf{R}'$ denote lattice vectors, and $i,j$ run over the electronic bands of the system. Downfolding methods
have been successfully applied to the description of charge-ordered systems  and charge density waves~\cite{Yoshimi2023,Schobert2024},
high-temperature superconductors~\cite{Arita2015,Hirayama2019,Ohgoe2020,Been2021,Schmid2023}, and beyond. However, the unfavorable scaling of classical methods for obtaining eigenstates of the Hamiltonian of Eq.\,\eqref{eq:Hamiltonian} has thus far prevented
downfolding methods from being used for large systems, which can result in finite-size effects~\cite{Aichhorn2006,holzmann2016theory,Qin2022}, and limits the number of bands
which may realistically be included in these models.

Quantum computers are a promising technology for the simulation of many-body quantum systems~\cite{Bernien2017,Fauseweh2024}, especially with regards to the number of qubits needed to simulate Hamiltonians of the form of eq.\,\eqref{eq:Hamiltonian}, as many algorithms can utilize a 1:1 correspondence between the qubits and the spin-orbitals of the system.  This constitutes the simulation of strongly-correlated materials and Hubbard-like models an ideal problem for solution on quantum
hardware~\cite{Wecker2015,Cade2020}, and the performance
of popular variational quantum algorithms in terms of producing an accurate ground state has recently been the subject of a detailed benchmark~\cite{alvertis2024classicalbenchmarksvariationalquantum}. The potential of emerging fault-tolerant quantum computers to obtain properties of extended Hubbard models
as that of Eq.\,\eqref{eq:Hamiltonian} has been discussed in
Ref.~\cite{agrawal2024quantifyingfaulttolerantsimulation}, where the necessary resources to access experimentally-relevant quantities have been estimated.  

The utility of quantum computers for materials simulation is currently limited by the fact that, as previously mentioned, 
the number of terms of the
full Hamiltonian of a typical material scales as $M^4$. This can in turn lead to extremely high resource estimates in terms of gates and run times. 
Here we circumvent this limitation by demonstrating
that current and near-term quantum algorithms can produce quantitatively and qualitatively accurate results for ground states of strongly-correlated materials, at a modest computational cost. We achieve this by utilizing compressed representations obtained via \emph{ab initio} downfolding, and classical tensor networks simulations of the variational
quantum eigensolver (VQE)~\cite{khan2023preoptimizing}. We show that our VQE energy 
for the ground state is quantitatively accurate when compared
to results obtained within density matrix renormalization group (DMRG), and that we obtain the correct behavior for
the ground state wavefunction and correlation functions across different scenarios,
and specifically the antiferromagnetic behavior of the quasi-1D cuprate $\text{Ca}_2\text{CuO}_3$~\cite{Rosner1997}, the excitonic insulating ground state of the two-dimensional material $\text{WTe}_2$~\cite{Sun2022,Jia2022}, and the charge-ordered state of the correlated metal $\text{SrVO}_3$~\cite{Aizaki2012,Zhang2016}. These results, combined with
demonstrations that quantum simulation of downfolded Hamiltonians can yield accurate excited state
properties in strongly correlated molecular systems~\cite{Bauman2019,chang2023downfolding,10.1063/5.0213525}, highlight the strong potential of current and emerging quantum computers for simulating the properties
of strongly-correlated materials at a modest computational cost, when combined with \emph{ab initio} downfolding. 

\begin{figure}[tb]
    \centering
    \includegraphics[width=0.8\linewidth]{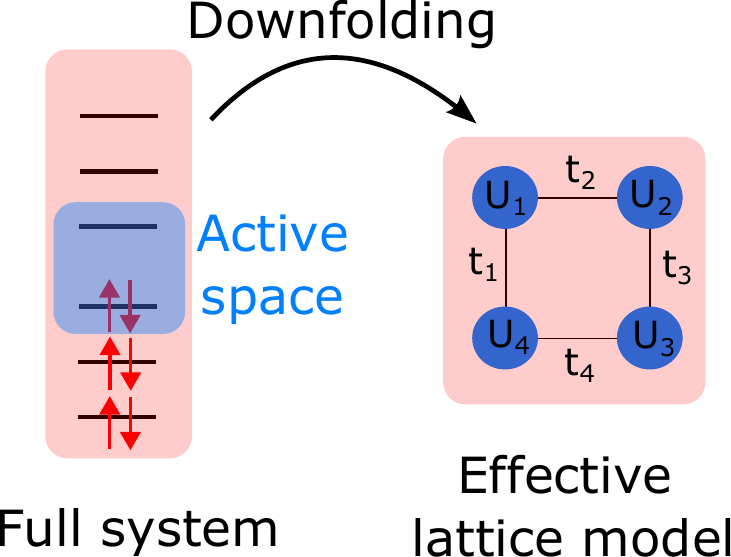}
    \caption{Schematic illustration of \emph{ab initio} downfolding. Starting from a low-level description of the full system of interest, an active space is identified, and a many-body lattice Hamiltonian is generated, representing the physics within the active space. A typical form for the downfolded Hamiltonian is the extended Hubbard form of eq.\,\eqref{eq:Hamiltonian}, where each lattice site can be associated with different hopping and Coulomb terms.}
    \label{fig:downfolding}
\end{figure}

\section{Methods}
\label{sec:Methods}

\subsection{\emph{Ab initio} downfolding}
The aim of \emph{ab initio} downfolding approaches is to 
generate a many-body Hamiltonian on a lattice, representing an active space
of interest for a given material, typically within the low-energy region in the vicinity of the Fermi level~\cite{Nakamura2008,Nakamura2010,Arita2015,Nakamura2021}. This is schematically illustrated in Fig.\,\ref{fig:downfolding}. Downfolding often utilizes a first-principles description such as DFT as a starting point. An exchange term may be included in downfolded Hamiltonians~\cite{Nakamura2021}, however we will
ignore it here given its small magnitude for our studied
systems, consistent with previous works~\cite{Yoshimi2023}. 
In order to describe periodic solids as a lattice model, we
work within the basis of maximally-localized Wannier functions~\cite{Marzari2012}, which we will denote as $\phi$. Here the Wannier functions are centered at the different lattice sites $\mathbf{R}$ appearing in Eq.\,\eqref{eq:Hamiltonian}, with $i,j$ indexing the $i^{\text{th}}$ and $j^{\text{th}}$ Wannier function respectively. Starting from a Kohn-Sham DFT calculation performed within Quantum Espresso~\cite{QE}, 
the Wannier representation of the studied systems within
the active space of interest is obtained using the Wannier90 code~\cite{Pizzi2020}. The Wannierization procedure yields the hopping
terms appearing in Eq.\,\eqref{eq:Hamiltonian} as
\begin{equation}
    \label{eq:hopping}
    t_{i\mathbf{R}j\mathbf{R}'}=\int_V d\mathbf{r}\phi_{i\mathbf{R}}^*H_{KS}\phi_{j\mathbf{R}'},
\end{equation}
where $H_{KS}$ the Kohn-Sham Hamiltonian and $V$ the volume of
the unit cell. For $\mathbf{R}=\mathbf{R}'$ this term represents the on-site potential of site $\mathbf{R}$.
Moreover, following the wannierization using Wannier90, we utilize the wan2respack~\cite{Kurita2023} and RESPACK~\cite{Nakamura2021}
software packages in order to obtain the two-body Coulomb integrals appearing in Eq.\,\eqref{eq:Hamiltonian} as
\begin{align}
    \label{eq:coulomb}
    U_{i\mathbf{R}j\mathbf{R}'}(\omega)=\nonumber \\ \int_V d\mathbf{r} \int_V d\mathbf{r}' \phi_{i\mathbf{R}}^*(\mathbf{r})\phi_{i\mathbf{R}}(\mathbf{r})W(\mathbf{r},\mathbf{r}',\omega)\phi_{j\mathbf{R}'}^*(\mathbf{r}')\phi_{j\mathbf{R}'}(\mathbf{r}),
\end{align}
with $W(\mathbf{r},\mathbf{r}',\omega)$ the screened Coulomb interaction. Terms with $\mathbf{R}=\mathbf{R}'$ represent
the on-site Coulomb repulsion between two electrons, whereas
the case with $\mathbf{R}\neq\mathbf{R}'$ encodes the magnitude of longer-range, off-site Coulomb interactions. In principle one could also include more general, four-index Coulomb terms of the form $\int_V d\mathbf{r} \int_V d\mathbf{r}' \phi_{i\mathbf{R}}^*(\mathbf{r})\phi_{j\mathbf{R}}(\mathbf{r})W(\mathbf{r},\mathbf{r}',\omega)\phi_{k\mathbf{R}'}^*(\mathbf{r}')\phi_{l\mathbf{R}'}(\mathbf{r})$ in the Hamiltonian representing the active space, and indeed \emph{ab initio} downfolding
allows one to compute these integrals as well. However, these terms have typically found to be small~\cite{Romanova2023} and we will ignore them here. Moreover, the Hamiltonian of Eq.\,\eqref{eq:Hamiltonian} obtained through \emph{ab initio} downfolding contains hopping
and Coulomb terms between all neighbors on a lattice, and indeed some of the longer-range terms,
particularly next-nearest neighbor coupling, can be important in certain cases~\cite{doi:10.1126/science.aal5304,PhysRevB.35.3359}. 
Here we include nearest-neighbor terms only
for the hopping and Coulomb terms, as these
dominate over longer-range terms in the
Wannier representation, due to the exponential decay of Wannier functions~\cite{PhysRevB.76.165108}.

The Coulomb integral
of Eq.\,\eqref{eq:coulomb} is frequency-dependent, however, it
is a common approximation to take the static limit $\omega=0$
in the Hamiltonian of Eq.\,\eqref{eq:Hamiltonian}. While this
approximation has been shown to lead to an over-screening of
the Coulomb interactions~\cite{Scott2024}, we will utilize it here,
reserving a more rigorous treatment of dynamical effects for
a future study, as these have been shown to be necessary to account for in order to achieve truly predictive accuracy 
for ground- and excited-state observables within \emph{ab initio} downfolding~\cite{Romanova2023,Scott2024}. 
The screened Coulomb interaction $W$ appearing in the integral of Eq.\,\eqref{eq:coulomb}, and by extension the Hubbard term $U_{i\mathbf{R}j\mathbf{R}'}$, is computed within the
constrained random phase approximation (cRPA)~\cite{PhysRevB.70.195104}. The cRPA 
excludes the screening
of the states within the active space, as their Coulomb interactions
are explicitly included in the Hamiltonian of Eq.\,\eqref{eq:Hamiltonian}. 

\subsection{Tensor network VQE simulation}

We simulate the VQE classically by representing the wavefunction as a matrix product state (MPS)~\cite{Orus2014} within a recently proposed variational tensor network eigensolver (VTNE) approach~\cite{khan2023preoptimizing}. Specifically, following Ref.~\cite{khan2023preoptimizing}, we start from an initial state $\ket{\psi_o}$, which is set to be the ground state of the non-interacting ($U=0$) case, and we
apply
a variational ansatz to generate a parameterized quantum state. We design a parameterized quantum circuit (PQC)
as
\begin{equation}
    \label{eq:PQC}
    \ket{\psi_{\mathrm{PQC}}(\bm{\theta})}=U_n(\bm{\theta}_n)...U_1(\bm{\theta}_1)\ket{\psi_o},
\end{equation}
where the precise form of the operators $U_n$ is determined by the choice of variational ansatz used in our simulations. Each of these operators takes
as arguments a set of parameters $\bm{\theta}_n$, which are initialized randomly (or in some cases with a classical heuristic). The PQC is applied to a wavefunction which is represented as an MPS $\ket{\psi_{\chi}(\bm{\theta})}$ with bond dimension $\chi$, and we can therefore compute the energy function
\begin{equation}
    \label{eq:energy}
    E_{\chi}(\bm{\theta})=\bra{\psi_{\chi}(\bm{\theta})}H\ket{\psi_{\chi}(\bm{\theta})},
\end{equation}
with the Hamiltonian $H$ represented as a matrix product operator (MPO)~\cite{Orus2014}. 
Within our optimization scheme, we vary the parameters $\bm{\theta}$ of the PQC
to minimize the energy in Eq.~\eqref{eq:energy}. In order to reduce the chance of becoming stuck in a local minimum,
we perform ten independent optimizations, each with different
random starting parameters $\bm{\theta}_n$, and we report
the minimum energy in each case.  This type of data analysis for exploring an energy surface with many local minima was performed in a previous study~\cite{alvertis2024classicalbenchmarksvariationalquantum}. Additionally, following
this energy minimization, we use the ten resulting approximate
representations of the ground state for an optimization
which instead minimizes the infidelity with respect to
the DMRG solution, which we consider as the ground truth:
\begin{equation}
    \label{eq:infidelity}
    \text{IF}=1-|\bra{\Psi_{VQE}}\ket{\Psi_{DMRG}}|^2.
\end{equation}
We find that this additional step leads to greatly improved
energies and fidelities. 

Unless otherwise explicitly stated, we use the maximum bond dimension $\chi_{\rm max}=2^{n_q/2}$, where $n_q$ the number of qubits.
Since this bond dimension is sufficient to exactly represent an arbitrary wavefunction on $n_q$ qubits, the MPS representation yields the energy expectation value with respect to the exact PQC. 
For all tensor operations in this work we have used the ITensor software package~\cite{itensor}. We provide details
on our optimization procedure in Section\,\ref{optimization} of the
Appendix. In preparing our PQC, 
we utilize two ans\"{a}tze in this work, a number-preserving (NP) ansatz~\cite{Cade2020}, which was designed specifically within the context of solving single-band
Hubbard models, and a more generic excitation-preserving (EP)~\cite{Qiskit} ansatz, which allows us to straightforwardly prepare our circuits in the case of multi-band Hubbard 
models. The performance of both ans\"{a}tze for producing
accurate ground states has been discussed in Ref.~\cite{alvertis2024classicalbenchmarksvariationalquantum}. We apply the ans\"{a}tze to the solution
of the non-interacting ($U=0$) Hubbard representation of our systems, prior to which we apply $R_z(\theta)$ gates to each qubit, which improves the optimization procedure.

\subsection{Hamiltonian compression and the measurement problem}

Our aim is to obtain the ground state of the Hamiltonian of Eq.~\eqref{eq:Hamiltonian}, for studied materials with an active space represented by different sets of $U_{i\mathbf{R}j\mathbf{R}'}$ and $t_{i\mathbf{R}j\mathbf{R}'}$ matrices. Before presenting our
results, it is worth discussing some aspects of the complexity of this problem. 
For a Hubbard model on an $N_x\times N_y$ square lattice, with $N_b$ electronic bands included in the active space of the studied system, $n_q=2 N_x N_y N_b$ is the number of qubits needed to simulate its properties, where the factor $2$ accounts for spin. Denoting \( N = N_x N_y \), the number of terms appearing in the Hamiltonian of Eq.~\eqref{eq:Hamiltonian} is classified as follows, considering only nearest-neighbor contributions:

\begin{itemize}
    \item \textbf{Intra-band hopping \( t_{i\mathbf{R}i\mathbf{R}'} \)}: The number of hopping terms within the same band is \( 2 \times N_b \times (N_x (N_y - 1) + N_y (N_x - 1)) \). This accounts for nearest-neighbor pairs along both horizontal and vertical directions.

    \item \textbf{Inter-band hopping \( t_{i\mathbf{R}j\mathbf{R}} \)}: These are the hopping terms between different bands at the same site, resulting in \( 2 \times N_b \times (N_b - 1) \times N \) terms.

    \item \textbf{On-site potential \( t_{i\mathbf{R}i\mathbf{R}} \)}: For the on-site energies, there are \( 2 \times N_b \times N \) terms.

    \item \textbf{On-site interaction \( U_{i\mathbf{R}i\mathbf{R}} \) and \( U_{i\mathbf{R}j\mathbf{R}} \)}: There are \( N_b \times N \) on-site interaction terms for each band, with an additional \( N_b \times (N_b - 1) \times N \) terms for inter-band interactions.

    \item \textbf{Off-site interaction \( U_{i\mathbf{R}i\mathbf{R}'} \) and \( U_{i\mathbf{R}j\mathbf{R}'} \)}: These terms include both intra-band and inter-band interactions between different sites, considering only nearest neighbors. The total is \( N_b^2 \times (N_x (N_y - 1) + N_y (N_x - 1)) \).

     \item \textbf{Total number of terms}:
     \begin{align}
         n_{\text{terms}}=N_b\times \bigg[(N_b+2)[N_x\times (N_y-1)+\nonumber \\N_y\times (N_x-1)]+3N_bN \bigg]\nonumber
     \end{align}
\end{itemize}
We do not include inter-band hopping between different sites to ensure a manageable computational cost for the systems studied here. Boundary effects have been implicitly considered by restricting hopping and off-site interactions to existing nearest-neighbor pairs within the lattice. It becomes
clear that the compressed representation arising
from \emph{ab initio} downfolding, leads to a much
improved scaling of $N_b^2N_xN_y$ to leading order, compared to the $(N_{b,f}N_xN_y)^4$ scaling of the full many-body Hamiltonian. Note that here $N_{b,f}$ is the full number of bands of the system, and not only the ones
within an active space, and the total number of
terms appearing in the full-many body Hamiltonian is equal to $2(N_xN_yN_{b,f})^2+4N_xN_yN_{b,f})^4$.
Therefore, the compression scales as $N_b^2/(N_{b,f}^4N_x^3N_y^3)$. For the specific examples we study below, we precisely quantify the degree of Hamiltonian
compression achieved through our downfolding procedure. The number of bands $N_{b,f}$ entering the full many-body Hamiltonian is set to be equal to the number of occupied bands, plus any empty states that enter the active space
of the material within the downfolded representation. To also
enable a more fair comparison, we
give the number of terms of the
full many-body Hamiltonian when it
is restricted within the active space of the compressed representation. 

The compressed representation of materials becomes
particularly important when considering the problem
of measuring expectation values of their Hamiltonian on quantum hardware. For a system described my $M$ qubits,
depending on the specific measurement strategy, such as
measuring qubit-wise commuting terms simultaneously, measuring a family of non-crossing (NC) terms simultaneously, or performing a basis rotation (BR) grouping,
the lower bound for the number
of measurements required in order to determine the energy
for a single VQE energy evaluation scales as $M^2-M^6$~\cite{PhysRevResearch.4.033154,Clinton2024}. Although a more
detailed investigation is warranted, these lower bounds
suggest that it is conceivable,
when using methods such as NC and BR groupings with an idealized scaling close to
$M^2$ for the number of measurements, to achieve accurate simulations of strongly-correlated materials on near-term quantum hardware. Even for less favorable scaling in real devices, the compressed representations of our systems (no more than $M=54$ qubits) could be particularly important in this direction. 

Finally, it is worth highlighting
that when performing a finite amount of measurements
for different observables on real
quantum hardware, the expectation
values necessarily have a statistical uncertainty associated with them. Our classical tensor network simulation of VQE does not currently include shot noise as part of the simulation, but rather it represents the ultimate
accuracy that could be expected
from the simulation of the
compressed material Hamiltonians.

\subsection{Downfolded Hamiltonian simulation with near-term and fault-tolerant hardware}

It is worth emphasizing that the quantum simulation of materials using representations as the Hamiltonian of Eq.\,\eqref{eq:Hamiltonian} has been the subject of detailed benchmarks, and found
to likely be feasible with near-term resources, without necessitating fully fault-tolerant quantum computers~\cite{Clinton2024}. Indeed, as we outline in Section\,\ref{results} for the different materials studied
here, the number of qubits and two-qubit gates are well within the thresholds set in Ref.~\cite{Clinton2024} as
what might be feasible on near-term hardware. Moreover,
as also detailed in Section\,\ref{results}, by assuming that the fidelity of two-qubit gates is $99.9$\%~\cite{IBM}, we find that for all studied systems here
our circuit fidelities are in the range of $52-75$\% thanks
to our compression of the Hamiltonian, which places
them well above the $10$\% threshold that has been identified
for obtaining useful results~\cite{O’Brien2023}. We emphasize that while this is only meant to provide a rough estimate, it clearly demonstrates the importance of the
compression for reducing the measurement problem. 

Nevertheless, as we aim to build increasing complexity
into the reduced Hamiltonians obtained through downfolding or related methods, and approach the full
many-body limit, it will become necessary to utilize fault-tolerant architectures~\cite{agrawal2024quantifyingfaulttolerantsimulation}. However, our work here is mostly agnostic to this. We can obtain a rough estimate of the resources
that would be needed for a fault-tolerant simulation
of the downfolded Hamiltonians using the ansatz considered here, through computing different metrics. For example,
one may compute the 1-norm of the Hamiltonian of eq.\,\eqref{eq:Hamiltonian}:
\begin{align}
    \label{eq:1-norm}
    ||H||_1= \sum_{ij}\sum_{\mathbf{R}\mathbf{R}'}|\tilde{t}_{i\mathbf{R}j\mathbf{R}'}|+ \frac{1}{2}\sum_{ij}\sum_{\mathbf{R}\mathbf{R}'}|\tilde{U}_{i\mathbf{R}j\mathbf{R}'}|,
\end{align} 
where $\tilde{t},\tilde{U}$ are dimensionless, obtained
through dividing the hopping and Coulomb matrices by the
value of the dominant intra-band nearest-neighbor hopping. 
Our definition for the 1-norm includes 
summations over all neighbors, despite the fact that within our VQE simulations we only include nearest neighbor terms. The localized nature of the Wannier functions results in terms beyond
nearest neighbor being small, and our values for the 1-norm should be considered as an upper bound for the Hamiltonians we actually simulate.
The 1-norm provides a bound for the Trotter error~\cite{PhysRevX.11.011020},
and smaller values for this quantity are associated with
more efficient quantum phase estimation~\cite{Rocca2024}. 

For the purpose of fault-tolerant simulation it is also interesting to estimate the number of $T$-gates which will be necessary in order
to implement our circuits, where the one-qubit and two-qubit operators
are approximated with a required accuracy $\epsilon$. As described previously, we apply an $R_z$
gate to each qubit prior to the application of the unitary operators associated with the variational ansatz. 
The approximation of a one-qubit gate has been found to be possible with $1.15\log_2(1/\epsilon)$ $T$-gates~\cite{PhysRevLett.114.080502}. 
From this we have $n_q\times1.15\log_2(1/\epsilon)$ $T$-gates associated with this step. Additionally, both the NP and EP ans\"atze involve the application of two-qubit gates with two parameters each. For a total of $n_{\text{params}}$ parameters in the variational ansatz we have $n_{\text{params}}/2$ two-qubit gates, each of which requires  $\mathcal{O}[16\log(1/\epsilon)+32]$ $T$-gates to represent~\cite{PhysRevA.107.042424}.
Therefore, we have a total of $n_q\times1.15\log_2(1/\epsilon)n+\frac{n_{\text{params}}}{2}\times\mathcal{O}[16\log(1/\epsilon)+32]$  $T$-gates required for the fault-tolerant simulation of our systems with accuracy $\epsilon$.

\section{Results}
\label{results}

\begin{figure}[tb]
    \centering
    \includegraphics[width=0.8\linewidth]{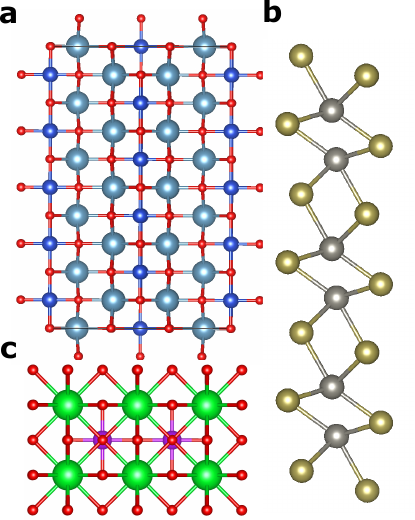}
    \caption{Structures of the systems studied in this work; $\text{Ca}_2\text{CuO}_3$ (panel \textbf{a}), $\text{WTe}_2$ (panel \textbf{b}), and $\text{SrVO}_3$ (panel \textbf{c}). Ca atoms are given in gray, Cu in blue, O in red, Sr in green, V in purple, W in silver and Te in gold.}
    \label{fig:structures_vqe}
\end{figure}

\begin{table*}[tb]
\centering
  \setlength{\tabcolsep}{6pt} 
\begin{tabular}{cccccc}
\hline
system & lattice size & $N_b$  & DMRG energy (eV) & VQE energy (eV)  & Fidelity  \\
\hline
$\text{Ca}_2\text{CuO}_3$ & $10\times 1$ & $1$ & $6.005$ & $6.028$ & $99.3\%$ \\
$\text{WTe}_2$ & $2\times 2$ & $4$ & $115.029$ & $115.097$ & $96.2\%$ \\
$\text{SrVO}_3$ & $3\times 3$ & $3$ & $-105.383$ & $-105.365$ & $31.8\%$ \\
\hline
\end{tabular}
\caption{Studied systems, simulated lattice sizes, number of bands in the subspace, DMRG energies, and best VQE energies and fidelities with respect to the DMRG solution. As outlined in the main text, we solve the Hubbard model representation of $\text{Ca}_2\text{CuO}_3$ at half-filling, of $\text{WTe}_2$ at full filling of the valence bands, and of $\text{SrVO}_3$ at half-filling of the lowest band.}
\label{table:results_vqe}
\end{table*}

\begin{table*}[tb]
\centering
  \setlength{\tabcolsep}{6pt} 
\begin{tabular}{cccccc}
\hline
system & $n_q$ & $n_{2q,G}$  & circuit fidelity & $||H||_1$  & $n_{\text{terms}}$  \\
\hline
$\text{Ca}_2\text{CuO}_3$ & $20$ & $290$ & $74.8$\% & $2.67\times10^2$ & $37$ \\
$\text{WTe}_2$ & $32$ & $652$ & $52.1$\% & $3.31\times10^2$ & $288$ \\
$\text{SrVO}_3$ & $54$ & $484$ & $55.8$\% & $2.315\times10^3$ & $423$ \\
\hline
\end{tabular}
\caption{Parameters relating to the near-term and fault-tolerant quantum simulation of the studied systems. Specifcally, we give the number of qubits ($n_q$) required to simulate the downfolded Hamiltonians, the number of two-qubit gates $n_{2q,G}$ in our circuits, and the respective circuit fidelity if one assumes a $99.9\%$ fidelity for the individual gates, as well as the 1-norm and the number of terms of the downfolded Hamiltonians.}
\label{table:resources_vqe}
\end{table*}

Fig.\,\ref{fig:structures_vqe} visualizes the structures
of the systems we study here. These are chosen to
demonstrate the capability of our approach to correctly
predict the ground state properties of diverse strongly-correlated materials; $\text{Ca}_2\text{CuO}_3$ is a quasi-1D cuprate known
to display antiferromagnetic behavior along the Cu-O chains~\cite{Rosner1997}; $\text{WTe}_2$
has been proposed to host an excitonic ground state, \emph{i.e.}, one where correlated electron-hole pairs form spontaneously~\cite{Sun2022,Jia2022}; and $\text{SrVO}_3$ is a correlated metal that exhibits substantial charge ordering~\cite{Aizaki2012,Zhang2016}. Table\,\ref{table:results_vqe} summarizes the number of bands included in the active space of each system, the lattice size on which
the downfolded Hamiltonian is solved, the DMRG energy, and the best value for the energy
and fidelity obtained from the ten independent VQE optimizations we perform. For $\text{Ca}_2\text{CuO}_3$ we only perform a single optimization, as the simpler energy landscape of this system makes this sufficient for finding a low-energy, high-fidelity solution. More details on the optimization for each system are given below. In Section\,\ref{optimization} of the Appendix we also give the VQE energy and fidelity values for all optimizations performed for $\text{WTe}_2$ and $\text{SrVO}_3$.
Moreover, we summarize details
of all DFT and cRPA calculations employed for these systems in Section\,\ref{computational_details} of the Appendix. We now discuss the properties of the
ground state wavefunctions of the materials of Fig.\,\ref{fig:structures_vqe}
as obtained within our combined \emph{ab initio} downfolding/VQE approach. 

\subsection{Antiferromagnetism in $\text{Ca}_2\text{CuO}_3$}

\begin{figure}[tb]
    \centering
    \includegraphics[width=\linewidth]{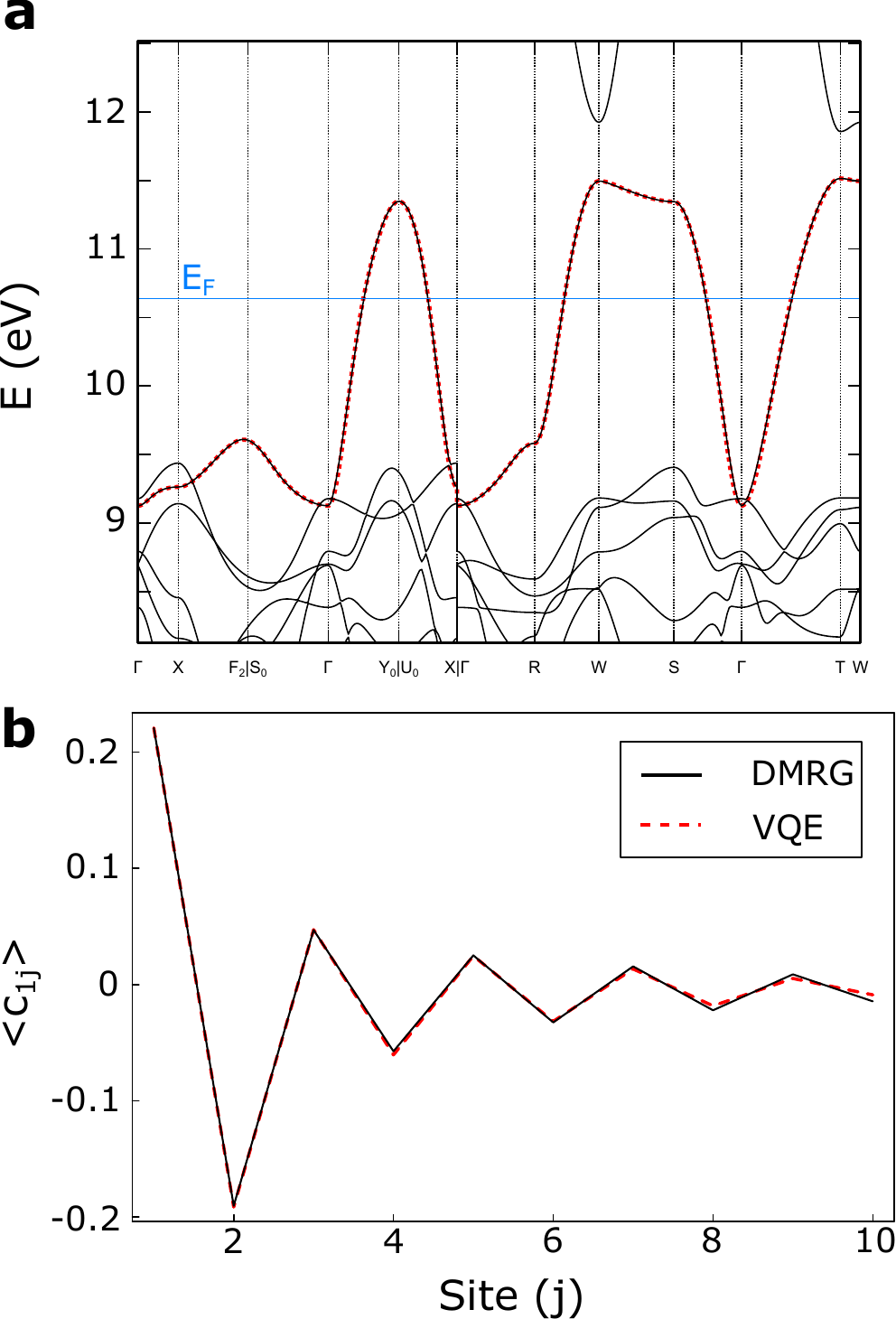}
    \caption{Band structure of $\text{Ca}_2\text{CuO}_3$ (panel \textbf{a}), with the DFT bands given in black and the Wannier-interpolated bands in red. The spin correlation function of the ground state of downfolded $\text{Ca}_2\text{CuO}_3$ (panel \textbf{b}) indicates clear antiferromagnetic behavior.}
    \label{fig:ca2cuo3}
\end{figure}

Within the bulk structure of $\text{Ca}_2\text{CuO}_3$,
Cu atoms form one-dimensional chains connected by O atoms,
as seen in Fig.\,\ref{fig:structures_vqe}a, which results
in well-known antiferromagnetic behavior.
The band structure of this system is visualized in Fig.\,\ref{fig:ca2cuo3}a, as obtained within DFT calculations at the PBE level~\cite{pbe}. We downfold the electronic
structure of this system onto the active space
consisting of the highest occupied bands, which
has  strong contributions from Cu $d$-orbitals. Wannier interpolation
yields a band structure within this active space (red), which
is in excellent agreement with the full DFT calculations. The derived Hubbard model parameters clearly demonstrate the
1D character of the system, with a dominant nearest-neighbor
hopping term of $t=-0.491$\,eV and nearest neighbor Coulomb repulsion $V=0.903$\,eV along
a single spatial direction.
The on-site Coulomb
repulsion was found to be $U=3.578$\,eV, in good agreement with previous estimates~\cite{Rosner1997}.

We solve for the electronic ground state of $\text{Ca}_2\text{CuO}_3$, by performing VQE simulations
of a one-dimensional, one-band Hubbard model using the above parameters, for a $10\times 1$ lattice at half-filling, \emph{i.e.} we perform a $20$-qubit simulation of a Hamiltonian with $37$ terms ($20$ terms corresponding to the
on-site potential of the single band are discarded since we
can define $t_{i\mathbf{R}i\mathbf{R}}=0$), compared to $2.83\times10^{10}$ terms in the full many-body Hamiltonian, and $4.02\times10^{4}$ terms in the many-body Hamiltonian, when restricted within the active space. 
We use ten layers of the NP ansatz to represent
our wavefunction, corresponding to $580$
variational parameters, corresponding to $290$ two-qubit gates. If we assume a $99.9$\% fidelity for two-qubit gates on near-term hardware, this suggests a $74.8$\% circuit fidelity for a single VQE energy measurement. The 1-norm of the downfolded Hamiltonian in this system is found to be $||H||_1=2.67\times10^2$, which is much lower than typical values of full many-body Hamiltonians~\cite{chamaki2022selfconsistentquantumiterativelysparsified},
thus indicating the scalability of our approach to fault-tolerant architectures. These parameters, which relate to
the near-term and fault-tolerant simulation of $\text{Ca}_2\text{CuO}_3$, and also of the other materials studied here, are summarized in Table\,\ref{table:resources_vqe}.

Our tensor network simulation of VQE leads to a ground state energy within $23$\,meV of the DMRG energy,
and with a fidelity $\mathcal{F}=|\bra{\Psi_{VQE}}\ket{\Psi_{DMRG}}|^2=99.3\%$.
Moreover, we compute the spin correlation function
\begin{equation}
    \langle C_{ij} \rangle = \langle S_i^z S_j^z \rangle - \langle S_i^z \rangle \langle S_j^z \rangle,
\end{equation}
and we plot $\langle C_{1j} \rangle$ in Fig.\,\ref{fig:ca2cuo3}b. The alternating sign of the
spin correlation function indicates clear antiferromagnetic
behavior for $\text{Ca}_2\text{CuO}_3$, as expected, and we see that
VQE is in near-perfect agreement to DMRG.

\subsection{Excitonic ground state in $\text{WTe}_2$}

\begin{figure}[tb]
    \centering
    \includegraphics[width=\linewidth]{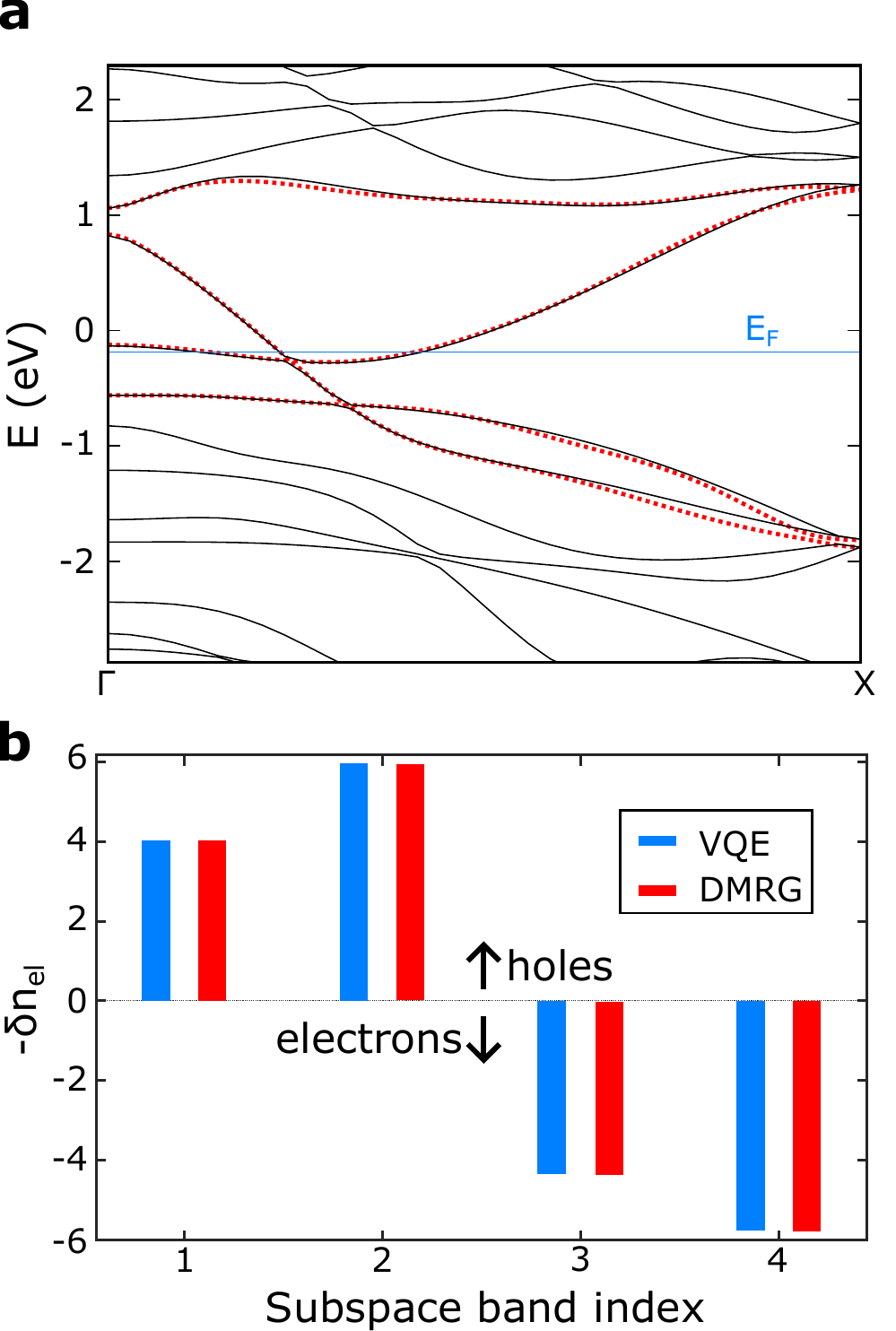}
    \caption{Band structure of $\text{WTe}_2$ (panel \textbf{a}), with the DFT bands given in black and the Wannier-interpolated bands in red. 
    Difference $-\delta n_{el}$ (panel \textbf{b}) in the total number of electrons within the bands of
    the strongly-correlated ground state within VQE and DMRG, from the
    number of electrons in the bands within the conventional band-insulating state. Positive values indicate hole formation, while negative values suggest excess electrons compared to the band insulating case.}
    \label{fig:wte2}
\end{figure}

Monolayer $\text{WTe}_2$ is a two-dimensional
system, visualized in Fig.\,\ref{fig:structures_vqe}b, which has been proposed to be an
excitonic insulator~\cite{Sun2022,Jia2022}, \emph{i.e.}, to
host correlated electron-hole pairs in its
ground state. The band structure of this
system as obtained within semi-local DFT is
visualized in Fig.\,\ref{fig:wte2}a, in good
agreement with previous reports~\cite{Sun2022}. Ref.~\cite{Sun2022} included a small amount of exact exchange
in the DFT functional in order to induce a small
gap around the Fermi level. Additionally, Ref.~\cite{PhysRevB.110.075133} demonstrated that a gap around the Fermi level opens
when a static $GW$ approximation is used. In both cases, the subsequent
solution of the so-called Bethe-Salpeter equation for excitonic states~\cite{Rohlfing1998,Rohlfing2000}
yields bound excitons of negative energy~\cite{Sun2022,PhysRevB.110.075133}, hence
suggesting that the band insulator ground state predicted by DFT and/or $GW$ is unstable towards an excitonic one.  

Here we downfold the electronic structure on
the subspace of four bands around the Fermi level (red in Fig.\,\ref{fig:wte2}a). We were
not able to perfectly reproduce the Kohn-Sham
band structure with our computed Wannier-interpolated bands, with small deviations persisting towards the band edges. However, since
this region is far from the Fermi level, it
does not affect the formation of bound excitons.
The crossing of the second and third bands
of the subspace near the Fermi level is
immediately suggestive of the possibility of
exciton formation. To verify this, we obtain
the ground state of the downfolded four-band
extended Hubbard model within this subspace,
the parameters for which are given in Section\,\ref{hamiltonians} of the Appendix. Given the insulating character of
this material, we solve our Hamiltonian at full-filling of the lower two states for a $2\times 2$ lattice, corresponding to a $2\times 2\times 4\times 2 = 32$-qubit simulation of
a Hamiltonian with $288$ terms, compared to $6.29\times 10^8$ terms in the full many-body Hamiltonian, and $2.63\times 10^5$ terms in the many-body Hamiltonian when restricted within the active space. 
We
limit the bond dimension to $\chi=512$, and we utilize twenty layers of the EP ansatz, which leads to $1,304$ variational optimization parameters, corresponding to $652$ two-qubit gates and a $52.1$\% circuit fidelity, if one were to assume a $99.9$\% fidelity for the individual gates. The 1-norm of the downfolded Hamiltonian is $||H||_1=3.31\times10^2$, which here too is significantly smaller than values of the order to $10^4$, which are obtained even for small chemical systems~\cite{chamaki2022selfconsistentquantumiterativelysparsified}. 
The obtained
ground state energy from VQE is within $68$\,meV of the DMRG value, while the ground state VQE wavefunction has a fidelity of $96.2\%$.
The difference $-\delta n_{el}$ in the total number of electrons within the bands of
the strongly-correlated ground state we obtain with VQE (and with DMRG), from the
number of electrons in the bands within the conventional band-insulating state (\emph{i.e.}, full-filling of the two lower-lying bands within the subspace, while the two higher-lying bands are empty), is visualized in Fig.\,\ref{fig:wte2}b.
The first observation to make here is that
the DMRG and the VQE predictions are in excellent agreement. Moreover, it becomes
evident that in our reduced system of $16$
electrons ($4$ lattice sites, times $2$ spin directions, times $2$ fully occupied electronic bands when we initialize), a significant lack of electrons occurs in the
two lower-lying bands, \emph{i.e.}, positively charged holes form in the ground
state. At the same time, a very substantial electronic
population resides in the bands above the Fermi
level (\emph{i.e.}, in the ``conduction'' bands).
This presence of electrons in the ``conduction'' bands and holes in the  ``valence'' bands is a hallmark of
excitonic insulating behavior. To further
quantify the formation of exciton pairs, we
compute the excitonic insulator order parameter~\cite{Kaneko2012}
\begin{equation}
\label{eq:EI_order_parameter}
    \Delta = \frac{U'}{N_x N_y} \sum_{x,y,v,c,\sigma}  \langle \psi \vert C^{\dagger}_{c, x, y, \sigma} C_{v, x, y, \sigma} \vert \psi \rangle
\end{equation}
where $v,c$ denote valence and conduction states
respectively, and $U'$ the inter-orbital, on-site
Coulomb repulsion between electrons. 
Here we define $U'$ as the average on-site, inter-orbital repulsive interaction between electrons, as these terms are given in Section\,\ref{hamiltonians} of the Appendix.
For a $2\times 2$ lattice we find $\Delta=0.379$, which is underestimated compared to the order parameter obtained using DMRG $\Delta_{DMRG}=0.640$ on
the same lattice. The fact that VQE perfectly reproduces the DMRG band occupations, however leads to a finite error
in the calculation of the order parameter is
in agreement with previous observations on the Hubbard model, where it was found that even high-fidelity solutions can struggle
to capture correlation functions~\cite{anselmemartinSimulatingStronglyInteracting2022}, although our additional step of an overlap-based optimization substantially improves their accuracy~\cite{alvertis2024classicalbenchmarksvariationalquantum}.  Overall, for a small-sized lattice representation of $\text{WTe}_2$, 
our combined \emph{ab initio} downfolding/VQE
approach captures the previously reported excitonic
ground state of this system, entirely from first principles.

\subsection{Charge-ordered state in $\text{SrVO}_3$}

\begin{figure}[tb]
    \centering
    \includegraphics[width=\linewidth]{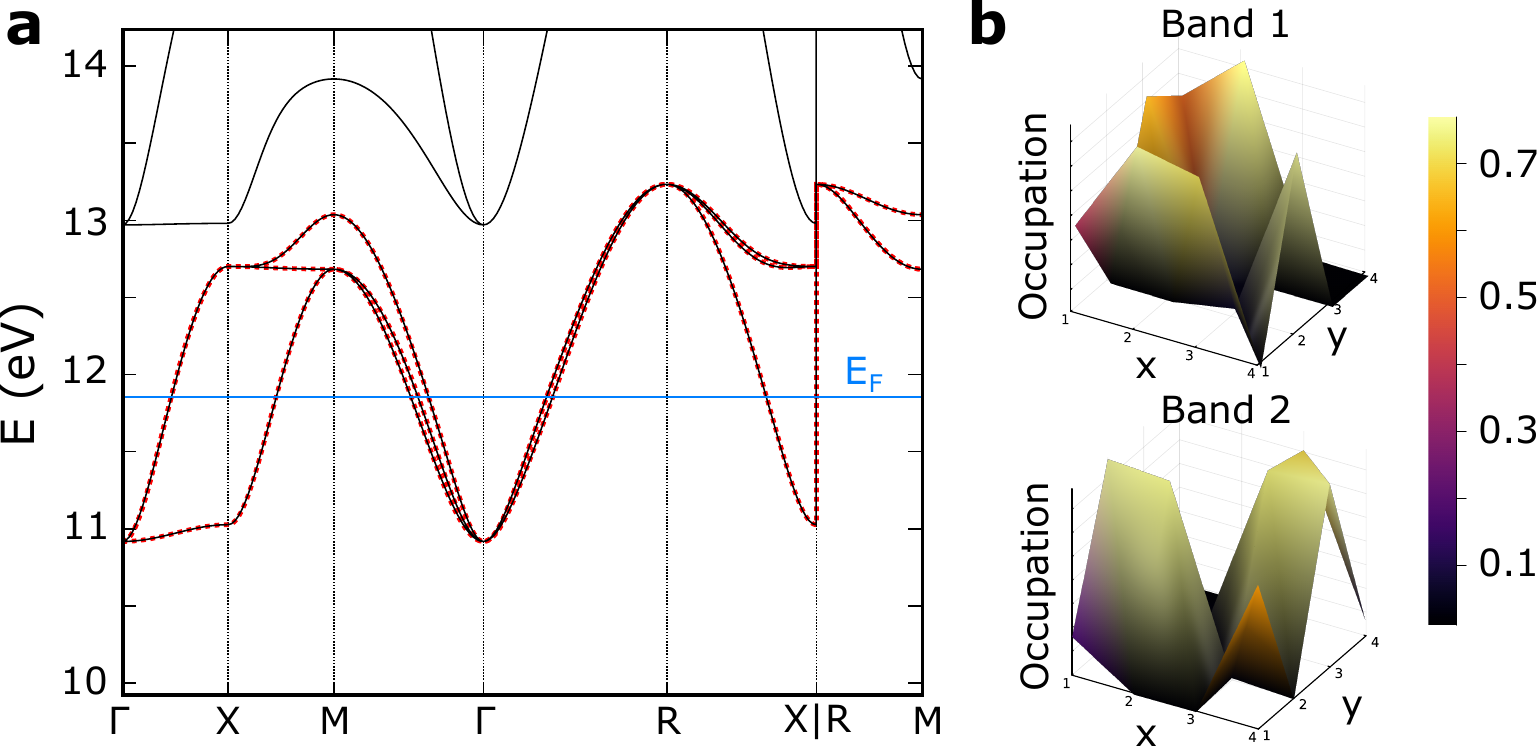}
    \caption{Band structure of $\text{SrVO}_3$ (panel \textbf{a}), with the DFT bands given in black and the Wannier-interpolated bands in red. The lattice site occupations in the first two bands of the $\text{SrVO}_3$ ground state (panel \textbf{b}) indicate a clear charge ordering.}
    \label{fig:srvo3}
\end{figure}

The band structure of $\text{SrVO}_3$ is visualized in Fig.\,\ref{fig:srvo3}a. The Fermi
level crosses the subspace of the three vanadium $d$-bands, and hence semi-local DFT
suggests that this system is a regular metal.
However, $\text{SrVO}_3$ is known to be a correlated metal and to exhibit charge order~\cite{Aizaki2012,Zhang2016}. In order to obtain these signatures of
strong correlations with our VQE approach, 
we downfold the electronic structure onto
the active space of the three bands crossing
the Fermi level (visualized in red in Fig.\,\ref{fig:srvo3}a), and we obtain a
three-band extended Hubbard model, the parameters for which are given in Section\,\ref{hamiltonians} of the Appendix, and which are in close agreement to
previous downfolding calculations for this system~\cite{Kurita2023}. 

We perform VQE simulations for a $3\times 3$
lattice of this three-band Hubbard Hamiltonian with $423$ terms, compared to $7.34\times10^9$ terms in the full many-body Hamiltonian, and $2.13\times10^6$ terms in the many-body Hamiltonian, when restricted to the active space. In the compressed representation, our system is described using $3\times3\times3\times2=54$ qubits. We apply ten layers of the EP ansatz, which requires to $1,168$ variational parameters, with a bond dimension of $\chi=512$. This corresponds to $584$ two-qubit gates, and assuming a fidelity of $99.9$\% for each one, suggests a $55.8$\% circuit fidelity on near-term hardware. The 1-norm of the downfolded Hamiltonian is $||H||_1=2.315\times10^3$, which makes this system somewhat more challenging compared to the ones compared previously in terms of its potential simulation on fault-tolerant hardware. 
While we were unable to obtain a VQE ground state of this complex system with a fidelity higher than $31.8\%$, our solution has an energy which is only $18$\,meV above that of the DMRG solution, and as we will see, qualitatively reproduces key manifestations of strong electronic interactions.  
In
order to quantify the charge ordering in the
ground state wavefunction, we compute the
charge disproportionation parameter
\begin{equation}
    \label{eq:CO_parameter}
    \Phi = \frac{\left|\sum_{A} - \sum_{B}\right|}{N_x \cdot N_y},
\end{equation}
where $\sum_{A},\sum_{B}$ the total charge
in two sublattices, which we define as $A$ and $B$ respectively. The total charge within a sublattice is obtained as $\sum_{A,B} = \sum_{b} \sum_{x} \sum_{y}\sum_{\sigma} n(x, y, b, \sigma)$, with $b$ a band index and a site belonging to sublattice $A$ if $(x + y) \mod 2 = 0$, and to sublattice $B$ otherwise. We find significant charge disproportionation with an order parameter $\Phi=0.21$, which is however overestimated compared to DMRG simulations on the same lattice ($\Phi_{DMRG}=0.12$). In Fig.\,\ref{fig:srvo3}b we visualize the
charge distribution of the ground state wavefunction obtained with DMRG on a $4\times4$
lattice, within the first two bands of the subspace, as this case most clearly illustrates
the significant charge ordering which occurs. However, even for the $3\times3$ lattices
studied within VQE, we make qualitatively
very similar observations.

\section{Conclusions and outlook}
In this work we have utilized \emph{ab initio} downfolding to generate compressed representations of strongly-correlated materials. We have shown that near-term quantum algorithms, such as the VQE, can produce the ground states of these compressed Hamiltonians and yields the expected behavior, at a modest computational cost and also in quantitative agreement with DMRG. We apply our approach to the antiferromagnetic quasi-1D cuprate $\text{Ca}_2\text{CuO}_3$, the excitonic insulator $\text{WTe}_2$, and the correlated metal $\text{SrVO}_3$, and we correctly predict the key physics of
these systems, while semi-quantitatively reproducing the ground state energy as 
obtained using DMRG. This highlights the potential of our
approach to utilize current and emerging quantum computing
technologies in order to accurately predict the ground-state
properties of diverse strongly-correlated materials, entirely from first quantum mechanical principles. Additionally, our classical simulation of VQE may be used as a pre-optimization step to the simulation of strongly-correlated materials on quantum hardware, hence providing an excellent starting point and minimizing the quantum resource cost~\cite{khan2023preoptimizing,gustafson2024surrogateoptimizationvariationalquantum}. 

Our work raises several questions and highlights the
importance of pursuing various avenues of research. The
variational approaches we employ here can struggle to quantitatively reproduce the
true ground state energy in the case of 
systems downfolded on two-dimensional lattices, it will therefore be important to
explore flexible ans\"{a}tze as ADAPT-VQE~\cite{Grimsley2019,mullinax2024classical} for such problems moving forward. This will be particularly important as we are moving towards the fault-tolerant era~\cite{agrawal2024quantifyingfaulttolerantsimulation} and
quantum computer architectures with more capabilities, 
making it possible to describe 
increasingly complex materials, using our and similar approaches. 
Moreover, while here
we have employed a static approximation for the Coulomb
interactions within the active space, it will be interesting to perform a rigorous treatment of the frequency-dependence, as has been described in recent works~\cite{Romanova2023,Scott2024}. Furthermore, 
here we have employed the DFT
generalized gradient approximation as a starting 
point for downfolding; it will be interesting
to explore the influence of the starting point moving
forward, as it has been shown previously that optimal DFT
starting points can result in improved descriptions of
correlated states with higher levels of electronic structure
calculations~\cite{Wing2021,Ohad2023}. Finally, while here we have accounted 
only for electronic screening on Coulomb interactions, 
lattice motions have recently been shown to be capable
of substantially modifying the screening, particularly in
polar materials~\cite{Filip2021,alvertis2023phonon}, which could lead to modified downfolded Hamiltonians and phase diagrams~\cite{Nomura2015,VanLoon2021,tubman2025theoryabinitiodownfolding,tubman2025phononmediatedelectronattractionsrtio3}; we reserve a detailed discussion
of these effects for a future study. 

\section{Acknowledgments}
This material is based upon work supported by the U.S. Department of Energy, Office of Science, National Quantum Information Science Research Centers, Superconducting Quantum Materials and Systems Center (SQMS) under contract No. DE-AC02-07CH11359.  We are grateful for support from NASA Ames Research Center.   This research used resources of the National Energy Research
Scientific Computing Center, a DOE Office of Science User Facility supported by the Office of Science of the U.S. Department of Energy under Contract No. DE-AC02-05CH11231 using NERSC awards HEP-ERCAP0029167 and DDR-ERCAP0029710.

\appendix
\section{Details of VQE optimization}
\label{optimization}
Following Ref.~\cite{khan2023preoptimizing}, 
we start by optimizing for the ground state
of the non-interacting system ($U=0$). We
use this state $\ket{\psi_o}$ to generate the parameterized quantum circuit
\begin{equation}
    \label{eq:PQC}
    \ket{\psi_{\mathrm{PQC}}(\bm{\theta})}=U_n(\bm{\theta}_n)...U_1(\bm{\theta}_1)\ket{\psi_o}
\end{equation}
as the starting point for the optimization of
the interacting case, representing the real
material of interest.  
The initial parameters $\boldsymbol{\theta}$ 
are obtained randomly from a Gaussian distribution 
$\mathcal{N}(0,10^{-5})$ with zero mean and 
variance $\sigma^2 = 10^{-5}$. 
We perform an energy minimization, until one
of three conditions is met: the energy 
tolerance 
(defined by the absolute difference between the 
energy at the final step and the
penultimate step)  reaches $10^{-7}$, the energy 
gradient reaches $10^{-6}$, or the optimization 
reaches $500$ steps. The optimizations are performed using the L-BFGS 
method~\cite{nocedal1999numerical}. 

In order to obtain ground states
with high fidelities, following the energy minimization described above, we use the resulting
wave functions as a starting point for an overlap-based optimization, where the optimizer minimizes
the loss function $f=\log_{10}(1-|\bra{\Psi_{VQE}}\ket{\Psi_{DMRG}}|^2)$ (the logarithm of the infidelity with respect to the DMRG ground state). We find that this hybrid optimization
strategy yields the overall best results in terms
of minimizing the energy and also producing a high-fidelity ground state. We perform
ten independent optimizations following the above hybrid energy-/overlap-based minimization, and we take
the state with the minimal energy among these ten
as the ground state, which prevents the system
from becoming stuck in local minima. 

An important factor in the ground state optimization is the electron filling of the
bands of the different materials studied here. $\text{Ca}_2\text{CuO}_3$ and $\text{SrVO}_3$ are
metallic at the DFT level of theory, and we
solve for their ground state at half-filling of 
the first electronic band included in the model. $\text{WTe}_2$ is predicted to be a conventional band insulator
at the Kohn-Sham level of theory, we therefore
solve for the ground state of this system with an initial state where the two lower-energy bands in the active space (see Fig.\,3 in main manuscript) are at full-filling, and
the two upper bands are empty. Since here we work in the basis of Wannier functions, we populate the
Wannier states with the greatest contribution
from the two lower-energy Kohn-Sham states, averaged across the Brillouin zone, as can be deduced from the Wannier rotation matrices obtained within Wannier90~\cite{Pizzi2020}. 

In Tables\,\ref{tab:WTe2_energies}-\ref{tab:SrVO3_fidelities} we give the VQE energies and fidelities obtained for $\text{WTe}_2$ and $\text{SrVO}_3$, ranked from best to worst from right to left.

\begin{table*}[tb]
\footnotesize
\centering
\begin{tabular}{|c|c|c|c|c|c|c|c|c|c|}
\hline
 VQE 1 & VQE 2 & VQE 3 & VQE 4 & VQE 5 & VQE 6 & VQE 7 & VQE 8 & VQE 9 & VQE 10 \\
\hline
$115.097$\,eV & $117.327$\,eV & $118.903$\,eV & $122.972$\,eV & $129.077$\,eV & $130.034$\,eV & $131.099$\,eV & $133.006$\,eV & $134.170$\,eV & $134.216$\,eV \\
\hline
\multicolumn{10}{|c|}{DMRG Energy: $115.029$\,eV} \\
\hline
\end{tabular}
\caption{VQE energies of $\text{WTe}_2$.}
\label{tab:WTe2_energies}
\end{table*}

\begin{table*}[tb]
\footnotesize
\centering
\begin{tabular}{|c|c|c|c|c|c|c|c|c|c|}
\hline
 VQE 1 & VQE 2 & VQE 3 & VQE 4 & VQE 5 & VQE 6 & VQE 7 & VQE 8 & VQE 9 & VQE 10 \\
\hline
$96.2\%$ & $0\%$ & $0\%$ & $0\%$ & $0\%$ & $0\%$ & $0\%$ & $0\%$ & $0\%$ & $0\%$\\
\hline
\end{tabular}
\caption{VQE fidelities of $\text{WTe}_2$ with respect to DMRG reference.}
\label{tab:WTe2_fidelities}
\end{table*}

\begin{table*}[tb]
\centering
\small
\begin{tabular}{|c|c|c|c|c|c|c|c|c|c|}
\hline
 VQE 1 & VQE 2 & VQE 3 & VQE 4 & VQE 5 & VQE 6 & VQE 7 & VQE 8 & VQE 9 & VQE 10 \\
\hline
$-105.365$\,eV & $-105.363$\,eV & $-105.298$\,eV & $-105.260$\,eV & $-105.167$\,eV & $-105.120$\,eV & $-101.833$\,eV & $-84.214$\,eV & $-84.094$\,eV & $-82.473$\,eV\\
\hline
\multicolumn{10}{|c|}{DMRG Energy: $-105.383$\,eV} \\
\hline
\end{tabular}
\caption{VQE energies of $\text{SrVO}_3$.}
\label{tab:SrVO3_energies}
\end{table*}

\begin{table*}[tb]
\footnotesize
\centering
\begin{tabular}{|c|c|c|c|c|c|c|c|c|c|}
\hline
 VQE 1 & VQE 2 & VQE 3 & VQE 4 & VQE 5 & VQE 6 & VQE 7 & VQE 8 & VQE 9 & VQE 10 \\
\hline
$31.8\%$ & $11.7\%$ & $0\%$ & $0\%$ & $0\%$ & $0\%$ & $0\%$ & $0\%$ & $0\%$ & $0\%$\\
\hline
\end{tabular}
\caption{VQE fidelities of $\text{SrVO}_3$ with respect to DMRG reference.}
\label{tab:SrVO3_fidelities}
\end{table*}

\section{Details of DFT and constrained RPA calculations}
\label{computational_details}
We perform all DFT calculations within the Quantum Espresso software package~\cite{QE}, within the generalized gradient approximation (GGA) of Perdew, Burke and Ernzerhof (PBE)~\cite{pbe}.
We utilize scalar-relativistic optimized norm-conserving Vanderbilt pseudopotentials (ONCV)~\cite{Hamann2013} with standard accuracy,
as these are given in Pseudo Dojo~\cite{VanSetten2018}.

For our DFT calculations on $\text{Ca}_2\text{CuO}_3$ we employ a wave function
cutoff of $80$\,Ry, and a $6\times 6\times 6$ k-grid. We compute the dielectric function and the Coulomb integrals of the
system within RESPACK~\cite{Nakamura2021}, using
a polarizability cutoff of $7$\,Ry and $100$ bands, excluding a single band crossing the Fermi level within cRPA. For $\text{WTe}_2$ we use a wave function
cutoff of $80$\,Ry, and a $6\times 6\times 1$ k-grid. We compute the dielectric function within cRPA by excluding the four bands around the Fermi level, with a polarizability cutoff of $5$\,Ry and $600$ bands. For $\text{SrVO}_3$ we use a wave function
cutoff of $81$\,Ry, and a $6\times 6\times 6$ k-grid. We compute the dielectric function within cRPA by excluding the three bands crossing the Fermi level, with a polarizability cutoff of $5$\,Ry and $600$ bands, yielding Coulomb parameters in close agreement to those reported previously~\cite{Nakamura2021}.

\section{Downfolded Hamiltonian parameters}
\label{hamiltonians}
Here we give the parameters resulting from
downfolding the electronic structure of the
different materials onto the extended Hubbard
Hamiltonian of eq.\,1 of the main manuscript. Here we give the nearest-neighbor terms along the crystallographic direction where the maximal coupling and interactions occur, for each system. 

\subsection{$\text{Ca}_2\text{CuO}_3$}
The hopping and Coulomb terms of this system
are dominant along the crystallographic direction which aligns with chains of Cu atoms. We therefore construct a one-dimensional, single-band Hubbard model with the following parameters resulting from the Wannierization and downfolding procedures: hopping integral of $t=-0.491$\,eV, on-site Coulomb interaction of $U=3.578$\,eV and off-site Coulomb repulsion of $V=0.903$\,eV.

\subsection{$\text{WTe}_2$}
Within the subspace 
of four bands around the Fermi level of $\text{WTe}_2$, we find
the hopping term (all values in eV)
\begin{equation}
    t_{i\mathbf{R}j\mathbf{R}'}=\begin{pmatrix}
-0.201 & 0.178 & -0.398 & -0.128 \\
0.108 & -0.144 & 0.072 & -0.071 \\
0.398 & 0.003 & 0.387 & 0.025 \\
0.019 & 0.071 & 0.057 & 0.124 \\
\end{pmatrix}.
\end{equation}
with $i=j$ the intra-band terms, and $i\neq j$ the
inter-band contributions, for $\mathbf{R}, \mathbf{R}'$ corresponding to nearest neighbors.

Similarly, for the on-site Coulomb interaction
\begin{equation}
    U_{ij}=\begin{pmatrix}
1.107 & 0.822 & 0.922 & 0.765 \\
0.822 & 1.095 & 0.760 & 0.684 \\
0.922 & 0.760 & 1.096 & 0.853 \\
0.765 & 0.684 & 0.853 & 1.174 \\
\end{pmatrix},
\end{equation}
and the nearest-neighbor off-site terms:
\begin{equation}
    V_{ij}=\begin{pmatrix}
0.924 & 0.822 & 0.841 & 0.765 \\
0.754 & 0.917 & 0.715 & 0.672 \\
0.841 & 0.760 & 0.855 & 0.853 \\
0.721 & 0.672 & 0.762 & 0.860 \\
\end{pmatrix}.
\end{equation}

\subsection{$\text{SrVO}_3$}
$\text{SrVO}_3$ has cubic symmetry, making the Hamiltonian parameters
identical along the three crystallographic axes. We find within the subspace of the three electronic bands 
crossing the Fermi level that we have the following
intra- and inter-band terms, where all values are given in eV. For the hopping term
\begin{equation}
    t_{i\mathbf{R}j\mathbf{R}'}=\begin{pmatrix}
-0.263 & 0 & 0 \\
0 & -0.263 & 0 \\
0 & 0 & -0.027 \\
\end{pmatrix}.
\end{equation}
The on-site Coulomb interaction
\begin{equation}
    U_{ij}=\begin{pmatrix}
3.527 & 2.349 & 2.349 \\
2.349 & 3.527 & 2.349 \\
2.349 & 2.349 & 3.527 \\
\end{pmatrix},
\end{equation}
and the nearest-neighbor off-site terms:
\begin{equation}
    V_{ij}=\begin{pmatrix}
0.649 & 0.635 & 0.555 \\
0.635 & 0.649 & 0.555 \\
0.555 & 0.555 & 0.492 \\
\end{pmatrix},
\end{equation}
where here too all values are given in eV.

\textbf{}
\bibliography{references}

\begin{thebibliography}{88}%
\makeatletter
\providecommand \@ifxundefined [1]{%
 \@ifx{#1\undefined}
}%
\providecommand \@ifnum [1]{%
 \ifnum #1\expandafter \@firstoftwo
 \else \expandafter \@secondoftwo
 \fi
}%
\providecommand \@ifx [1]{%
 \ifx #1\expandafter \@firstoftwo
 \else \expandafter \@secondoftwo
 \fi
}%
\providecommand \natexlab [1]{#1}%
\providecommand \enquote  [1]{``#1''}%
\providecommand \bibnamefont  [1]{#1}%
\providecommand \bibfnamefont [1]{#1}%
\providecommand \citenamefont [1]{#1}%
\providecommand \href@noop [0]{\@secondoftwo}%
\providecommand \href [0]{\begingroup \@sanitize@url \@href}%
\providecommand \@href[1]{\@@startlink{#1}\@@href}%
\providecommand \@@href[1]{\endgroup#1\@@endlink}%
\providecommand \@sanitize@url [0]{\catcode `\\12\catcode `\$12\catcode `\&12\catcode `\#12\catcode `\^12\catcode `\_12\catcode `\%12\relax}%
\providecommand \@@startlink[1]{}%
\providecommand \@@endlink[0]{}%
\providecommand \url  [0]{\begingroup\@sanitize@url \@url }%
\providecommand \@url [1]{\endgroup\@href {#1}{\urlprefix }}%
\providecommand \urlprefix  [0]{URL }%
\providecommand \Eprint [0]{\href }%
\providecommand \doibase [0]{https://doi.org/}%
\providecommand \selectlanguage [0]{\@gobble}%
\providecommand \bibinfo  [0]{\@secondoftwo}%
\providecommand \bibfield  [0]{\@secondoftwo}%
\providecommand \translation [1]{[#1]}%
\providecommand \BibitemOpen [0]{}%
\providecommand \bibitemStop [0]{}%
\providecommand \bibitemNoStop [0]{.\EOS\space}%
\providecommand \EOS [0]{\spacefactor3000\relax}%
\providecommand \BibitemShut  [1]{\csname bibitem#1\endcsname}%
\let\auto@bib@innerbib\@empty
\bibitem [{\citenamefont {Capone}\ \emph {et~al.}(2002)\citenamefont {Capone}, \citenamefont {Fabrizio}, \citenamefont {Castellani},\ and\ \citenamefont {Tosatti}}]{doi:10.1126/science.1071122}%
  \BibitemOpen
  \bibfield  {author} {\bibinfo {author} {\bibfnamefont {M.}~\bibnamefont {Capone}}, \bibinfo {author} {\bibfnamefont {M.}~\bibnamefont {Fabrizio}}, \bibinfo {author} {\bibfnamefont {C.}~\bibnamefont {Castellani}},\ and\ \bibinfo {author} {\bibfnamefont {E.}~\bibnamefont {Tosatti}},\ }\bibfield  {title} {\bibinfo {title} {{Strongly Correlated Superconductivity}},\ }\href {https://doi.org/10.1126/science.1071122} {\bibfield  {journal} {\bibinfo  {journal} {Science}\ }\textbf {\bibinfo {volume} {296}},\ \bibinfo {pages} {2364} (\bibinfo {year} {2002})}\BibitemShut {NoStop}%
\bibitem [{\citenamefont {Li}\ \emph {et~al.}(2019)\citenamefont {Li}, \citenamefont {Lee}, \citenamefont {Wang}, \citenamefont {Osada}, \citenamefont {Crossley}, \citenamefont {Lee}, \citenamefont {Cui}, \citenamefont {Hikita},\ and\ \citenamefont {Hwang}}]{Li2019}%
  \BibitemOpen
  \bibfield  {author} {\bibinfo {author} {\bibfnamefont {D.}~\bibnamefont {Li}}, \bibinfo {author} {\bibfnamefont {K.}~\bibnamefont {Lee}}, \bibinfo {author} {\bibfnamefont {B.~Y.}\ \bibnamefont {Wang}}, \bibinfo {author} {\bibfnamefont {M.}~\bibnamefont {Osada}}, \bibinfo {author} {\bibfnamefont {S.}~\bibnamefont {Crossley}}, \bibinfo {author} {\bibfnamefont {H.~R.}\ \bibnamefont {Lee}}, \bibinfo {author} {\bibfnamefont {Y.}~\bibnamefont {Cui}}, \bibinfo {author} {\bibfnamefont {Y.}~\bibnamefont {Hikita}},\ and\ \bibinfo {author} {\bibfnamefont {H.~Y.}\ \bibnamefont {Hwang}},\ }\bibfield  {title} {\bibinfo {title} {{Superconductivity in an infinite-layer nickelate}},\ }\href {https://doi.org/10.1038/s41586-019-1496-5} {\bibfield  {journal} {\bibinfo  {journal} {Nature}\ }\textbf {\bibinfo {volume} {572}},\ \bibinfo {pages} {624} (\bibinfo {year} {2019})}\BibitemShut {NoStop}%
\bibitem [{\citenamefont {Goto}\ and\ \citenamefont {L{\"{u}}thi}(2003)}]{doi:10.1080/0001873021000057114}%
  \BibitemOpen
  \bibfield  {author} {\bibinfo {author} {\bibfnamefont {T.}~\bibnamefont {Goto}}\ and\ \bibinfo {author} {\bibfnamefont {B.}~\bibnamefont {L{\"{u}}thi}},\ }\bibfield  {title} {\bibinfo {title} {{Charge ordering, charge fluctuations and lattice effects in strongly correlated electron systems}},\ }\href {https://doi.org/10.1080/0001873021000057114} {\bibfield  {journal} {\bibinfo  {journal} {Advances in Physics}\ }\textbf {\bibinfo {volume} {52}},\ \bibinfo {pages} {67} (\bibinfo {year} {2003})}\BibitemShut {NoStop}%
\bibitem [{\citenamefont {Clay}\ and\ \citenamefont {Mazumdar}(2019)}]{Clay2019}%
  \BibitemOpen
  \bibfield  {author} {\bibinfo {author} {\bibfnamefont {R.~T.}\ \bibnamefont {Clay}}\ and\ \bibinfo {author} {\bibfnamefont {S.}~\bibnamefont {Mazumdar}},\ }\bibfield  {title} {\bibinfo {title} {{From charge- and spin-ordering to superconductivity in the organic charge-transfer solids}},\ }\href {https://doi.org/10.1016/j.physrep.2018.10.006} {\bibfield  {journal} {\bibinfo  {journal} {Physics Reports}\ }\textbf {\bibinfo {volume} {788}},\ \bibinfo {pages} {1} (\bibinfo {year} {2019})},\ \Eprint {https://arxiv.org/abs/1802.01551} {arXiv:1802.01551} \BibitemShut {NoStop}%
\bibitem [{\citenamefont {Lee}\ \emph {et~al.}(2018)\citenamefont {Lee}, \citenamefont {Chung}, \citenamefont {Shi}, \citenamefont {Kim}, \citenamefont {Campbell}, \citenamefont {Xue}, \citenamefont {Song}, \citenamefont {Choi}, \citenamefont {Podkaminer}, \citenamefont {Kim}, \citenamefont {Ryan}, \citenamefont {Kim}, \citenamefont {Paudel}, \citenamefont {Kang}, \citenamefont {Spinuzzi}, \citenamefont {Tenne}, \citenamefont {Tsymbal}, \citenamefont {Rzchowski}, \citenamefont {Chen}, \citenamefont {Lee},\ and\ \citenamefont {Eom}}]{doi:10.1126/science.aam9189}%
  \BibitemOpen
  \bibfield  {author} {\bibinfo {author} {\bibfnamefont {D.}~\bibnamefont {Lee}}, \bibinfo {author} {\bibfnamefont {B.}~\bibnamefont {Chung}}, \bibinfo {author} {\bibfnamefont {Y.}~\bibnamefont {Shi}}, \bibinfo {author} {\bibfnamefont {G.-Y.}\ \bibnamefont {Kim}}, \bibinfo {author} {\bibfnamefont {N.}~\bibnamefont {Campbell}}, \bibinfo {author} {\bibfnamefont {F.}~\bibnamefont {Xue}}, \bibinfo {author} {\bibfnamefont {K.}~\bibnamefont {Song}}, \bibinfo {author} {\bibfnamefont {S.-Y.}\ \bibnamefont {Choi}}, \bibinfo {author} {\bibfnamefont {J.~P.}\ \bibnamefont {Podkaminer}}, \bibinfo {author} {\bibfnamefont {T.~H.}\ \bibnamefont {Kim}}, \bibinfo {author} {\bibfnamefont {P.~J.}\ \bibnamefont {Ryan}}, \bibinfo {author} {\bibfnamefont {J.-W.}\ \bibnamefont {Kim}}, \bibinfo {author} {\bibfnamefont {T.~R.}\ \bibnamefont {Paudel}}, \bibinfo {author} {\bibfnamefont {J.-H.}\ \bibnamefont {Kang}}, \bibinfo {author} {\bibfnamefont {J.~W.}\ \bibnamefont {Spinuzzi}}, \bibinfo {author} {\bibfnamefont {D.~A.}\ \bibnamefont
  {Tenne}}, \bibinfo {author} {\bibfnamefont {E.~Y.}\ \bibnamefont {Tsymbal}}, \bibinfo {author} {\bibfnamefont {M.~S.}\ \bibnamefont {Rzchowski}}, \bibinfo {author} {\bibfnamefont {L.~Q.}\ \bibnamefont {Chen}}, \bibinfo {author} {\bibfnamefont {J.}~\bibnamefont {Lee}},\ and\ \bibinfo {author} {\bibfnamefont {C.~B.}\ \bibnamefont {Eom}},\ }\bibfield  {title} {\bibinfo {title} {{Isostructural metal-insulator transition in VO$_2$}},\ }\href {https://doi.org/10.1126/science.aam9189} {\bibfield  {journal} {\bibinfo  {journal} {Science}\ }\textbf {\bibinfo {volume} {362}},\ \bibinfo {pages} {1037} (\bibinfo {year} {2018})}\BibitemShut {NoStop}%
\bibitem [{\citenamefont {Grytsiuk}\ \emph {et~al.}(2024)\citenamefont {Grytsiuk}, \citenamefont {Katsnelson}, \citenamefont {van Loon},\ and\ \citenamefont {R{\"{o}}sner}}]{Grytsiuk2024}%
  \BibitemOpen
  \bibfield  {author} {\bibinfo {author} {\bibfnamefont {S.}~\bibnamefont {Grytsiuk}}, \bibinfo {author} {\bibfnamefont {M.~I.}\ \bibnamefont {Katsnelson}}, \bibinfo {author} {\bibfnamefont {E.~G. C.~P.}\ \bibnamefont {van Loon}},\ and\ \bibinfo {author} {\bibfnamefont {M.}~\bibnamefont {R{\"{o}}sner}},\ }\bibfield  {title} {\bibinfo {title} {{Nb3Cl8: a prototypical layered Mott-Hubbard insulator}},\ }\href {https://doi.org/10.1038/s41535-024-00619-5} {\bibfield  {journal} {\bibinfo  {journal} {npj Quantum Materials}\ }\textbf {\bibinfo {volume} {9}},\ \bibinfo {pages} {8} (\bibinfo {year} {2024})}\BibitemShut {NoStop}%
\bibitem [{\citenamefont {Jia}\ \emph {et~al.}(2022)\citenamefont {Jia}, \citenamefont {Wang}, \citenamefont {Chiu}, \citenamefont {Song}, \citenamefont {Yu}, \citenamefont {J{\"{a}}ck}, \citenamefont {Lei}, \citenamefont {Klemenz}, \citenamefont {Cevallos}, \citenamefont {Onyszczak}, \citenamefont {Fishchenko}, \citenamefont {Liu}, \citenamefont {Farahi}, \citenamefont {Xie}, \citenamefont {Xu}, \citenamefont {Watanabe}, \citenamefont {Taniguchi}, \citenamefont {Bernevig}, \citenamefont {Cava}, \citenamefont {Schoop}, \citenamefont {Yazdani},\ and\ \citenamefont {Wu}}]{Jia2022}%
  \BibitemOpen
  \bibfield  {author} {\bibinfo {author} {\bibfnamefont {Y.}~\bibnamefont {Jia}}, \bibinfo {author} {\bibfnamefont {P.}~\bibnamefont {Wang}}, \bibinfo {author} {\bibfnamefont {C.-L.}\ \bibnamefont {Chiu}}, \bibinfo {author} {\bibfnamefont {Z.}~\bibnamefont {Song}}, \bibinfo {author} {\bibfnamefont {G.}~\bibnamefont {Yu}}, \bibinfo {author} {\bibfnamefont {B.}~\bibnamefont {J{\"{a}}ck}}, \bibinfo {author} {\bibfnamefont {S.}~\bibnamefont {Lei}}, \bibinfo {author} {\bibfnamefont {S.}~\bibnamefont {Klemenz}}, \bibinfo {author} {\bibfnamefont {F.~A.}\ \bibnamefont {Cevallos}}, \bibinfo {author} {\bibfnamefont {M.}~\bibnamefont {Onyszczak}}, \bibinfo {author} {\bibfnamefont {N.}~\bibnamefont {Fishchenko}}, \bibinfo {author} {\bibfnamefont {X.}~\bibnamefont {Liu}}, \bibinfo {author} {\bibfnamefont {G.}~\bibnamefont {Farahi}}, \bibinfo {author} {\bibfnamefont {F.}~\bibnamefont {Xie}}, \bibinfo {author} {\bibfnamefont {Y.}~\bibnamefont {Xu}}, \bibinfo {author} {\bibfnamefont {K.}~\bibnamefont {Watanabe}}, \bibinfo
  {author} {\bibfnamefont {T.}~\bibnamefont {Taniguchi}}, \bibinfo {author} {\bibfnamefont {B.~A.}\ \bibnamefont {Bernevig}}, \bibinfo {author} {\bibfnamefont {R.~J.}\ \bibnamefont {Cava}}, \bibinfo {author} {\bibfnamefont {L.~M.}\ \bibnamefont {Schoop}}, \bibinfo {author} {\bibfnamefont {A.}~\bibnamefont {Yazdani}},\ and\ \bibinfo {author} {\bibfnamefont {S.}~\bibnamefont {Wu}},\ }\bibfield  {title} {\bibinfo {title} {{Evidence for a monolayer excitonic insulator}},\ }\href {https://doi.org/10.1038/s41567-021-01422-w} {\bibfield  {journal} {\bibinfo  {journal} {Nature Physics}\ }\textbf {\bibinfo {volume} {18}},\ \bibinfo {pages} {87} (\bibinfo {year} {2022})}\BibitemShut {NoStop}%
\bibitem [{\citenamefont {Ma}\ \emph {et~al.}(2021)\citenamefont {Ma}, \citenamefont {Nguyen}, \citenamefont {Wang}, \citenamefont {Zeng}, \citenamefont {Watanabe}, \citenamefont {Taniguchi}, \citenamefont {MacDonald}, \citenamefont {Mak},\ and\ \citenamefont {Shan}}]{Ma2021}%
  \BibitemOpen
  \bibfield  {author} {\bibinfo {author} {\bibfnamefont {L.}~\bibnamefont {Ma}}, \bibinfo {author} {\bibfnamefont {P.~X.}\ \bibnamefont {Nguyen}}, \bibinfo {author} {\bibfnamefont {Z.}~\bibnamefont {Wang}}, \bibinfo {author} {\bibfnamefont {Y.}~\bibnamefont {Zeng}}, \bibinfo {author} {\bibfnamefont {K.}~\bibnamefont {Watanabe}}, \bibinfo {author} {\bibfnamefont {T.}~\bibnamefont {Taniguchi}}, \bibinfo {author} {\bibfnamefont {A.~H.}\ \bibnamefont {MacDonald}}, \bibinfo {author} {\bibfnamefont {K.~F.}\ \bibnamefont {Mak}},\ and\ \bibinfo {author} {\bibfnamefont {J.}~\bibnamefont {Shan}},\ }\bibfield  {title} {\bibinfo {title} {{Strongly correlated excitonic insulator in atomic double layers}},\ }\href {https://doi.org/10.1038/s41586-021-03947-9} {\bibfield  {journal} {\bibinfo  {journal} {Nature}\ }\textbf {\bibinfo {volume} {598}},\ \bibinfo {pages} {585} (\bibinfo {year} {2021})}\BibitemShut {NoStop}%
\bibitem [{\citenamefont {Haunschild}\ \emph {et~al.}(2016)\citenamefont {Haunschild}, \citenamefont {Barth},\ and\ \citenamefont {Marx}}]{Haunschild2016}%
  \BibitemOpen
  \bibfield  {author} {\bibinfo {author} {\bibfnamefont {R.}~\bibnamefont {Haunschild}}, \bibinfo {author} {\bibfnamefont {A.}~\bibnamefont {Barth}},\ and\ \bibinfo {author} {\bibfnamefont {W.}~\bibnamefont {Marx}},\ }\bibfield  {title} {\bibinfo {title} {{Evolution of DFT studies in view of a scientometric perspective}},\ }\href {https://doi.org/10.1186/s13321-016-0166-y} {\bibfield  {journal} {\bibinfo  {journal} {Journal of Cheminformatics}\ }\textbf {\bibinfo {volume} {8}},\ \bibinfo {pages} {1} (\bibinfo {year} {2016})}\BibitemShut {NoStop}%
\bibitem [{\citenamefont {Perdew}\ \emph {et~al.}(2009)\citenamefont {Perdew}, \citenamefont {Ruzsinszky}, \citenamefont {Constantin}, \citenamefont {Sun},\ and\ \citenamefont {Csonka}}]{Perdew2009}%
  \BibitemOpen
  \bibfield  {author} {\bibinfo {author} {\bibfnamefont {J.~P.}\ \bibnamefont {Perdew}}, \bibinfo {author} {\bibfnamefont {A.}~\bibnamefont {Ruzsinszky}}, \bibinfo {author} {\bibfnamefont {L.~A.}\ \bibnamefont {Constantin}}, \bibinfo {author} {\bibfnamefont {J.}~\bibnamefont {Sun}},\ and\ \bibinfo {author} {\bibfnamefont {G.~I.}\ \bibnamefont {Csonka}},\ }\bibfield  {title} {\bibinfo {title} {{Some fundamental issues in ground-state density functional theory: A guide for the perplexed}},\ }\href {https://doi.org/10.1021/ct800531s} {\bibfield  {journal} {\bibinfo  {journal} {Journal of Chemical Theory and Computation}\ }\textbf {\bibinfo {volume} {5}},\ \bibinfo {pages} {902} (\bibinfo {year} {2009})}\BibitemShut {NoStop}%
\bibitem [{\citenamefont {Nakamura}\ \emph {et~al.}(2008)\citenamefont {Nakamura}, \citenamefont {Yoshimoto}, \citenamefont {Arita}, \citenamefont {Tsuneyuki},\ and\ \citenamefont {Imada}}]{Nakamura2008}%
  \BibitemOpen
  \bibfield  {author} {\bibinfo {author} {\bibfnamefont {K.}~\bibnamefont {Nakamura}}, \bibinfo {author} {\bibfnamefont {Y.}~\bibnamefont {Yoshimoto}}, \bibinfo {author} {\bibfnamefont {R.}~\bibnamefont {Arita}}, \bibinfo {author} {\bibfnamefont {S.}~\bibnamefont {Tsuneyuki}},\ and\ \bibinfo {author} {\bibfnamefont {M.}~\bibnamefont {Imada}},\ }\bibfield  {title} {\bibinfo {title} {{Optical absorption study by ab initio downfolding approach: Application to GaAs}},\ }\href {https://doi.org/10.1103/PhysRevB.77.195126} {\bibfield  {journal} {\bibinfo  {journal} {Physical Review B - Condensed Matter and Materials Physics}\ }\textbf {\bibinfo {volume} {77}},\ \bibinfo {pages} {1} (\bibinfo {year} {2008})},\ \Eprint {https://arxiv.org/abs/0710.4371} {arXiv:0710.4371} \BibitemShut {NoStop}%
\bibitem [{\citenamefont {Nakamura}\ \emph {et~al.}(2010)\citenamefont {Nakamura}, \citenamefont {Yoshimoto}, \citenamefont {Nohara},\ and\ \citenamefont {Imada}}]{Nakamura2010}%
  \BibitemOpen
  \bibfield  {author} {\bibinfo {author} {\bibfnamefont {K.}~\bibnamefont {Nakamura}}, \bibinfo {author} {\bibfnamefont {Y.}~\bibnamefont {Yoshimoto}}, \bibinfo {author} {\bibfnamefont {Y.}~\bibnamefont {Nohara}},\ and\ \bibinfo {author} {\bibfnamefont {M.}~\bibnamefont {Imada}},\ }\bibfield  {title} {\bibinfo {title} {{Ab initio low-dimensional physics opened up by dimensional downfolding: Application to LaFeAsO}},\ }\href {https://doi.org/10.1143/JPSJ.79.123708} {\bibfield  {journal} {\bibinfo  {journal} {Journal of the Physical Society of Japan}\ }\textbf {\bibinfo {volume} {79}},\ \bibinfo {pages} {12} (\bibinfo {year} {2010})},\ \Eprint {https://arxiv.org/abs/1007.4429} {1007.4429} \BibitemShut {NoStop}%
\bibitem [{\citenamefont {Arita}\ \emph {et~al.}(2015)\citenamefont {Arita}, \citenamefont {Ikeda}, \citenamefont {Sakai},\ and\ \citenamefont {Suzuki}}]{Arita2015}%
  \BibitemOpen
  \bibfield  {author} {\bibinfo {author} {\bibfnamefont {R.}~\bibnamefont {Arita}}, \bibinfo {author} {\bibfnamefont {H.}~\bibnamefont {Ikeda}}, \bibinfo {author} {\bibfnamefont {S.}~\bibnamefont {Sakai}},\ and\ \bibinfo {author} {\bibfnamefont {M.~T.}\ \bibnamefont {Suzuki}},\ }\bibfield  {title} {\bibinfo {title} {{Ab initio downfolding study of the iron-based ladder superconductor BaFe2 S3}},\ }\href {https://doi.org/10.1103/PhysRevB.92.054515} {\bibfield  {journal} {\bibinfo  {journal} {Physical Review B - Condensed Matter and Materials Physics}\ }\textbf {\bibinfo {volume} {92}},\ \bibinfo {pages} {1} (\bibinfo {year} {2015})},\ \Eprint {https://arxiv.org/abs/1507.05715} {1507.05715} \BibitemShut {NoStop}%
\bibitem [{\citenamefont {Zheng}\ \emph {et~al.}(2018)\citenamefont {Zheng}, \citenamefont {Changlani}, \citenamefont {Williams}, \citenamefont {Busemeyer},\ and\ \citenamefont {Wagner}}]{Zheng2018}%
  \BibitemOpen
  \bibfield  {author} {\bibinfo {author} {\bibfnamefont {H.}~\bibnamefont {Zheng}}, \bibinfo {author} {\bibfnamefont {H.~J.}\ \bibnamefont {Changlani}}, \bibinfo {author} {\bibfnamefont {K.~T.}\ \bibnamefont {Williams}}, \bibinfo {author} {\bibfnamefont {B.}~\bibnamefont {Busemeyer}},\ and\ \bibinfo {author} {\bibfnamefont {L.~K.}\ \bibnamefont {Wagner}},\ }\bibfield  {title} {\bibinfo {title} {{From real materials to model Hamiltonians with density matrix downfolding}},\ }\href {https://doi.org/10.3389/fphy.2018.00043} {\bibfield  {journal} {\bibinfo  {journal} {Frontiers in Physics}\ }\textbf {\bibinfo {volume} {6}},\ \bibinfo {pages} {1} (\bibinfo {year} {2018})},\ \Eprint {https://arxiv.org/abs/1712.00477} {1712.00477} \BibitemShut {NoStop}%
\bibitem [{\citenamefont {Yoshimi}\ \emph {et~al.}(2021)\citenamefont {Yoshimi}, \citenamefont {Tsumuraya},\ and\ \citenamefont {Misawa}}]{Yoshimi2021}%
  \BibitemOpen
  \bibfield  {author} {\bibinfo {author} {\bibfnamefont {K.}~\bibnamefont {Yoshimi}}, \bibinfo {author} {\bibfnamefont {T.}~\bibnamefont {Tsumuraya}},\ and\ \bibinfo {author} {\bibfnamefont {T.}~\bibnamefont {Misawa}},\ }\bibfield  {title} {\bibinfo {title} {{Ab initio derivation and exact diagonalization analysis of low-energy effective Hamiltonians for $\beta$- X[Pd(dmit)2]2}},\ }\href {https://doi.org/10.1103/PhysRevResearch.3.043224} {\bibfield  {journal} {\bibinfo  {journal} {Physical Review Research}\ }\textbf {\bibinfo {volume} {3}},\ \bibinfo {pages} {1} (\bibinfo {year} {2021})},\ \Eprint {https://arxiv.org/abs/2109.10542} {2109.10542} \BibitemShut {NoStop}%
\bibitem [{\citenamefont {Botzung}\ and\ \citenamefont {Nataf}(2024)}]{Botzung2024}%
  \BibitemOpen
  \bibfield  {author} {\bibinfo {author} {\bibfnamefont {T.}~\bibnamefont {Botzung}}\ and\ \bibinfo {author} {\bibfnamefont {P.}~\bibnamefont {Nataf}},\ }\bibfield  {title} {\bibinfo {title} {{Exact Diagonalization of SU (N) Fermi-Hubbard Models}},\ }\href {https://doi.org/10.1103/PhysRevLett.132.153001} {\bibfield  {journal} {\bibinfo  {journal} {Physical Review Letters}\ }\textbf {\bibinfo {volume} {132}},\ \bibinfo {pages} {153001} (\bibinfo {year} {2024})}\BibitemShut {NoStop}%
\bibitem [{\citenamefont {Tubman}\ \emph {et~al.}(2016)\citenamefont {Tubman}, \citenamefont {Lee}, \citenamefont {Takeshita}, \citenamefont {Head-Gordon},\ and\ \citenamefont {Whaley}}]{tubman2016deterministic}%
  \BibitemOpen
  \bibfield  {author} {\bibinfo {author} {\bibfnamefont {N.~M.}\ \bibnamefont {Tubman}}, \bibinfo {author} {\bibfnamefont {J.}~\bibnamefont {Lee}}, \bibinfo {author} {\bibfnamefont {T.~Y.}\ \bibnamefont {Takeshita}}, \bibinfo {author} {\bibfnamefont {M.}~\bibnamefont {Head-Gordon}},\ and\ \bibinfo {author} {\bibfnamefont {K.~B.}\ \bibnamefont {Whaley}},\ }\bibfield  {title} {\bibinfo {title} {A deterministic alternative to the full configuration interaction quantum monte carlo method},\ }\href@noop {} {\bibfield  {journal} {\bibinfo  {journal} {The Journal of chemical physics}\ }\textbf {\bibinfo {volume} {145}} (\bibinfo {year} {2016})}\BibitemShut {NoStop}%
\bibitem [{\citenamefont {Tubman}\ \emph {et~al.}(2020)\citenamefont {Tubman}, \citenamefont {Freeman}, \citenamefont {Levine}, \citenamefont {Hait}, \citenamefont {Head-Gordon},\ and\ \citenamefont {Whaley}}]{tubman2020modern}%
  \BibitemOpen
  \bibfield  {author} {\bibinfo {author} {\bibfnamefont {N.~M.}\ \bibnamefont {Tubman}}, \bibinfo {author} {\bibfnamefont {C.~D.}\ \bibnamefont {Freeman}}, \bibinfo {author} {\bibfnamefont {D.~S.}\ \bibnamefont {Levine}}, \bibinfo {author} {\bibfnamefont {D.}~\bibnamefont {Hait}}, \bibinfo {author} {\bibfnamefont {M.}~\bibnamefont {Head-Gordon}},\ and\ \bibinfo {author} {\bibfnamefont {K.~B.}\ \bibnamefont {Whaley}},\ }\bibfield  {title} {\bibinfo {title} {Modern approaches to exact diagonalization and selected configuration interaction with the adaptive sampling ci method},\ }\href@noop {} {\bibfield  {journal} {\bibinfo  {journal} {Journal of chemical theory and computation}\ }\textbf {\bibinfo {volume} {16}},\ \bibinfo {pages} {2139} (\bibinfo {year} {2020})}\BibitemShut {NoStop}%
\bibitem [{\citenamefont {Ma}\ \emph {et~al.}(2015)\citenamefont {Ma}, \citenamefont {Purwanto}, \citenamefont {Zhang},\ and\ \citenamefont {Krakauer}}]{Ma2015}%
  \BibitemOpen
  \bibfield  {author} {\bibinfo {author} {\bibfnamefont {F.}~\bibnamefont {Ma}}, \bibinfo {author} {\bibfnamefont {W.}~\bibnamefont {Purwanto}}, \bibinfo {author} {\bibfnamefont {S.}~\bibnamefont {Zhang}},\ and\ \bibinfo {author} {\bibfnamefont {H.}~\bibnamefont {Krakauer}},\ }\bibfield  {title} {\bibinfo {title} {{Quantum monte carlo calculations in solids with downfolded hamiltonians}},\ }\href {https://doi.org/10.1103/PhysRevLett.114.226401} {\bibfield  {journal} {\bibinfo  {journal} {Physical Review Letters}\ }\textbf {\bibinfo {volume} {114}},\ \bibinfo {pages} {1} (\bibinfo {year} {2015})},\ \Eprint {https://arxiv.org/abs/1412.0322} {arXiv:1412.0322} \BibitemShut {NoStop}%
\bibitem [{\citenamefont {Kim}\ \emph {et~al.}(2018)\citenamefont {Kim}, \citenamefont {Baczewski}, \citenamefont {Beaudet}, \citenamefont {Benali}, \citenamefont {Bennett}, \citenamefont {Berrill}, \citenamefont {Blunt}, \citenamefont {Borda}, \citenamefont {Casula}, \citenamefont {Ceperley} \emph {et~al.}}]{kim2018qmcpack}%
  \BibitemOpen
  \bibfield  {author} {\bibinfo {author} {\bibfnamefont {J.}~\bibnamefont {Kim}}, \bibinfo {author} {\bibfnamefont {A.~D.}\ \bibnamefont {Baczewski}}, \bibinfo {author} {\bibfnamefont {T.~D.}\ \bibnamefont {Beaudet}}, \bibinfo {author} {\bibfnamefont {A.}~\bibnamefont {Benali}}, \bibinfo {author} {\bibfnamefont {M.~C.}\ \bibnamefont {Bennett}}, \bibinfo {author} {\bibfnamefont {M.~A.}\ \bibnamefont {Berrill}}, \bibinfo {author} {\bibfnamefont {N.~S.}\ \bibnamefont {Blunt}}, \bibinfo {author} {\bibfnamefont {E.~J.~L.}\ \bibnamefont {Borda}}, \bibinfo {author} {\bibfnamefont {M.}~\bibnamefont {Casula}}, \bibinfo {author} {\bibfnamefont {D.~M.}\ \bibnamefont {Ceperley}}, \emph {et~al.},\ }\bibfield  {title} {\bibinfo {title} {Qmcpack: an open source ab initio quantum monte carlo package for the electronic structure of atoms, molecules and solids},\ }\href@noop {} {\bibfield  {journal} {\bibinfo  {journal} {Journal of Physics: Condensed Matter}\ }\textbf {\bibinfo {volume} {30}},\ \bibinfo {pages} {195901}
  (\bibinfo {year} {2018})}\BibitemShut {NoStop}%
\bibitem [{\citenamefont {Tubman}\ \emph {et~al.}(2015)\citenamefont {Tubman}, \citenamefont {Liberatore}, \citenamefont {Pierleoni}, \citenamefont {Holzmann},\ and\ \citenamefont {Ceperley}}]{tubman2015molecular}%
  \BibitemOpen
  \bibfield  {author} {\bibinfo {author} {\bibfnamefont {N.~M.}\ \bibnamefont {Tubman}}, \bibinfo {author} {\bibfnamefont {E.}~\bibnamefont {Liberatore}}, \bibinfo {author} {\bibfnamefont {C.}~\bibnamefont {Pierleoni}}, \bibinfo {author} {\bibfnamefont {M.}~\bibnamefont {Holzmann}},\ and\ \bibinfo {author} {\bibfnamefont {D.~M.}\ \bibnamefont {Ceperley}},\ }\bibfield  {title} {\bibinfo {title} {Molecular-atomic transition along the deuterium hugoniot curve with coupled electron-ion monte carlo simulations},\ }\href@noop {} {\bibfield  {journal} {\bibinfo  {journal} {Physical review letters}\ }\textbf {\bibinfo {volume} {115}},\ \bibinfo {pages} {045301} (\bibinfo {year} {2015})}\BibitemShut {NoStop}%
\bibitem [{\citenamefont {Zhu}\ \emph {et~al.}(2019)\citenamefont {Zhu}, \citenamefont {Jim\'enez-Hoyos}, \citenamefont {McClain}, \citenamefont {Berkelbach},\ and\ \citenamefont {Chan}}]{PhysRevB.100.115154}%
  \BibitemOpen
  \bibfield  {author} {\bibinfo {author} {\bibfnamefont {T.}~\bibnamefont {Zhu}}, \bibinfo {author} {\bibfnamefont {C.~A.}\ \bibnamefont {Jim\'enez-Hoyos}}, \bibinfo {author} {\bibfnamefont {J.}~\bibnamefont {McClain}}, \bibinfo {author} {\bibfnamefont {T.~C.}\ \bibnamefont {Berkelbach}},\ and\ \bibinfo {author} {\bibfnamefont {G.~K.-L.}\ \bibnamefont {Chan}},\ }\bibfield  {title} {\bibinfo {title} {Coupled-cluster impurity solvers for dynamical mean-field theory},\ }\href {https://doi.org/10.1103/PhysRevB.100.115154} {\bibfield  {journal} {\bibinfo  {journal} {Phys. Rev. B}\ }\textbf {\bibinfo {volume} {100}},\ \bibinfo {pages} {115154} (\bibinfo {year} {2019})}\BibitemShut {NoStop}%
\bibitem [{\citenamefont {Nakamura@article{tubman2016deterministic, title={A deterministic alternative to the full configuration interaction quantum Monte Carlo method}, author={Tubman, Norm M and Lee, Joonho and Takeshita, Tyler Y and Head-Gordon, Martin and Whaley, K Birgitta}, journal={The Journal of chemical physics}, volume={145}, number={4}, year={2016}, publisher={AIP Publishing} }}\ \emph {et~al.}(2021)\citenamefont {Nakamura@article{tubman2016deterministic, title={A deterministic alternative to the full configuration interaction quantum Monte Carlo method}, author={Tubman, Norm M and Lee, Joonho and Takeshita, Tyler Y and Head-Gordon, Martin and Whaley, K Birgitta}, journal={The Journal of chemical physics}, volume={145}, number={4}, year={2016}, publisher={AIP Publishing} }}, \citenamefont {Yoshimoto}, \citenamefont {Nomura}, \citenamefont {Tadano}, \citenamefont {Kawamura}, \citenamefont {Kosugi}, \citenamefont {Yoshimi}, \citenamefont {Misawa},\ and\ \citenamefont {Motoyama}}]{Nakamura2021}%
  \BibitemOpen
  \bibfield  {author} {\bibinfo {author} {\bibfnamefont {K.}~\bibnamefont {Nakamura@article{tubman2016deterministic, title={A deterministic alternative to the full configuration interaction quantum Monte Carlo method}, author={Tubman, Norm M and Lee, Joonho and Takeshita, Tyler Y and Head-Gordon, Martin and Whaley, K Birgitta}, journal={The Journal of chemical physics}, volume={145}, number={4}, year={2016}, publisher={AIP Publishing} }}}, \bibinfo {author} {\bibfnamefont {Y.}~\bibnamefont {Yoshimoto}}, \bibinfo {author} {\bibfnamefont {Y.}~\bibnamefont {Nomura}}, \bibinfo {author} {\bibfnamefont {T.}~\bibnamefont {Tadano}}, \bibinfo {author} {\bibfnamefont {M.}~\bibnamefont {Kawamura}}, \bibinfo {author} {\bibfnamefont {T.}~\bibnamefont {Kosugi}}, \bibinfo {author} {\bibfnamefont {K.}~\bibnamefont {Yoshimi}}, \bibinfo {author} {\bibfnamefont {T.}~\bibnamefont {Misawa}},\ and\ \bibinfo {author} {\bibfnamefont {Y.}~\bibnamefont {Motoyama}},\ }\bibfield  {title} {\bibinfo {title} {{RESPACK: An ab initio tool
  for derivation of effective low-energy model of material}},\ }\href {https://doi.org/10.1016/j.cpc.2020.107781} {\bibfield  {journal} {\bibinfo  {journal} {Computer Physics Communications}\ }\textbf {\bibinfo {volume} {261}},\ \bibinfo {pages} {107781} (\bibinfo {year} {2021})},\ \Eprint {https://arxiv.org/abs/2001.02351} {arXiv:2001.02351} \BibitemShut {NoStop}%
\bibitem [{\citenamefont {Yoshimi}\ \emph {et~al.}(2023)\citenamefont {Yoshimi}, \citenamefont {Misawa}, \citenamefont {Tsumuraya},\ and\ \citenamefont {Seo}}]{Yoshimi2023}%
  \BibitemOpen
  \bibfield  {author} {\bibinfo {author} {\bibfnamefont {K.}~\bibnamefont {Yoshimi}}, \bibinfo {author} {\bibfnamefont {T.}~\bibnamefont {Misawa}}, \bibinfo {author} {\bibfnamefont {T.}~\bibnamefont {Tsumuraya}},\ and\ \bibinfo {author} {\bibfnamefont {H.}~\bibnamefont {Seo}},\ }\bibfield  {title} {\bibinfo {title} {{Comprehensive Ab Initio Investigation of the Phase Diagram of Quasi-One-Dimensional Molecular Solids}},\ }\href {https://doi.org/10.1103/PhysRevLett.131.036401} {\bibfield  {journal} {\bibinfo  {journal} {Physical Review Letters}\ }\textbf {\bibinfo {volume} {131}},\ \bibinfo {pages} {36401} (\bibinfo {year} {2023})}\BibitemShut {NoStop}%
\bibitem [{\citenamefont {Schobert}\ \emph {et~al.}(2024)\citenamefont {Schobert}, \citenamefont {Berges}, \citenamefont {van Loon}, \citenamefont {Sentef}, \citenamefont {Brener}, \citenamefont {Rossi},\ and\ \citenamefont {Wehling}}]{Schobert2024}%
  \BibitemOpen
  \bibfield  {author} {\bibinfo {author} {\bibfnamefont {A.}~\bibnamefont {Schobert}}, \bibinfo {author} {\bibfnamefont {J.}~\bibnamefont {Berges}}, \bibinfo {author} {\bibfnamefont {E.~G.}\ \bibnamefont {van Loon}}, \bibinfo {author} {\bibfnamefont {M.~A.}\ \bibnamefont {Sentef}}, \bibinfo {author} {\bibfnamefont {S.}~\bibnamefont {Brener}}, \bibinfo {author} {\bibfnamefont {M.}~\bibnamefont {Rossi}},\ and\ \bibinfo {author} {\bibfnamefont {T.~O.}\ \bibnamefont {Wehling}},\ }\bibfield  {title} {\bibinfo {title} {{Ab initio electron-lattice downfolding: Potential energy landscapes, anharmonicity, and molecular dynamics in charge density wave materials}},\ }\href {https://doi.org/10.21468/SciPostPhys.16.2.046} {\bibfield  {journal} {\bibinfo  {journal} {SciPost Physics}\ }\textbf {\bibinfo {volume} {16}},\ \bibinfo {pages} {1} (\bibinfo {year} {2024})},\ \Eprint {https://arxiv.org/abs/2303.07261} {arXiv:2303.07261} \BibitemShut {NoStop}%
\bibitem [{\citenamefont {Hirayama}\ \emph {et~al.}(2019)\citenamefont {Hirayama}, \citenamefont {Misawa}, \citenamefont {Ohgoe}, \citenamefont {Yamaji},\ and\ \citenamefont {Imada}}]{Hirayama2019}%
  \BibitemOpen
  \bibfield  {author} {\bibinfo {author} {\bibfnamefont {M.}~\bibnamefont {Hirayama}}, \bibinfo {author} {\bibfnamefont {T.}~\bibnamefont {Misawa}}, \bibinfo {author} {\bibfnamefont {T.}~\bibnamefont {Ohgoe}}, \bibinfo {author} {\bibfnamefont {Y.}~\bibnamefont {Yamaji}},\ and\ \bibinfo {author} {\bibfnamefont {M.}~\bibnamefont {Imada}},\ }\bibfield  {title} {\bibinfo {title} {{Effective Hamiltonian for cuprate superconductors derived from multiscale ab initio scheme with level renormalization}},\ }\href {https://doi.org/10.1103/PhysRevB.99.245155} {\bibfield  {journal} {\bibinfo  {journal} {Physical Review B}\ }\textbf {\bibinfo {volume} {99}},\ \bibinfo {pages} {1} (\bibinfo {year} {2019})},\ \Eprint {https://arxiv.org/abs/1901.00763} {1901.00763} \BibitemShut {NoStop}%
\bibitem [{\citenamefont {Ohgoe}\ \emph {et~al.}(2020)\citenamefont {Ohgoe}, \citenamefont {Hirayama}, \citenamefont {Misawa}, \citenamefont {Ido}, \citenamefont {Yamaji},\ and\ \citenamefont {Imada}}]{Ohgoe2020}%
  \BibitemOpen
  \bibfield  {author} {\bibinfo {author} {\bibfnamefont {T.}~\bibnamefont {Ohgoe}}, \bibinfo {author} {\bibfnamefont {M.}~\bibnamefont {Hirayama}}, \bibinfo {author} {\bibfnamefont {T.}~\bibnamefont {Misawa}}, \bibinfo {author} {\bibfnamefont {K.}~\bibnamefont {Ido}}, \bibinfo {author} {\bibfnamefont {Y.}~\bibnamefont {Yamaji}},\ and\ \bibinfo {author} {\bibfnamefont {M.}~\bibnamefont {Imada}},\ }\bibfield  {title} {\bibinfo {title} {{Ab initio study of superconductivity and inhomogeneity in a Hg-based cuprate superconductor}},\ }\href {https://doi.org/10.1103/PhysRevB.101.045124} {\bibfield  {journal} {\bibinfo  {journal} {Physical Review B}\ }\textbf {\bibinfo {volume} {101}},\ \bibinfo {pages} {1} (\bibinfo {year} {2020})},\ \Eprint {https://arxiv.org/abs/1902.00122} {arXiv:1902.00122} \BibitemShut {NoStop}%
\bibitem [{\citenamefont {Been}\ \emph {et~al.}(2021)\citenamefont {Been}, \citenamefont {Lee}, \citenamefont {Hwang}, \citenamefont {Cui}, \citenamefont {Zaanen}, \citenamefont {Devereaux}, \citenamefont {Moritz},\ and\ \citenamefont {Jia}}]{Been2021}%
  \BibitemOpen
  \bibfield  {author} {\bibinfo {author} {\bibfnamefont {E.}~\bibnamefont {Been}}, \bibinfo {author} {\bibfnamefont {W.~S.}\ \bibnamefont {Lee}}, \bibinfo {author} {\bibfnamefont {H.~Y.}\ \bibnamefont {Hwang}}, \bibinfo {author} {\bibfnamefont {Y.}~\bibnamefont {Cui}}, \bibinfo {author} {\bibfnamefont {J.}~\bibnamefont {Zaanen}}, \bibinfo {author} {\bibfnamefont {T.}~\bibnamefont {Devereaux}}, \bibinfo {author} {\bibfnamefont {B.}~\bibnamefont {Moritz}},\ and\ \bibinfo {author} {\bibfnamefont {C.}~\bibnamefont {Jia}},\ }\bibfield  {title} {\bibinfo {title} {{Electronic Structure Trends across the Rare-Earth Series in Superconducting Infinite-Layer Nickelates}},\ }\href {https://doi.org/10.1103/PhysRevX.11.011050} {\bibfield  {journal} {\bibinfo  {journal} {Physical Review X}\ }\textbf {\bibinfo {volume} {11}},\ \bibinfo {pages} {11050} (\bibinfo {year} {2021})},\ \Eprint {https://arxiv.org/abs/2002.12300} {arXiv:2002.12300} \BibitemShut {NoStop}%
\bibitem [{\citenamefont {Schmid}\ \emph {et~al.}(2023)\citenamefont {Schmid}, \citenamefont {Mor{\'{e}}e}, \citenamefont {Kaneko}, \citenamefont {Yamaji},\ and\ \citenamefont {Imada}}]{Schmid2023}%
  \BibitemOpen
  \bibfield  {author} {\bibinfo {author} {\bibfnamefont {M.~T.}\ \bibnamefont {Schmid}}, \bibinfo {author} {\bibfnamefont {J.~B.}\ \bibnamefont {Mor{\'{e}}e}}, \bibinfo {author} {\bibfnamefont {R.}~\bibnamefont {Kaneko}}, \bibinfo {author} {\bibfnamefont {Y.}~\bibnamefont {Yamaji}},\ and\ \bibinfo {author} {\bibfnamefont {M.}~\bibnamefont {Imada}},\ }\bibfield  {title} {\bibinfo {title} {{Superconductivity Studied by Solving Ab Initio Low-Energy Effective Hamiltonians for Carrier Doped CaCuO2, Bi2Sr2CuO6, Bi2Sr2CaCu2 O8, and HgBa2CuO4}},\ }\href {https://doi.org/10.1103/PhysRevX.13.041036} {\bibfield  {journal} {\bibinfo  {journal} {Physical Review X}\ }\textbf {\bibinfo {volume} {13}},\ \bibinfo {pages} {1} (\bibinfo {year} {2023})}\BibitemShut {NoStop}%
\bibitem [{\citenamefont {Aichhorn}\ \emph {et~al.}(2006)\citenamefont {Aichhorn}, \citenamefont {Arrigoni}, \citenamefont {Potthoff},\ and\ \citenamefont {Hanke}}]{Aichhorn2006}%
  \BibitemOpen
  \bibfield  {author} {\bibinfo {author} {\bibfnamefont {M.}~\bibnamefont {Aichhorn}}, \bibinfo {author} {\bibfnamefont {E.}~\bibnamefont {Arrigoni}}, \bibinfo {author} {\bibfnamefont {M.}~\bibnamefont {Potthoff}},\ and\ \bibinfo {author} {\bibfnamefont {W.}~\bibnamefont {Hanke}},\ }\bibfield  {title} {\bibinfo {title} {{Variational cluster approach to the Hubbard model: Phase-separation tendency and finite-size effects}},\ }\href {https://doi.org/10.1103/PhysRevB.74.235117} {\bibfield  {journal} {\bibinfo  {journal} {Physical Review B - Condensed Matter and Materials Physics}\ }\textbf {\bibinfo {volume} {74}},\ \bibinfo {pages} {1} (\bibinfo {year} {2006})},\ \Eprint {https://arxiv.org/abs/0607271} {0607271 [cond-mat]} \BibitemShut {NoStop}%
\bibitem [{\citenamefont {Holzmann}\ \emph {et~al.}(2016)\citenamefont {Holzmann}, \citenamefont {Clay~III}, \citenamefont {Morales}, \citenamefont {Tubman}, \citenamefont {Ceperley},\ and\ \citenamefont {Pierleoni}}]{holzmann2016theory}%
  \BibitemOpen
  \bibfield  {author} {\bibinfo {author} {\bibfnamefont {M.}~\bibnamefont {Holzmann}}, \bibinfo {author} {\bibfnamefont {R.~C.}\ \bibnamefont {Clay~III}}, \bibinfo {author} {\bibfnamefont {M.~A.}\ \bibnamefont {Morales}}, \bibinfo {author} {\bibfnamefont {N.~M.}\ \bibnamefont {Tubman}}, \bibinfo {author} {\bibfnamefont {D.~M.}\ \bibnamefont {Ceperley}},\ and\ \bibinfo {author} {\bibfnamefont {C.}~\bibnamefont {Pierleoni}},\ }\bibfield  {title} {\bibinfo {title} {Theory of finite size effects for electronic quantum monte carlo calculations of liquids and solids},\ }\href@noop {} {\bibfield  {journal} {\bibinfo  {journal} {Physical Review B}\ }\textbf {\bibinfo {volume} {94}},\ \bibinfo {pages} {035126} (\bibinfo {year} {2016})}\BibitemShut {NoStop}%
\bibitem [{\citenamefont {Qin}\ \emph {et~al.}(2022)\citenamefont {Qin}, \citenamefont {Schafer}, \citenamefont {Andergassen}, \citenamefont {Corboz},\ and\ \citenamefont {Gull}}]{Qin2022}%
  \BibitemOpen
  \bibfield  {author} {\bibinfo {author} {\bibfnamefont {M.}~\bibnamefont {Qin}}, \bibinfo {author} {\bibfnamefont {T.}~\bibnamefont {Schafer}}, \bibinfo {author} {\bibfnamefont {S.}~\bibnamefont {Andergassen}}, \bibinfo {author} {\bibfnamefont {P.}~\bibnamefont {Corboz}},\ and\ \bibinfo {author} {\bibfnamefont {E.}~\bibnamefont {Gull}},\ }\bibfield  {title} {\bibinfo {title} {{The Hubbard Model: A Computational Perspective}},\ }\href {https://doi.org/10.1146/annurev-conmatphys-090921-033948} {\bibfield  {journal} {\bibinfo  {journal} {Annual Review of Condensed Matter Physics}\ }\textbf {\bibinfo {volume} {13}},\ \bibinfo {pages} {275} (\bibinfo {year} {2022})},\ \Eprint {https://arxiv.org/abs/2104.00064} {2104.00064} \BibitemShut {NoStop}%
\bibitem [{\citenamefont {Bernien}\ \emph {et~al.}(2017)\citenamefont {Bernien}, \citenamefont {Schwartz}, \citenamefont {Keesling}, \citenamefont {Levine}, \citenamefont {Omran}, \citenamefont {Pichler}, \citenamefont {Choi}, \citenamefont {Zibrov}, \citenamefont {Endres}, \citenamefont {Greiner}, \citenamefont {Vuleti{\'{c}}},\ and\ \citenamefont {Lukin}}]{Bernien2017}%
  \BibitemOpen
  \bibfield  {author} {\bibinfo {author} {\bibfnamefont {H.}~\bibnamefont {Bernien}}, \bibinfo {author} {\bibfnamefont {S.}~\bibnamefont {Schwartz}}, \bibinfo {author} {\bibfnamefont {A.}~\bibnamefont {Keesling}}, \bibinfo {author} {\bibfnamefont {H.}~\bibnamefont {Levine}}, \bibinfo {author} {\bibfnamefont {A.}~\bibnamefont {Omran}}, \bibinfo {author} {\bibfnamefont {H.}~\bibnamefont {Pichler}}, \bibinfo {author} {\bibfnamefont {S.}~\bibnamefont {Choi}}, \bibinfo {author} {\bibfnamefont {A.~S.}\ \bibnamefont {Zibrov}}, \bibinfo {author} {\bibfnamefont {M.}~\bibnamefont {Endres}}, \bibinfo {author} {\bibfnamefont {M.}~\bibnamefont {Greiner}}, \bibinfo {author} {\bibfnamefont {V.}~\bibnamefont {Vuleti{\'{c}}}},\ and\ \bibinfo {author} {\bibfnamefont {M.~D.}\ \bibnamefont {Lukin}},\ }\bibfield  {title} {\bibinfo {title} {{Probing many-body dynamics on a 51-atom quantum simulator}},\ }\href {https://doi.org/10.1038/nature24622} {\bibfield  {journal} {\bibinfo  {journal} {Nature}\ }\textbf {\bibinfo {volume}
  {551}},\ \bibinfo {pages} {579} (\bibinfo {year} {2017})}\BibitemShut {NoStop}%
\bibitem [{\citenamefont {Fauseweh}(2024)}]{Fauseweh2024}%
  \BibitemOpen
  \bibfield  {author} {\bibinfo {author} {\bibfnamefont {B.}~\bibnamefont {Fauseweh}},\ }\bibfield  {title} {\bibinfo {title} {{Quantum many-body simulations on digital quantum computers: State-of-the-art and future challenges}},\ }\bibfield  {journal} {\bibinfo  {journal} {Nature Communications}\ }\textbf {\bibinfo {volume} {15}},\ \href {https://doi.org/10.1038/s41467-024-46402-9} {10.1038/s41467-024-46402-9} (\bibinfo {year} {2024})\BibitemShut {NoStop}%
\bibitem [{\citenamefont {Wecker}\ \emph {et~al.}(2015)\citenamefont {Wecker}, \citenamefont {Hastings}, \citenamefont {Wiebe}, \citenamefont {Clark}, \citenamefont {Nayak},\ and\ \citenamefont {Troyer}}]{Wecker2015}%
  \BibitemOpen
  \bibfield  {author} {\bibinfo {author} {\bibfnamefont {D.}~\bibnamefont {Wecker}}, \bibinfo {author} {\bibfnamefont {M.~B.}\ \bibnamefont {Hastings}}, \bibinfo {author} {\bibfnamefont {N.}~\bibnamefont {Wiebe}}, \bibinfo {author} {\bibfnamefont {B.~K.}\ \bibnamefont {Clark}}, \bibinfo {author} {\bibfnamefont {C.}~\bibnamefont {Nayak}},\ and\ \bibinfo {author} {\bibfnamefont {M.}~\bibnamefont {Troyer}},\ }\bibfield  {title} {\bibinfo {title} {{Solving strongly correlated electron models on a quantum computer}},\ }\href {https://doi.org/10.1103/PhysRevA.92.062318} {\bibfield  {journal} {\bibinfo  {journal} {Physical Review A - Atomic, Molecular, and Optical Physics}\ }\textbf {\bibinfo {volume} {92}},\ \bibinfo {pages} {1} (\bibinfo {year} {2015})},\ \Eprint {https://arxiv.org/abs/1506.05135} {arXiv:1506.05135} \BibitemShut {NoStop}%
\bibitem [{\citenamefont {Cade}\ \emph {et~al.}(2020)\citenamefont {Cade}, \citenamefont {Mineh}, \citenamefont {Montanaro},\ and\ \citenamefont {Stanisic}}]{Cade2020}%
  \BibitemOpen
  \bibfield  {author} {\bibinfo {author} {\bibfnamefont {C.}~\bibnamefont {Cade}}, \bibinfo {author} {\bibfnamefont {L.}~\bibnamefont {Mineh}}, \bibinfo {author} {\bibfnamefont {A.}~\bibnamefont {Montanaro}},\ and\ \bibinfo {author} {\bibfnamefont {S.}~\bibnamefont {Stanisic}},\ }\bibfield  {title} {\bibinfo {title} {{Strategies for solving the Fermi-Hubbard model on near-term quantum computers}},\ }\href {https://doi.org/10.1103/PhysRevB.102.235122} {\bibfield  {journal} {\bibinfo  {journal} {Physical Review B}\ }\textbf {\bibinfo {volume} {102}},\ \bibinfo {pages} {1} (\bibinfo {year} {2020})},\ \Eprint {https://arxiv.org/abs/1912.06007} {arXiv:1912.06007} \BibitemShut {NoStop}%
\bibitem [{\citenamefont {Alvertis}\ \emph {et~al.}(2024{\natexlab{a}})\citenamefont {Alvertis}, \citenamefont {Khan}, \citenamefont {Iadecola}, \citenamefont {Orth},\ and\ \citenamefont {Tubman}}]{alvertis2024classicalbenchmarksvariationalquantum}%
  \BibitemOpen
  \bibfield  {author} {\bibinfo {author} {\bibfnamefont {A.~M.}\ \bibnamefont {Alvertis}}, \bibinfo {author} {\bibfnamefont {A.}~\bibnamefont {Khan}}, \bibinfo {author} {\bibfnamefont {T.}~\bibnamefont {Iadecola}}, \bibinfo {author} {\bibfnamefont {P.~P.}\ \bibnamefont {Orth}},\ and\ \bibinfo {author} {\bibfnamefont {N.}~\bibnamefont {Tubman}},\ }\href {https://arxiv.org/abs/2408.00836} {\bibinfo {title} {Classical benchmarks for variational quantum eigensolver simulations of the hubbard model}} (\bibinfo {year} {2024}{\natexlab{a}}),\ \Eprint {https://arxiv.org/abs/2408.00836} {arXiv:2408.00836 [quant-ph]} \BibitemShut {NoStop}%
\bibitem [{\citenamefont {Agrawal}\ \emph {et~al.}(2024)\citenamefont {Agrawal}, \citenamefont {Job}, \citenamefont {Wilson}, \citenamefont {Saadatmand}, \citenamefont {Hodson}, \citenamefont {Mutus}, \citenamefont {Caesura}, \citenamefont {Johnson}, \citenamefont {Elenewski}, \citenamefont {Morrell},\ and\ \citenamefont {Kemper}}]{agrawal2024quantifyingfaulttolerantsimulation}%
  \BibitemOpen
  \bibfield  {author} {\bibinfo {author} {\bibfnamefont {A.~A.}\ \bibnamefont {Agrawal}}, \bibinfo {author} {\bibfnamefont {J.}~\bibnamefont {Job}}, \bibinfo {author} {\bibfnamefont {T.~L.}\ \bibnamefont {Wilson}}, \bibinfo {author} {\bibfnamefont {S.~N.}\ \bibnamefont {Saadatmand}}, \bibinfo {author} {\bibfnamefont {M.~J.}\ \bibnamefont {Hodson}}, \bibinfo {author} {\bibfnamefont {J.~Y.}\ \bibnamefont {Mutus}}, \bibinfo {author} {\bibfnamefont {A.}~\bibnamefont {Caesura}}, \bibinfo {author} {\bibfnamefont {P.~D.}\ \bibnamefont {Johnson}}, \bibinfo {author} {\bibfnamefont {J.~E.}\ \bibnamefont {Elenewski}}, \bibinfo {author} {\bibfnamefont {K.~J.}\ \bibnamefont {Morrell}},\ and\ \bibinfo {author} {\bibfnamefont {A.~F.}\ \bibnamefont {Kemper}},\ }\href {https://arxiv.org/abs/2406.06511} {\bibinfo {title} {Quantifying fault tolerant simulation of strongly correlated systems using the fermi-hubbard model}} (\bibinfo {year} {2024}),\ \Eprint {https://arxiv.org/abs/2406.06511} {arXiv:2406.06511 [quant-ph]}
  \BibitemShut {NoStop}%
\bibitem [{\citenamefont {Khan}\ \emph {et~al.}(2023)\citenamefont {Khan}, \citenamefont {Clark},\ and\ \citenamefont {Tubman}}]{khan2023preoptimizing}%
  \BibitemOpen
  \bibfield  {author} {\bibinfo {author} {\bibfnamefont {A.}~\bibnamefont {Khan}}, \bibinfo {author} {\bibfnamefont {B.~K.}\ \bibnamefont {Clark}},\ and\ \bibinfo {author} {\bibfnamefont {N.~M.}\ \bibnamefont {Tubman}},\ }\href@noop {} {\bibinfo {title} {Pre-optimizing variational quantum eigensolvers with tensor networks}} (\bibinfo {year} {2023}),\ \Eprint {https://arxiv.org/abs/2310.12965} {arXiv:2310.12965 [quant-ph]} \BibitemShut {NoStop}%
\bibitem [{\citenamefont {Rosner}\ \emph {et~al.}(1997)\citenamefont {Rosner}, \citenamefont {Eschrig}, \citenamefont {Hayn}, \citenamefont {Drechsler},\ and\ \citenamefont {M{\'{a}}lek}}]{Rosner1997}%
  \BibitemOpen
  \bibfield  {author} {\bibinfo {author} {\bibfnamefont {H.}~\bibnamefont {Rosner}}, \bibinfo {author} {\bibfnamefont {H.}~\bibnamefont {Eschrig}}, \bibinfo {author} {\bibfnamefont {R.}~\bibnamefont {Hayn}}, \bibinfo {author} {\bibfnamefont {S.}~\bibnamefont {Drechsler}},\ and\ \bibinfo {author} {\bibfnamefont {J.}~\bibnamefont {M{\'{a}}lek}},\ }\bibfield  {title} {\bibinfo {title} {{Electronic structure and magnetic properties of the linear chain cuprates Sr2CuO3 and Ca2CuO3}},\ }\href {https://doi.org/10.1103/PhysRevB.56.3402} {\bibfield  {journal} {\bibinfo  {journal} {Physical Review B - Condensed Matter and Materials Physics}\ }\textbf {\bibinfo {volume} {56}},\ \bibinfo {pages} {3402} (\bibinfo {year} {1997})}\BibitemShut {NoStop}%
\bibitem [{\citenamefont {Sun}\ \emph {et~al.}(2022)\citenamefont {Sun}, \citenamefont {Zhao}, \citenamefont {Palomaki}, \citenamefont {Fei}, \citenamefont {Runburg}, \citenamefont {Malinowski}, \citenamefont {Huang}, \citenamefont {Cenker}, \citenamefont {Cui}, \citenamefont {Chu}, \citenamefont {Xu}, \citenamefont {Ataei}, \citenamefont {Varsano}, \citenamefont {Palummo}, \citenamefont {Molinari}, \citenamefont {Rontani},\ and\ \citenamefont {Cobden}}]{Sun2022}%
  \BibitemOpen
  \bibfield  {author} {\bibinfo {author} {\bibfnamefont {B.}~\bibnamefont {Sun}}, \bibinfo {author} {\bibfnamefont {W.}~\bibnamefont {Zhao}}, \bibinfo {author} {\bibfnamefont {T.}~\bibnamefont {Palomaki}}, \bibinfo {author} {\bibfnamefont {Z.}~\bibnamefont {Fei}}, \bibinfo {author} {\bibfnamefont {E.}~\bibnamefont {Runburg}}, \bibinfo {author} {\bibfnamefont {P.}~\bibnamefont {Malinowski}}, \bibinfo {author} {\bibfnamefont {X.}~\bibnamefont {Huang}}, \bibinfo {author} {\bibfnamefont {J.}~\bibnamefont {Cenker}}, \bibinfo {author} {\bibfnamefont {Y.~T.}\ \bibnamefont {Cui}}, \bibinfo {author} {\bibfnamefont {J.~H.}\ \bibnamefont {Chu}}, \bibinfo {author} {\bibfnamefont {X.}~\bibnamefont {Xu}}, \bibinfo {author} {\bibfnamefont {S.~S.}\ \bibnamefont {Ataei}}, \bibinfo {author} {\bibfnamefont {D.}~\bibnamefont {Varsano}}, \bibinfo {author} {\bibfnamefont {M.}~\bibnamefont {Palummo}}, \bibinfo {author} {\bibfnamefont {E.}~\bibnamefont {Molinari}}, \bibinfo {author} {\bibfnamefont {M.}~\bibnamefont {Rontani}},\ and\
  \bibinfo {author} {\bibfnamefont {D.~H.}\ \bibnamefont {Cobden}},\ }\bibfield  {title} {\bibinfo {title} {{Evidence for equilibrium exciton condensation in monolayer WTe2}},\ }\href {https://doi.org/10.1038/s41567-021-01427-5} {\bibfield  {journal} {\bibinfo  {journal} {Nature Physics}\ }\textbf {\bibinfo {volume} {18}},\ \bibinfo {pages} {94} (\bibinfo {year} {2022})}\BibitemShut {NoStop}%
\bibitem [{\citenamefont {Aizaki}\ \emph {et~al.}(2012)\citenamefont {Aizaki}, \citenamefont {Yoshida}, \citenamefont {Yoshimatsu}, \citenamefont {Takizawa}, \citenamefont {Minohara}, \citenamefont {Ideta}, \citenamefont {Fujimori}, \citenamefont {Gupta}, \citenamefont {Mahadevan}, \citenamefont {Horiba}, \citenamefont {Kumigashira},\ and\ \citenamefont {Oshima}}]{Aizaki2012}%
  \BibitemOpen
  \bibfield  {author} {\bibinfo {author} {\bibfnamefont {S.}~\bibnamefont {Aizaki}}, \bibinfo {author} {\bibfnamefont {T.}~\bibnamefont {Yoshida}}, \bibinfo {author} {\bibfnamefont {K.}~\bibnamefont {Yoshimatsu}}, \bibinfo {author} {\bibfnamefont {M.}~\bibnamefont {Takizawa}}, \bibinfo {author} {\bibfnamefont {M.}~\bibnamefont {Minohara}}, \bibinfo {author} {\bibfnamefont {S.}~\bibnamefont {Ideta}}, \bibinfo {author} {\bibfnamefont {A.}~\bibnamefont {Fujimori}}, \bibinfo {author} {\bibfnamefont {K.}~\bibnamefont {Gupta}}, \bibinfo {author} {\bibfnamefont {P.}~\bibnamefont {Mahadevan}}, \bibinfo {author} {\bibfnamefont {K.}~\bibnamefont {Horiba}}, \bibinfo {author} {\bibfnamefont {H.}~\bibnamefont {Kumigashira}},\ and\ \bibinfo {author} {\bibfnamefont {M.}~\bibnamefont {Oshima}},\ }\bibfield  {title} {\bibinfo {title} {{Self-energy on the low- to high-energy electronic structure of correlated metal SrVO 3}},\ }\href {https://doi.org/10.1103/PhysRevLett.109.056401} {\bibfield  {journal} {\bibinfo  {journal}
  {Physical Review Letters}\ }\textbf {\bibinfo {volume} {109}},\ \bibinfo {pages} {1} (\bibinfo {year} {2012})}\BibitemShut {NoStop}%
\bibitem [{\citenamefont {Zhang}\ \emph {et~al.}(2016)\citenamefont {Zhang}, \citenamefont {Zhou}, \citenamefont {Guo}, \citenamefont {Zhao}, \citenamefont {Barnes}, \citenamefont {Zhang}, \citenamefont {Eaton}, \citenamefont {Zheng}, \citenamefont {Brahlek}, \citenamefont {Haneef}, \citenamefont {Podraza}, \citenamefont {Chan}, \citenamefont {Gopalan}, \citenamefont {Rabe},\ and\ \citenamefont {Engel-Herbert}}]{Zhang2016}%
  \BibitemOpen
  \bibfield  {author} {\bibinfo {author} {\bibfnamefont {L.}~\bibnamefont {Zhang}}, \bibinfo {author} {\bibfnamefont {Y.}~\bibnamefont {Zhou}}, \bibinfo {author} {\bibfnamefont {L.}~\bibnamefont {Guo}}, \bibinfo {author} {\bibfnamefont {W.}~\bibnamefont {Zhao}}, \bibinfo {author} {\bibfnamefont {A.}~\bibnamefont {Barnes}}, \bibinfo {author} {\bibfnamefont {H.~T.}\ \bibnamefont {Zhang}}, \bibinfo {author} {\bibfnamefont {C.}~\bibnamefont {Eaton}}, \bibinfo {author} {\bibfnamefont {Y.}~\bibnamefont {Zheng}}, \bibinfo {author} {\bibfnamefont {M.}~\bibnamefont {Brahlek}}, \bibinfo {author} {\bibfnamefont {H.~F.}\ \bibnamefont {Haneef}}, \bibinfo {author} {\bibfnamefont {N.~J.}\ \bibnamefont {Podraza}}, \bibinfo {author} {\bibfnamefont {M.~H.}\ \bibnamefont {Chan}}, \bibinfo {author} {\bibfnamefont {V.}~\bibnamefont {Gopalan}}, \bibinfo {author} {\bibfnamefont {K.~M.}\ \bibnamefont {Rabe}},\ and\ \bibinfo {author} {\bibfnamefont {R.}~\bibnamefont {Engel-Herbert}},\ }\bibfield  {title} {\bibinfo {title}
  {{Correlated metals as transparent conductors}},\ }\href {https://doi.org/10.1038/nmat4493} {\bibfield  {journal} {\bibinfo  {journal} {Nature Materials}\ }\textbf {\bibinfo {volume} {15}},\ \bibinfo {pages} {204} (\bibinfo {year} {2016})}\BibitemShut {NoStop}%
\bibitem [{\citenamefont {Bauman}\ \emph {et~al.}(2019)\citenamefont {Bauman}, \citenamefont {Low},\ and\ \citenamefont {Kowalski}}]{Bauman2019}%
  \BibitemOpen
  \bibfield  {author} {\bibinfo {author} {\bibfnamefont {N.~P.}\ \bibnamefont {Bauman}}, \bibinfo {author} {\bibfnamefont {G.~H.}\ \bibnamefont {Low}},\ and\ \bibinfo {author} {\bibfnamefont {K.}~\bibnamefont {Kowalski}},\ }\bibfield  {title} {\bibinfo {title} {{Quantum simulations of excited states with active-space downfolded Hamiltonians}},\ }\bibfield  {journal} {\bibinfo  {journal} {Journal of Chemical Physics}\ }\textbf {\bibinfo {volume} {151}},\ \href {https://doi.org/10.1063/1.5128103} {10.1063/1.5128103} (\bibinfo {year} {2019}),\ \Eprint {https://arxiv.org/abs/1909.06404} {arXiv:1909.06404} \BibitemShut {NoStop}%
\bibitem [{\citenamefont {Chang}\ \emph {et~al.}(2023)\citenamefont {Chang}, \citenamefont {van Loon}, \citenamefont {Eskridge}, \citenamefont {Busemeyer}, \citenamefont {Morales}, \citenamefont {Dreyer}, \citenamefont {Millis}, \citenamefont {Zhang}, \citenamefont {Wehling}, \citenamefont {Wagner},\ and\ \citenamefont {Rösner}}]{chang2023downfolding}%
  \BibitemOpen
  \bibfield  {author} {\bibinfo {author} {\bibfnamefont {Y.}~\bibnamefont {Chang}}, \bibinfo {author} {\bibfnamefont {E.~G. C.~P.}\ \bibnamefont {van Loon}}, \bibinfo {author} {\bibfnamefont {B.}~\bibnamefont {Eskridge}}, \bibinfo {author} {\bibfnamefont {B.}~\bibnamefont {Busemeyer}}, \bibinfo {author} {\bibfnamefont {M.~A.}\ \bibnamefont {Morales}}, \bibinfo {author} {\bibfnamefont {C.~E.}\ \bibnamefont {Dreyer}}, \bibinfo {author} {\bibfnamefont {A.~J.}\ \bibnamefont {Millis}}, \bibinfo {author} {\bibfnamefont {S.}~\bibnamefont {Zhang}}, \bibinfo {author} {\bibfnamefont {T.~O.}\ \bibnamefont {Wehling}}, \bibinfo {author} {\bibfnamefont {L.~K.}\ \bibnamefont {Wagner}},\ and\ \bibinfo {author} {\bibfnamefont {M.}~\bibnamefont {Rösner}},\ }\href@noop {} {\bibinfo {title} {Downfolding from ab initio to interacting model hamiltonians: Comprehensive analysis and benchmarking}} (\bibinfo {year} {2023}),\ \Eprint {https://arxiv.org/abs/2311.05987} {arXiv:2311.05987 [cond-mat.str-el]} \BibitemShut {NoStop}%
\bibitem [{\citenamefont {Yoshida}\ \emph {et~al.}(2024)\citenamefont {Yoshida}, \citenamefont {Takemori},\ and\ \citenamefont {Mizukami}}]{10.1063/5.0213525}%
  \BibitemOpen
  \bibfield  {author} {\bibinfo {author} {\bibfnamefont {Y.}~\bibnamefont {Yoshida}}, \bibinfo {author} {\bibfnamefont {N.}~\bibnamefont {Takemori}},\ and\ \bibinfo {author} {\bibfnamefont {W.}~\bibnamefont {Mizukami}},\ }\bibfield  {title} {\bibinfo {title} {{Ab initio extended Hubbard model of short polyenes for efficient quantum computing}},\ }\href {https://doi.org/10.1063/5.0213525} {\bibfield  {journal} {\bibinfo  {journal} {The Journal of Chemical Physics}\ }\textbf {\bibinfo {volume} {161}},\ \bibinfo {pages} {084303} (\bibinfo {year} {2024})},\ \Eprint {https://arxiv.org/abs/https://pubs.aip.org/aip/jcp/article-pdf/doi/10.1063/5.0213525/20131956/084303\_1\_5.0213525.pdf} {https://pubs.aip.org/aip/jcp/article-pdf/doi/10.1063/5.0213525/20131956/084303\_1\_5.0213525.pdf} \BibitemShut {NoStop}%
\bibitem [{\citenamefont {Marzari}\ \emph {et~al.}(2012)\citenamefont {Marzari}, \citenamefont {Mostofi}, \citenamefont {Yates}, \citenamefont {Souza},\ and\ \citenamefont {Vanderbilt}}]{Marzari2012}%
  \BibitemOpen
  \bibfield  {author} {\bibinfo {author} {\bibfnamefont {N.}~\bibnamefont {Marzari}}, \bibinfo {author} {\bibfnamefont {A.~A.}\ \bibnamefont {Mostofi}}, \bibinfo {author} {\bibfnamefont {J.~R.}\ \bibnamefont {Yates}}, \bibinfo {author} {\bibfnamefont {I.}~\bibnamefont {Souza}},\ and\ \bibinfo {author} {\bibfnamefont {D.}~\bibnamefont {Vanderbilt}},\ }\bibfield  {title} {\bibinfo {title} {{Maximally localized Wannier functions: Theory and applications}},\ }\href {https://doi.org/10.1103/RevModPhys.84.1419} {\bibfield  {journal} {\bibinfo  {journal} {Reviews of Modern Physics}\ }\textbf {\bibinfo {volume} {84}},\ \bibinfo {pages} {1419} (\bibinfo {year} {2012})},\ \Eprint {https://arxiv.org/abs/1112.5411} {1112.5411} \BibitemShut {NoStop}%
\bibitem [{\citenamefont {Giannozzi}\ \emph {et~al.}(2009)\citenamefont {Giannozzi}, \citenamefont {Baroni}, \citenamefont {Bonini}, \citenamefont {Calandra}, \citenamefont {Car}, \citenamefont {Cavazzoni}, \citenamefont {Ceresoli}, \citenamefont {Chiarotti}, \citenamefont {Cococcioni}, \citenamefont {Dabo}, \citenamefont {{Dal Corso}}, \citenamefont {Fabris}, \citenamefont {Fratesi}, \citenamefont {{de Gironcoli}}, \citenamefont {Gebauer}, \citenamefont {Gerstmann}, \citenamefont {Gougoussis}, \citenamefont {Kokalj}, \citenamefont {Lazzeri}, \citenamefont {Martin-Samos}, \citenamefont {Marzari}, \citenamefont {Mauri}, \citenamefont {Mazzarello}, \citenamefont {Paolini}, \citenamefont {Pasquarello}, \citenamefont {Paulatto}, \citenamefont {Sbraccia}, \citenamefont {Scandolo}, \citenamefont {Sclauzero}, \citenamefont {Seitsonen}, \citenamefont {Smogunov}, \citenamefont {Umari},\ and\ \citenamefont {Wentzcovitch}}]{QE}%
  \BibitemOpen
  \bibfield  {author} {\bibinfo {author} {\bibfnamefont {P.}~\bibnamefont {Giannozzi}}, \bibinfo {author} {\bibfnamefont {S.}~\bibnamefont {Baroni}}, \bibinfo {author} {\bibfnamefont {N.}~\bibnamefont {Bonini}}, \bibinfo {author} {\bibfnamefont {M.}~\bibnamefont {Calandra}}, \bibinfo {author} {\bibfnamefont {R.}~\bibnamefont {Car}}, \bibinfo {author} {\bibfnamefont {C.}~\bibnamefont {Cavazzoni}}, \bibinfo {author} {\bibfnamefont {D.}~\bibnamefont {Ceresoli}}, \bibinfo {author} {\bibfnamefont {G.~L.}\ \bibnamefont {Chiarotti}}, \bibinfo {author} {\bibfnamefont {M.}~\bibnamefont {Cococcioni}}, \bibinfo {author} {\bibfnamefont {I.}~\bibnamefont {Dabo}}, \bibinfo {author} {\bibfnamefont {A.}~\bibnamefont {{Dal Corso}}}, \bibinfo {author} {\bibfnamefont {S.}~\bibnamefont {Fabris}}, \bibinfo {author} {\bibfnamefont {G.}~\bibnamefont {Fratesi}}, \bibinfo {author} {\bibfnamefont {S.}~\bibnamefont {{de Gironcoli}}}, \bibinfo {author} {\bibfnamefont {R.}~\bibnamefont {Gebauer}}, \bibinfo {author} {\bibfnamefont
  {U.}~\bibnamefont {Gerstmann}}, \bibinfo {author} {\bibfnamefont {C.}~\bibnamefont {Gougoussis}}, \bibinfo {author} {\bibfnamefont {A.}~\bibnamefont {Kokalj}}, \bibinfo {author} {\bibfnamefont {M.}~\bibnamefont {Lazzeri}}, \bibinfo {author} {\bibfnamefont {L.}~\bibnamefont {Martin-Samos}}, \bibinfo {author} {\bibfnamefont {N.}~\bibnamefont {Marzari}}, \bibinfo {author} {\bibfnamefont {F.}~\bibnamefont {Mauri}}, \bibinfo {author} {\bibfnamefont {R.}~\bibnamefont {Mazzarello}}, \bibinfo {author} {\bibfnamefont {S.}~\bibnamefont {Paolini}}, \bibinfo {author} {\bibfnamefont {A.}~\bibnamefont {Pasquarello}}, \bibinfo {author} {\bibfnamefont {L.}~\bibnamefont {Paulatto}}, \bibinfo {author} {\bibfnamefont {C.}~\bibnamefont {Sbraccia}}, \bibinfo {author} {\bibfnamefont {S.}~\bibnamefont {Scandolo}}, \bibinfo {author} {\bibfnamefont {G.}~\bibnamefont {Sclauzero}}, \bibinfo {author} {\bibfnamefont {A.~P.}\ \bibnamefont {Seitsonen}}, \bibinfo {author} {\bibfnamefont {A.}~\bibnamefont {Smogunov}}, \bibinfo {author}
  {\bibfnamefont {P.}~\bibnamefont {Umari}},\ and\ \bibinfo {author} {\bibfnamefont {R.~M.}\ \bibnamefont {Wentzcovitch}},\ }\bibfield  {title} {\bibinfo {title} {{QUANTUM ESPRESSO: a modular and open-source software project for quantum simulations of materials}},\ }\href@noop {} {\bibfield  {journal} {\bibinfo  {journal} {Journal of Physics: Condensed Matter}\ }\textbf {\bibinfo {volume} {21}},\ \bibinfo {pages} {395502} (\bibinfo {year} {2009})}\BibitemShut {NoStop}%
\bibitem [{\citenamefont {Pizzi}\ \emph {et~al.}(2020)\citenamefont {Pizzi}, \citenamefont {Vitale}, \citenamefont {Arita}, \citenamefont {Bl{\"{u}}gel}, \citenamefont {Freimuth}, \citenamefont {G{\'{e}}ranton}, \citenamefont {Gibertini}, \citenamefont {Gresch}, \citenamefont {Johnson}, \citenamefont {Koretsune}, \citenamefont {Ibanez-Azpiroz}, \citenamefont {Lee}, \citenamefont {Lihm}, \citenamefont {Marchand}, \citenamefont {Marrazzo}, \citenamefont {Mokrousov}, \citenamefont {Mustafa}, \citenamefont {Nohara}, \citenamefont {Nomura}, \citenamefont {Paulatto}, \citenamefont {Ponc{\'{e}}}, \citenamefont {Ponweiser}, \citenamefont {Qiao}, \citenamefont {Th{\"{o}}le}, \citenamefont {Tsirkin}, \citenamefont {Wierzbowska}, \citenamefont {Marzari}, \citenamefont {Vanderbilt}, \citenamefont {Souza}, \citenamefont {Mostofi},\ and\ \citenamefont {Yates}}]{Pizzi2020}%
  \BibitemOpen
  \bibfield  {author} {\bibinfo {author} {\bibfnamefont {G.}~\bibnamefont {Pizzi}}, \bibinfo {author} {\bibfnamefont {V.}~\bibnamefont {Vitale}}, \bibinfo {author} {\bibfnamefont {R.}~\bibnamefont {Arita}}, \bibinfo {author} {\bibfnamefont {S.}~\bibnamefont {Bl{\"{u}}gel}}, \bibinfo {author} {\bibfnamefont {F.}~\bibnamefont {Freimuth}}, \bibinfo {author} {\bibfnamefont {G.}~\bibnamefont {G{\'{e}}ranton}}, \bibinfo {author} {\bibfnamefont {M.}~\bibnamefont {Gibertini}}, \bibinfo {author} {\bibfnamefont {D.}~\bibnamefont {Gresch}}, \bibinfo {author} {\bibfnamefont {C.}~\bibnamefont {Johnson}}, \bibinfo {author} {\bibfnamefont {T.}~\bibnamefont {Koretsune}}, \bibinfo {author} {\bibfnamefont {J.}~\bibnamefont {Ibanez-Azpiroz}}, \bibinfo {author} {\bibfnamefont {H.}~\bibnamefont {Lee}}, \bibinfo {author} {\bibfnamefont {J.~M.}\ \bibnamefont {Lihm}}, \bibinfo {author} {\bibfnamefont {D.}~\bibnamefont {Marchand}}, \bibinfo {author} {\bibfnamefont {A.}~\bibnamefont {Marrazzo}}, \bibinfo {author} {\bibfnamefont
  {Y.}~\bibnamefont {Mokrousov}}, \bibinfo {author} {\bibfnamefont {J.~I.}\ \bibnamefont {Mustafa}}, \bibinfo {author} {\bibfnamefont {Y.}~\bibnamefont {Nohara}}, \bibinfo {author} {\bibfnamefont {Y.}~\bibnamefont {Nomura}}, \bibinfo {author} {\bibfnamefont {L.}~\bibnamefont {Paulatto}}, \bibinfo {author} {\bibfnamefont {S.}~\bibnamefont {Ponc{\'{e}}}}, \bibinfo {author} {\bibfnamefont {T.}~\bibnamefont {Ponweiser}}, \bibinfo {author} {\bibfnamefont {J.}~\bibnamefont {Qiao}}, \bibinfo {author} {\bibfnamefont {F.}~\bibnamefont {Th{\"{o}}le}}, \bibinfo {author} {\bibfnamefont {S.~S.}\ \bibnamefont {Tsirkin}}, \bibinfo {author} {\bibfnamefont {M.}~\bibnamefont {Wierzbowska}}, \bibinfo {author} {\bibfnamefont {N.}~\bibnamefont {Marzari}}, \bibinfo {author} {\bibfnamefont {D.}~\bibnamefont {Vanderbilt}}, \bibinfo {author} {\bibfnamefont {I.}~\bibnamefont {Souza}}, \bibinfo {author} {\bibfnamefont {A.~A.}\ \bibnamefont {Mostofi}},\ and\ \bibinfo {author} {\bibfnamefont {J.~R.}\ \bibnamefont {Yates}},\ }\bibfield
  {title} {\bibinfo {title} {{Wannier90 as a community code: New features and applications}},\ }\bibfield  {journal} {\bibinfo  {journal} {Journal of Physics Condensed Matter}\ }\textbf {\bibinfo {volume} {32}},\ \href {https://doi.org/10.1088/1361-648X/ab51ff} {10.1088/1361-648X/ab51ff} (\bibinfo {year} {2020})\BibitemShut {NoStop}%
\bibitem [{\citenamefont {Kurita}\ \emph {et~al.}(2023)\citenamefont {Kurita}, \citenamefont {Misawa}, \citenamefont {Yoshimi}, \citenamefont {Ido},\ and\ \citenamefont {Koretsune}}]{Kurita2023}%
  \BibitemOpen
  \bibfield  {author} {\bibinfo {author} {\bibfnamefont {K.}~\bibnamefont {Kurita}}, \bibinfo {author} {\bibfnamefont {T.}~\bibnamefont {Misawa}}, \bibinfo {author} {\bibfnamefont {K.}~\bibnamefont {Yoshimi}}, \bibinfo {author} {\bibfnamefont {K.}~\bibnamefont {Ido}},\ and\ \bibinfo {author} {\bibfnamefont {T.}~\bibnamefont {Koretsune}},\ }\bibfield  {title} {\bibinfo {title} {{Interface tool from Wannier90 to RESPACK: wan2respack}},\ }\href {https://doi.org/10.1016/j.cpc.2023.108854} {\bibfield  {journal} {\bibinfo  {journal} {Computer Physics Communications}\ }\textbf {\bibinfo {volume} {292}},\ \bibinfo {pages} {108854} (\bibinfo {year} {2023})},\ \Eprint {https://arxiv.org/abs/2302.13531} {2302.13531} \BibitemShut {NoStop}%
\bibitem [{\citenamefont {Romanova}\ \emph {et~al.}(2023)\citenamefont {Romanova}, \citenamefont {Weng}, \citenamefont {Apelian},\ and\ \citenamefont {Vl{\v{c}}ek}}]{Romanova2023}%
  \BibitemOpen
  \bibfield  {author} {\bibinfo {author} {\bibfnamefont {M.}~\bibnamefont {Romanova}}, \bibinfo {author} {\bibfnamefont {G.}~\bibnamefont {Weng}}, \bibinfo {author} {\bibfnamefont {A.}~\bibnamefont {Apelian}},\ and\ \bibinfo {author} {\bibfnamefont {V.}~\bibnamefont {Vl{\v{c}}ek}},\ }\bibfield  {title} {\bibinfo {title} {{Dynamical downfolding for localized quantum states}},\ }\bibfield  {journal} {\bibinfo  {journal} {npj Computational Materials}\ }\textbf {\bibinfo {volume} {9}},\ \href {https://doi.org/10.1038/s41524-023-01078-5} {10.1038/s41524-023-01078-5} (\bibinfo {year} {2023})\BibitemShut {NoStop}%
\bibitem [{\citenamefont {Jiang}\ and\ \citenamefont {Devereaux}(2019)}]{doi:10.1126/science.aal5304}%
  \BibitemOpen
  \bibfield  {author} {\bibinfo {author} {\bibfnamefont {H.-C.}\ \bibnamefont {Jiang}}\ and\ \bibinfo {author} {\bibfnamefont {T.~P.}\ \bibnamefont {Devereaux}},\ }\bibfield  {title} {\bibinfo {title} {{Superconductivity in the doped Hubbard model and its interplay with next-nearest hopping <i>t</i>\&\#x2032;}},\ }\href {https://doi.org/10.1126/science.aal5304} {\bibfield  {journal} {\bibinfo  {journal} {Science}\ }\textbf {\bibinfo {volume} {365}},\ \bibinfo {pages} {1424} (\bibinfo {year} {2019})}\BibitemShut {NoStop}%
\bibitem [{\citenamefont {Lin}\ and\ \citenamefont {Hirsch}(1987)}]{PhysRevB.35.3359}%
  \BibitemOpen
  \bibfield  {author} {\bibinfo {author} {\bibfnamefont {H.~Q.}\ \bibnamefont {Lin}}\ and\ \bibinfo {author} {\bibfnamefont {J.~E.}\ \bibnamefont {Hirsch}},\ }\bibfield  {title} {\bibinfo {title} {Two-dimensional hubbard model with nearest- and next-nearest-neighbor hopping},\ }\href {https://doi.org/10.1103/PhysRevB.35.3359} {\bibfield  {journal} {\bibinfo  {journal} {Phys. Rev. B}\ }\textbf {\bibinfo {volume} {35}},\ \bibinfo {pages} {3359} (\bibinfo {year} {1987})}\BibitemShut {NoStop}%
\bibitem [{\citenamefont {Giustino}\ \emph {et~al.}(2007)\citenamefont {Giustino}, \citenamefont {Cohen},\ and\ \citenamefont {Louie}}]{PhysRevB.76.165108}%
  \BibitemOpen
  \bibfield  {author} {\bibinfo {author} {\bibfnamefont {F.}~\bibnamefont {Giustino}}, \bibinfo {author} {\bibfnamefont {M.~L.}\ \bibnamefont {Cohen}},\ and\ \bibinfo {author} {\bibfnamefont {S.~G.}\ \bibnamefont {Louie}},\ }\bibfield  {title} {\bibinfo {title} {Electron-phonon interaction using wannier functions},\ }\href {https://doi.org/10.1103/PhysRevB.76.165108} {\bibfield  {journal} {\bibinfo  {journal} {Phys. Rev. B}\ }\textbf {\bibinfo {volume} {76}},\ \bibinfo {pages} {165108} (\bibinfo {year} {2007})}\BibitemShut {NoStop}%
\bibitem [{\citenamefont {Scott}\ and\ \citenamefont {Booth}(2024)}]{Scott2024}%
  \BibitemOpen
  \bibfield  {author} {\bibinfo {author} {\bibfnamefont {C.~J.}\ \bibnamefont {Scott}}\ and\ \bibinfo {author} {\bibfnamefont {G.~H.}\ \bibnamefont {Booth}},\ }\bibfield  {title} {\bibinfo {title} {{Rigorous Screened Interactions for Realistic Correlated Electron Systems}},\ }\href {https://doi.org/10.1103/PhysRevLett.132.076401} {\bibfield  {journal} {\bibinfo  {journal} {Physical Review Letters}\ }\textbf {\bibinfo {volume} {132}},\ \bibinfo {pages} {76401} (\bibinfo {year} {2024})},\ \Eprint {https://arxiv.org/abs/2307.13584} {2307.13584} \BibitemShut {NoStop}%
\bibitem [{\citenamefont {Aryasetiawan}\ \emph {et~al.}(2004)\citenamefont {Aryasetiawan}, \citenamefont {Imada}, \citenamefont {Georges}, \citenamefont {Kotliar}, \citenamefont {Biermann},\ and\ \citenamefont {Lichtenstein}}]{PhysRevB.70.195104}%
  \BibitemOpen
  \bibfield  {author} {\bibinfo {author} {\bibfnamefont {F.}~\bibnamefont {Aryasetiawan}}, \bibinfo {author} {\bibfnamefont {M.}~\bibnamefont {Imada}}, \bibinfo {author} {\bibfnamefont {A.}~\bibnamefont {Georges}}, \bibinfo {author} {\bibfnamefont {G.}~\bibnamefont {Kotliar}}, \bibinfo {author} {\bibfnamefont {S.}~\bibnamefont {Biermann}},\ and\ \bibinfo {author} {\bibfnamefont {A.~I.}\ \bibnamefont {Lichtenstein}},\ }\bibfield  {title} {\bibinfo {title} {Frequency-dependent local interactions and low-energy effective models from electronic structure calculations},\ }\href {https://doi.org/10.1103/PhysRevB.70.195104} {\bibfield  {journal} {\bibinfo  {journal} {Phys. Rev. B}\ }\textbf {\bibinfo {volume} {70}},\ \bibinfo {pages} {195104} (\bibinfo {year} {2004})}\BibitemShut {NoStop}%
\bibitem [{\citenamefont {Or{\'{u}}s}(2014)}]{Orus2014}%
  \BibitemOpen
  \bibfield  {author} {\bibinfo {author} {\bibfnamefont {R.}~\bibnamefont {Or{\'{u}}s}},\ }\bibfield  {title} {\bibinfo {title} {{A practical introduction to tensor networks: Matrix product states and projected entangled pair states}},\ }\href {https://doi.org/10.1016/j.aop.2014.06.013} {\bibfield  {journal} {\bibinfo  {journal} {Annals of Physics}\ }\textbf {\bibinfo {volume} {349}},\ \bibinfo {pages} {117} (\bibinfo {year} {2014})},\ \Eprint {https://arxiv.org/abs/1306.2164} {1306.2164} \BibitemShut {NoStop}%
\bibitem [{\citenamefont {Fishman}\ \emph {et~al.}(2022)\citenamefont {Fishman}, \citenamefont {White},\ and\ \citenamefont {Stoudenmire}}]{itensor}%
  \BibitemOpen
  \bibfield  {author} {\bibinfo {author} {\bibfnamefont {M.}~\bibnamefont {Fishman}}, \bibinfo {author} {\bibfnamefont {S.~R.}\ \bibnamefont {White}},\ and\ \bibinfo {author} {\bibfnamefont {E.~M.}\ \bibnamefont {Stoudenmire}},\ }\bibfield  {title} {\bibinfo {title} {{The ITensor Software Library for Tensor Network Calculations}},\ }\href {https://doi.org/10.21468/SciPostPhysCodeb.4} {\bibfield  {journal} {\bibinfo  {journal} {SciPost Phys. Codebases}\ ,\ \bibinfo {pages} {4}} (\bibinfo {year} {2022})}\BibitemShut {NoStop}%
\bibitem [{\citenamefont {{Qiskit contributors}}(2023)}]{Qiskit}%
  \BibitemOpen
  \bibfield  {author} {\bibinfo {author} {\bibnamefont {{Qiskit contributors}}},\ }\href {https://doi.org/10.5281/zenodo.2573505} {\bibinfo {title} {Qiskit: An open-source framework for quantum computing}} (\bibinfo {year} {2023})\BibitemShut {NoStop}%
\bibitem [{\citenamefont {Gonthier}\ \emph {et~al.}(2022)\citenamefont {Gonthier}, \citenamefont {Radin}, \citenamefont {Buda}, \citenamefont {Doskocil}, \citenamefont {Abuan},\ and\ \citenamefont {Romero}}]{PhysRevResearch.4.033154}%
  \BibitemOpen
  \bibfield  {author} {\bibinfo {author} {\bibfnamefont {J.~F.}\ \bibnamefont {Gonthier}}, \bibinfo {author} {\bibfnamefont {M.~D.}\ \bibnamefont {Radin}}, \bibinfo {author} {\bibfnamefont {C.}~\bibnamefont {Buda}}, \bibinfo {author} {\bibfnamefont {E.~J.}\ \bibnamefont {Doskocil}}, \bibinfo {author} {\bibfnamefont {C.~M.}\ \bibnamefont {Abuan}},\ and\ \bibinfo {author} {\bibfnamefont {J.}~\bibnamefont {Romero}},\ }\bibfield  {title} {\bibinfo {title} {Measurements as a roadblock to near-term practical quantum advantage in chemistry: Resource analysis},\ }\href {https://doi.org/10.1103/PhysRevResearch.4.033154} {\bibfield  {journal} {\bibinfo  {journal} {Phys. Rev. Res.}\ }\textbf {\bibinfo {volume} {4}},\ \bibinfo {pages} {033154} (\bibinfo {year} {2022})}\BibitemShut {NoStop}%
\bibitem [{\citenamefont {Clinton}\ \emph {et~al.}(2024)\citenamefont {Clinton}, \citenamefont {Cubitt}, \citenamefont {Flynn}, \citenamefont {Gambetta}, \citenamefont {Klassen}, \citenamefont {Montanaro}, \citenamefont {Piddock}, \citenamefont {Santos},\ and\ \citenamefont {Sheridan}}]{Clinton2024}%
  \BibitemOpen
  \bibfield  {author} {\bibinfo {author} {\bibfnamefont {L.}~\bibnamefont {Clinton}}, \bibinfo {author} {\bibfnamefont {T.}~\bibnamefont {Cubitt}}, \bibinfo {author} {\bibfnamefont {B.}~\bibnamefont {Flynn}}, \bibinfo {author} {\bibfnamefont {F.~M.}\ \bibnamefont {Gambetta}}, \bibinfo {author} {\bibfnamefont {J.}~\bibnamefont {Klassen}}, \bibinfo {author} {\bibfnamefont {A.}~\bibnamefont {Montanaro}}, \bibinfo {author} {\bibfnamefont {S.}~\bibnamefont {Piddock}}, \bibinfo {author} {\bibfnamefont {R.~A.}\ \bibnamefont {Santos}},\ and\ \bibinfo {author} {\bibfnamefont {E.}~\bibnamefont {Sheridan}},\ }\bibfield  {title} {\bibinfo {title} {Towards near-term quantum simulation of materials},\ }\href {https://doi.org/10.1038/s41467-023-43479-6} {\bibfield  {journal} {\bibinfo  {journal} {Nature Communications}\ }\textbf {\bibinfo {volume} {15}},\ \bibinfo {pages} {211} (\bibinfo {year} {2024})}\BibitemShut {NoStop}%
\bibitem [{IBM()}]{IBM}%
  \BibitemOpen
  \href@noop {} {\bibinfo {title} {Pushing quantum performance forward with our highest quantum volume yet}},\ \bibinfo {howpublished} {\url{hhttps://www.ibm.com/quantum/blog/quantum-volume-256}},\ \bibinfo {note} {accessed: 2025-02-12}\BibitemShut {NoStop}%
\bibitem [{\citenamefont {O'Brien}\ \emph {et~al.}(2023)\citenamefont {O'Brien}, \citenamefont {Anselmetti}, \citenamefont {Gkritsis}, \citenamefont {Elfving}, \citenamefont {Polla}, \citenamefont {Huggins}, \citenamefont {Oumarou}, \citenamefont {Kechedzhi}, \citenamefont {Abanin}, \citenamefont {Acharya}, \citenamefont {Aleiner}, \citenamefont {Allen}, \citenamefont {Andersen}, \citenamefont {Anderson}, \citenamefont {Ansmann}, \citenamefont {Arute}, \citenamefont {Arya}, \citenamefont {Asfaw}, \citenamefont {Atalaya}, \citenamefont {Bardin}, \citenamefont {Bengtsson}, \citenamefont {Bortoli}, \citenamefont {Bourassa}, \citenamefont {Bovaird}, \citenamefont {Brill}, \citenamefont {Broughton}, \citenamefont {Buckley}, \citenamefont {Buell}, \citenamefont {Burger}, \citenamefont {Burkett}, \citenamefont {Bushnell}, \citenamefont {Campero}, \citenamefont {Chen}, \citenamefont {Chiaro}, \citenamefont {Chik}, \citenamefont {Cogan}, \citenamefont {Collins}, \citenamefont {Conner}, \citenamefont {Courtney},
  \citenamefont {Crook}, \citenamefont {Curtin}, \citenamefont {Debroy}, \citenamefont {Demura}, \citenamefont {Drozdov}, \citenamefont {Dunsworth}, \citenamefont {Erickson}, \citenamefont {Faoro}, \citenamefont {Farhi}, \citenamefont {Fatemi}, \citenamefont {Ferreira}, \citenamefont {Flores~Burgos}, \citenamefont {Forati}, \citenamefont {Fowler}, \citenamefont {Foxen}, \citenamefont {Giang}, \citenamefont {Gidney}, \citenamefont {Gilboa}, \citenamefont {Giustina}, \citenamefont {Gosula}, \citenamefont {Grajales~Dau}, \citenamefont {Gross}, \citenamefont {Habegger}, \citenamefont {Hamilton}, \citenamefont {Hansen}, \citenamefont {Harrigan}, \citenamefont {Harrington}, \citenamefont {Heu}, \citenamefont {Hoffmann}, \citenamefont {Hong}, \citenamefont {Huang}, \citenamefont {Huff}, \citenamefont {Ioffe}, \citenamefont {Isakov}, \citenamefont {Iveland}, \citenamefont {Jeffrey}, \citenamefont {Jiang}, \citenamefont {Jones}, \citenamefont {Juhas}, \citenamefont {Kafri}, \citenamefont {Khattar}, \citenamefont
  {Khezri}, \citenamefont {Kieferov{\'a}}, \citenamefont {Kim}, \citenamefont {Klimov}, \citenamefont {Klots}, \citenamefont {Korotkov}, \citenamefont {Kostritsa}, \citenamefont {Kreikebaum}, \citenamefont {Landhuis}, \citenamefont {Laptev}, \citenamefont {Lau}, \citenamefont {Laws}, \citenamefont {Lee}, \citenamefont {Lee}, \citenamefont {Lester}, \citenamefont {Lill}, \citenamefont {Liu}, \citenamefont {Livingston}, \citenamefont {Locharla}, \citenamefont {Malone}, \citenamefont {Mandr{\`a}}, \citenamefont {Martin}, \citenamefont {Martin}, \citenamefont {McClean}, \citenamefont {McCourt}, \citenamefont {McEwen}, \citenamefont {Mi}, \citenamefont {Mieszala}, \citenamefont {Miao}, \citenamefont {Mohseni}, \citenamefont {Montazeri}, \citenamefont {Morvan}, \citenamefont {Movassagh}, \citenamefont {Mruczkiewicz}, \citenamefont {Naaman}, \citenamefont {Neeley}, \citenamefont {Neill}, \citenamefont {Nersisyan}, \citenamefont {Newman}, \citenamefont {Ng}, \citenamefont {Nguyen}, \citenamefont {Nguyen},
  \citenamefont {Niu}, \citenamefont {Omonije}, \citenamefont {Opremcak}, \citenamefont {Petukhov}, \citenamefont {Potter}, \citenamefont {Pryadko}, \citenamefont {Quintana}, \citenamefont {Rocque}, \citenamefont {Roushan}, \citenamefont {Saei}, \citenamefont {Sank}, \citenamefont {Sankaragomathi}, \citenamefont {Satzinger}, \citenamefont {Schurkus}, \citenamefont {Schuster}, \citenamefont {Shearn}, \citenamefont {Shorter}, \citenamefont {Shutty}, \citenamefont {Shvarts}, \citenamefont {Skruzny}, \citenamefont {Smith}, \citenamefont {Somma}, \citenamefont {Sterling}, \citenamefont {Strain}, \citenamefont {Szalay}, \citenamefont {Thor}, \citenamefont {Torres}, \citenamefont {Vidal}, \citenamefont {Villalonga}, \citenamefont {Vollgraff~Heidweiller}, \citenamefont {White}, \citenamefont {Woo}, \citenamefont {Xing}, \citenamefont {Yao}, \citenamefont {Yeh}, \citenamefont {Yoo}, \citenamefont {Young}, \citenamefont {Zalcman}, \citenamefont {Zhang}, \citenamefont {Zhu}, \citenamefont {Zobrist}, \citenamefont
  {Bacon}, \citenamefont {Boixo}, \citenamefont {Chen}, \citenamefont {Hilton}, \citenamefont {Kelly}, \citenamefont {Lucero}, \citenamefont {Megrant}, \citenamefont {Neven}, \citenamefont {Smelyanskiy}, \citenamefont {Gogolin}, \citenamefont {Babbush},\ and\ \citenamefont {Rubin}}]{O’Brien2023}%
  \BibitemOpen
  \bibfield  {author} {\bibinfo {author} {\bibfnamefont {T.~E.}\ \bibnamefont {O'Brien}}, \bibinfo {author} {\bibfnamefont {G.}~\bibnamefont {Anselmetti}}, \bibinfo {author} {\bibfnamefont {F.}~\bibnamefont {Gkritsis}}, \bibinfo {author} {\bibfnamefont {V.~E.}\ \bibnamefont {Elfving}}, \bibinfo {author} {\bibfnamefont {S.}~\bibnamefont {Polla}}, \bibinfo {author} {\bibfnamefont {W.~J.}\ \bibnamefont {Huggins}}, \bibinfo {author} {\bibfnamefont {O.}~\bibnamefont {Oumarou}}, \bibinfo {author} {\bibfnamefont {K.}~\bibnamefont {Kechedzhi}}, \bibinfo {author} {\bibfnamefont {D.}~\bibnamefont {Abanin}}, \bibinfo {author} {\bibfnamefont {R.}~\bibnamefont {Acharya}}, \bibinfo {author} {\bibfnamefont {I.}~\bibnamefont {Aleiner}}, \bibinfo {author} {\bibfnamefont {R.}~\bibnamefont {Allen}}, \bibinfo {author} {\bibfnamefont {T.~I.}\ \bibnamefont {Andersen}}, \bibinfo {author} {\bibfnamefont {K.}~\bibnamefont {Anderson}}, \bibinfo {author} {\bibfnamefont {M.}~\bibnamefont {Ansmann}}, \bibinfo {author} {\bibfnamefont
  {F.}~\bibnamefont {Arute}}, \bibinfo {author} {\bibfnamefont {K.}~\bibnamefont {Arya}}, \bibinfo {author} {\bibfnamefont {A.}~\bibnamefont {Asfaw}}, \bibinfo {author} {\bibfnamefont {J.}~\bibnamefont {Atalaya}}, \bibinfo {author} {\bibfnamefont {J.~C.}\ \bibnamefont {Bardin}}, \bibinfo {author} {\bibfnamefont {A.}~\bibnamefont {Bengtsson}}, \bibinfo {author} {\bibfnamefont {G.}~\bibnamefont {Bortoli}}, \bibinfo {author} {\bibfnamefont {A.}~\bibnamefont {Bourassa}}, \bibinfo {author} {\bibfnamefont {J.}~\bibnamefont {Bovaird}}, \bibinfo {author} {\bibfnamefont {L.}~\bibnamefont {Brill}}, \bibinfo {author} {\bibfnamefont {M.}~\bibnamefont {Broughton}}, \bibinfo {author} {\bibfnamefont {B.}~\bibnamefont {Buckley}}, \bibinfo {author} {\bibfnamefont {D.~A.}\ \bibnamefont {Buell}}, \bibinfo {author} {\bibfnamefont {T.}~\bibnamefont {Burger}}, \bibinfo {author} {\bibfnamefont {B.}~\bibnamefont {Burkett}}, \bibinfo {author} {\bibfnamefont {N.}~\bibnamefont {Bushnell}}, \bibinfo {author} {\bibfnamefont
  {J.}~\bibnamefont {Campero}}, \bibinfo {author} {\bibfnamefont {Z.}~\bibnamefont {Chen}}, \bibinfo {author} {\bibfnamefont {B.}~\bibnamefont {Chiaro}}, \bibinfo {author} {\bibfnamefont {D.}~\bibnamefont {Chik}}, \bibinfo {author} {\bibfnamefont {J.}~\bibnamefont {Cogan}}, \bibinfo {author} {\bibfnamefont {R.}~\bibnamefont {Collins}}, \bibinfo {author} {\bibfnamefont {P.}~\bibnamefont {Conner}}, \bibinfo {author} {\bibfnamefont {W.}~\bibnamefont {Courtney}}, \bibinfo {author} {\bibfnamefont {A.~L.}\ \bibnamefont {Crook}}, \bibinfo {author} {\bibfnamefont {B.}~\bibnamefont {Curtin}}, \bibinfo {author} {\bibfnamefont {D.~M.}\ \bibnamefont {Debroy}}, \bibinfo {author} {\bibfnamefont {S.}~\bibnamefont {Demura}}, \bibinfo {author} {\bibfnamefont {I.}~\bibnamefont {Drozdov}}, \bibinfo {author} {\bibfnamefont {A.}~\bibnamefont {Dunsworth}}, \bibinfo {author} {\bibfnamefont {C.}~\bibnamefont {Erickson}}, \bibinfo {author} {\bibfnamefont {L.}~\bibnamefont {Faoro}}, \bibinfo {author} {\bibfnamefont {E.}~\bibnamefont
  {Farhi}}, \bibinfo {author} {\bibfnamefont {R.}~\bibnamefont {Fatemi}}, \bibinfo {author} {\bibfnamefont {V.~S.}\ \bibnamefont {Ferreira}}, \bibinfo {author} {\bibfnamefont {L.}~\bibnamefont {Flores~Burgos}}, \bibinfo {author} {\bibfnamefont {E.}~\bibnamefont {Forati}}, \bibinfo {author} {\bibfnamefont {A.~G.}\ \bibnamefont {Fowler}}, \bibinfo {author} {\bibfnamefont {B.}~\bibnamefont {Foxen}}, \bibinfo {author} {\bibfnamefont {W.}~\bibnamefont {Giang}}, \bibinfo {author} {\bibfnamefont {C.}~\bibnamefont {Gidney}}, \bibinfo {author} {\bibfnamefont {D.}~\bibnamefont {Gilboa}}, \bibinfo {author} {\bibfnamefont {M.}~\bibnamefont {Giustina}}, \bibinfo {author} {\bibfnamefont {R.}~\bibnamefont {Gosula}}, \bibinfo {author} {\bibfnamefont {A.}~\bibnamefont {Grajales~Dau}}, \bibinfo {author} {\bibfnamefont {J.~A.}\ \bibnamefont {Gross}}, \bibinfo {author} {\bibfnamefont {S.}~\bibnamefont {Habegger}}, \bibinfo {author} {\bibfnamefont {M.~C.}\ \bibnamefont {Hamilton}}, \bibinfo {author} {\bibfnamefont
  {M.}~\bibnamefont {Hansen}}, \bibinfo {author} {\bibfnamefont {M.~P.}\ \bibnamefont {Harrigan}}, \bibinfo {author} {\bibfnamefont {S.~D.}\ \bibnamefont {Harrington}}, \bibinfo {author} {\bibfnamefont {P.}~\bibnamefont {Heu}}, \bibinfo {author} {\bibfnamefont {M.~R.}\ \bibnamefont {Hoffmann}}, \bibinfo {author} {\bibfnamefont {S.}~\bibnamefont {Hong}}, \bibinfo {author} {\bibfnamefont {T.}~\bibnamefont {Huang}}, \bibinfo {author} {\bibfnamefont {A.}~\bibnamefont {Huff}}, \bibinfo {author} {\bibfnamefont {L.~B.}\ \bibnamefont {Ioffe}}, \bibinfo {author} {\bibfnamefont {S.~V.}\ \bibnamefont {Isakov}}, \bibinfo {author} {\bibfnamefont {J.}~\bibnamefont {Iveland}}, \bibinfo {author} {\bibfnamefont {E.}~\bibnamefont {Jeffrey}}, \bibinfo {author} {\bibfnamefont {Z.}~\bibnamefont {Jiang}}, \bibinfo {author} {\bibfnamefont {C.}~\bibnamefont {Jones}}, \bibinfo {author} {\bibfnamefont {P.}~\bibnamefont {Juhas}}, \bibinfo {author} {\bibfnamefont {D.}~\bibnamefont {Kafri}}, \bibinfo {author} {\bibfnamefont
  {T.}~\bibnamefont {Khattar}}, \bibinfo {author} {\bibfnamefont {M.}~\bibnamefont {Khezri}}, \bibinfo {author} {\bibfnamefont {M.}~\bibnamefont {Kieferov{\'a}}}, \bibinfo {author} {\bibfnamefont {S.}~\bibnamefont {Kim}}, \bibinfo {author} {\bibfnamefont {P.~V.}\ \bibnamefont {Klimov}}, \bibinfo {author} {\bibfnamefont {A.~R.}\ \bibnamefont {Klots}}, \bibinfo {author} {\bibfnamefont {A.~N.}\ \bibnamefont {Korotkov}}, \bibinfo {author} {\bibfnamefont {F.}~\bibnamefont {Kostritsa}}, \bibinfo {author} {\bibfnamefont {J.~M.}\ \bibnamefont {Kreikebaum}}, \bibinfo {author} {\bibfnamefont {D.}~\bibnamefont {Landhuis}}, \bibinfo {author} {\bibfnamefont {P.}~\bibnamefont {Laptev}}, \bibinfo {author} {\bibfnamefont {K.-M.}\ \bibnamefont {Lau}}, \bibinfo {author} {\bibfnamefont {L.}~\bibnamefont {Laws}}, \bibinfo {author} {\bibfnamefont {J.}~\bibnamefont {Lee}}, \bibinfo {author} {\bibfnamefont {K.}~\bibnamefont {Lee}}, \bibinfo {author} {\bibfnamefont {B.~J.}\ \bibnamefont {Lester}}, \bibinfo {author} {\bibfnamefont
  {A.~T.}\ \bibnamefont {Lill}}, \bibinfo {author} {\bibfnamefont {W.}~\bibnamefont {Liu}}, \bibinfo {author} {\bibfnamefont {W.~P.}\ \bibnamefont {Livingston}}, \bibinfo {author} {\bibfnamefont {A.}~\bibnamefont {Locharla}}, \bibinfo {author} {\bibfnamefont {F.~D.}\ \bibnamefont {Malone}}, \bibinfo {author} {\bibfnamefont {S.}~\bibnamefont {Mandr{\`a}}}, \bibinfo {author} {\bibfnamefont {O.}~\bibnamefont {Martin}}, \bibinfo {author} {\bibfnamefont {S.}~\bibnamefont {Martin}}, \bibinfo {author} {\bibfnamefont {J.~R.}\ \bibnamefont {McClean}}, \bibinfo {author} {\bibfnamefont {T.}~\bibnamefont {McCourt}}, \bibinfo {author} {\bibfnamefont {M.}~\bibnamefont {McEwen}}, \bibinfo {author} {\bibfnamefont {X.}~\bibnamefont {Mi}}, \bibinfo {author} {\bibfnamefont {A.}~\bibnamefont {Mieszala}}, \bibinfo {author} {\bibfnamefont {K.~C.}\ \bibnamefont {Miao}}, \bibinfo {author} {\bibfnamefont {M.}~\bibnamefont {Mohseni}}, \bibinfo {author} {\bibfnamefont {S.}~\bibnamefont {Montazeri}}, \bibinfo {author} {\bibfnamefont
  {A.}~\bibnamefont {Morvan}}, \bibinfo {author} {\bibfnamefont {R.}~\bibnamefont {Movassagh}}, \bibinfo {author} {\bibfnamefont {W.}~\bibnamefont {Mruczkiewicz}}, \bibinfo {author} {\bibfnamefont {O.}~\bibnamefont {Naaman}}, \bibinfo {author} {\bibfnamefont {M.}~\bibnamefont {Neeley}}, \bibinfo {author} {\bibfnamefont {C.}~\bibnamefont {Neill}}, \bibinfo {author} {\bibfnamefont {A.}~\bibnamefont {Nersisyan}}, \bibinfo {author} {\bibfnamefont {M.}~\bibnamefont {Newman}}, \bibinfo {author} {\bibfnamefont {J.~H.}\ \bibnamefont {Ng}}, \bibinfo {author} {\bibfnamefont {A.}~\bibnamefont {Nguyen}}, \bibinfo {author} {\bibfnamefont {M.}~\bibnamefont {Nguyen}}, \bibinfo {author} {\bibfnamefont {M.~Y.}\ \bibnamefont {Niu}}, \bibinfo {author} {\bibfnamefont {S.}~\bibnamefont {Omonije}}, \bibinfo {author} {\bibfnamefont {A.}~\bibnamefont {Opremcak}}, \bibinfo {author} {\bibfnamefont {A.}~\bibnamefont {Petukhov}}, \bibinfo {author} {\bibfnamefont {R.}~\bibnamefont {Potter}}, \bibinfo {author} {\bibfnamefont {L.~P.}\
  \bibnamefont {Pryadko}}, \bibinfo {author} {\bibfnamefont {C.}~\bibnamefont {Quintana}}, \bibinfo {author} {\bibfnamefont {C.}~\bibnamefont {Rocque}}, \bibinfo {author} {\bibfnamefont {P.}~\bibnamefont {Roushan}}, \bibinfo {author} {\bibfnamefont {N.}~\bibnamefont {Saei}}, \bibinfo {author} {\bibfnamefont {D.}~\bibnamefont {Sank}}, \bibinfo {author} {\bibfnamefont {K.}~\bibnamefont {Sankaragomathi}}, \bibinfo {author} {\bibfnamefont {K.~J.}\ \bibnamefont {Satzinger}}, \bibinfo {author} {\bibfnamefont {H.~F.}\ \bibnamefont {Schurkus}}, \bibinfo {author} {\bibfnamefont {C.}~\bibnamefont {Schuster}}, \bibinfo {author} {\bibfnamefont {M.~J.}\ \bibnamefont {Shearn}}, \bibinfo {author} {\bibfnamefont {A.}~\bibnamefont {Shorter}}, \bibinfo {author} {\bibfnamefont {N.}~\bibnamefont {Shutty}}, \bibinfo {author} {\bibfnamefont {V.}~\bibnamefont {Shvarts}}, \bibinfo {author} {\bibfnamefont {J.}~\bibnamefont {Skruzny}}, \bibinfo {author} {\bibfnamefont {W.~C.}\ \bibnamefont {Smith}}, \bibinfo {author} {\bibfnamefont
  {R.~D.}\ \bibnamefont {Somma}}, \bibinfo {author} {\bibfnamefont {G.}~\bibnamefont {Sterling}}, \bibinfo {author} {\bibfnamefont {D.}~\bibnamefont {Strain}}, \bibinfo {author} {\bibfnamefont {M.}~\bibnamefont {Szalay}}, \bibinfo {author} {\bibfnamefont {D.}~\bibnamefont {Thor}}, \bibinfo {author} {\bibfnamefont {A.}~\bibnamefont {Torres}}, \bibinfo {author} {\bibfnamefont {G.}~\bibnamefont {Vidal}}, \bibinfo {author} {\bibfnamefont {B.}~\bibnamefont {Villalonga}}, \bibinfo {author} {\bibfnamefont {C.}~\bibnamefont {Vollgraff~Heidweiller}}, \bibinfo {author} {\bibfnamefont {T.}~\bibnamefont {White}}, \bibinfo {author} {\bibfnamefont {B.~W.~K.}\ \bibnamefont {Woo}}, \bibinfo {author} {\bibfnamefont {C.}~\bibnamefont {Xing}}, \bibinfo {author} {\bibfnamefont {Z.~J.}\ \bibnamefont {Yao}}, \bibinfo {author} {\bibfnamefont {P.}~\bibnamefont {Yeh}}, \bibinfo {author} {\bibfnamefont {J.}~\bibnamefont {Yoo}}, \bibinfo {author} {\bibfnamefont {G.}~\bibnamefont {Young}}, \bibinfo {author} {\bibfnamefont
  {A.}~\bibnamefont {Zalcman}}, \bibinfo {author} {\bibfnamefont {Y.}~\bibnamefont {Zhang}}, \bibinfo {author} {\bibfnamefont {N.}~\bibnamefont {Zhu}}, \bibinfo {author} {\bibfnamefont {N.}~\bibnamefont {Zobrist}}, \bibinfo {author} {\bibfnamefont {D.}~\bibnamefont {Bacon}}, \bibinfo {author} {\bibfnamefont {S.}~\bibnamefont {Boixo}}, \bibinfo {author} {\bibfnamefont {Y.}~\bibnamefont {Chen}}, \bibinfo {author} {\bibfnamefont {J.}~\bibnamefont {Hilton}}, \bibinfo {author} {\bibfnamefont {J.}~\bibnamefont {Kelly}}, \bibinfo {author} {\bibfnamefont {E.}~\bibnamefont {Lucero}}, \bibinfo {author} {\bibfnamefont {A.}~\bibnamefont {Megrant}}, \bibinfo {author} {\bibfnamefont {H.}~\bibnamefont {Neven}}, \bibinfo {author} {\bibfnamefont {V.}~\bibnamefont {Smelyanskiy}}, \bibinfo {author} {\bibfnamefont {C.}~\bibnamefont {Gogolin}}, \bibinfo {author} {\bibfnamefont {R.}~\bibnamefont {Babbush}},\ and\ \bibinfo {author} {\bibfnamefont {N.~C.}\ \bibnamefont {Rubin}},\ }\bibfield  {title} {\bibinfo {title}
  {Purification-based quantum error mitigation of pair-correlated electron simulations},\ }\href {https://doi.org/10.1038/s41567-023-02240-y} {\bibfield  {journal} {\bibinfo  {journal} {Nature Physics}\ }\textbf {\bibinfo {volume} {19}},\ \bibinfo {pages} {1787} (\bibinfo {year} {2023})}\BibitemShut {NoStop}%
\bibitem [{\citenamefont {Childs}\ \emph {et~al.}(2021)\citenamefont {Childs}, \citenamefont {Su}, \citenamefont {Tran}, \citenamefont {Wiebe},\ and\ \citenamefont {Zhu}}]{PhysRevX.11.011020}%
  \BibitemOpen
  \bibfield  {author} {\bibinfo {author} {\bibfnamefont {A.~M.}\ \bibnamefont {Childs}}, \bibinfo {author} {\bibfnamefont {Y.}~\bibnamefont {Su}}, \bibinfo {author} {\bibfnamefont {M.~C.}\ \bibnamefont {Tran}}, \bibinfo {author} {\bibfnamefont {N.}~\bibnamefont {Wiebe}},\ and\ \bibinfo {author} {\bibfnamefont {S.}~\bibnamefont {Zhu}},\ }\bibfield  {title} {\bibinfo {title} {Theory of trotter error with commutator scaling},\ }\href {https://doi.org/10.1103/PhysRevX.11.011020} {\bibfield  {journal} {\bibinfo  {journal} {Phys. Rev. X}\ }\textbf {\bibinfo {volume} {11}},\ \bibinfo {pages} {011020} (\bibinfo {year} {2021})}\BibitemShut {NoStop}%
\bibitem [{\citenamefont {Rocca}\ \emph {et~al.}(2024)\citenamefont {Rocca}, \citenamefont {Cortes}, \citenamefont {Gonthier}, \citenamefont {Ollitrault}, \citenamefont {Parrish}, \citenamefont {Anselmetti}, \citenamefont {Degroote}, \citenamefont {Moll}, \citenamefont {Santagati},\ and\ \citenamefont {Streif}}]{Rocca2024}%
  \BibitemOpen
  \bibfield  {author} {\bibinfo {author} {\bibfnamefont {D.}~\bibnamefont {Rocca}}, \bibinfo {author} {\bibfnamefont {C.~L.}\ \bibnamefont {Cortes}}, \bibinfo {author} {\bibfnamefont {J.~F.}\ \bibnamefont {Gonthier}}, \bibinfo {author} {\bibfnamefont {P.~J.}\ \bibnamefont {Ollitrault}}, \bibinfo {author} {\bibfnamefont {R.~M.}\ \bibnamefont {Parrish}}, \bibinfo {author} {\bibfnamefont {G.-L.}\ \bibnamefont {Anselmetti}}, \bibinfo {author} {\bibfnamefont {M.}~\bibnamefont {Degroote}}, \bibinfo {author} {\bibfnamefont {N.}~\bibnamefont {Moll}}, \bibinfo {author} {\bibfnamefont {R.}~\bibnamefont {Santagati}},\ and\ \bibinfo {author} {\bibfnamefont {M.}~\bibnamefont {Streif}},\ }\bibfield  {title} {\bibinfo {title} {Reducing the runtime of fault-tolerant quantum simulations in chemistry through symmetry-compressed double factorization},\ }\href {https://doi.org/10.1021/acs.jctc.4c00352} {\bibfield  {journal} {\bibinfo  {journal} {Journal of Chemical Theory and Computation}\ }\textbf {\bibinfo {volume} {20}},\
  \bibinfo {pages} {4639} (\bibinfo {year} {2024})}\BibitemShut {NoStop}%
\bibitem [{\citenamefont {Bocharov}\ \emph {et~al.}(2015)\citenamefont {Bocharov}, \citenamefont {Roetteler},\ and\ \citenamefont {Svore}}]{PhysRevLett.114.080502}%
  \BibitemOpen
  \bibfield  {author} {\bibinfo {author} {\bibfnamefont {A.}~\bibnamefont {Bocharov}}, \bibinfo {author} {\bibfnamefont {M.}~\bibnamefont {Roetteler}},\ and\ \bibinfo {author} {\bibfnamefont {K.~M.}\ \bibnamefont {Svore}},\ }\bibfield  {title} {\bibinfo {title} {Efficient synthesis of universal repeat-until-success quantum circuits},\ }\href {https://doi.org/10.1103/PhysRevLett.114.080502} {\bibfield  {journal} {\bibinfo  {journal} {Phys. Rev. Lett.}\ }\textbf {\bibinfo {volume} {114}},\ \bibinfo {pages} {080502} (\bibinfo {year} {2015})}\BibitemShut {NoStop}%
\bibitem [{\citenamefont {Li}\ \emph {et~al.}(2023)\citenamefont {Li}, \citenamefont {Guo}, \citenamefont {Li},\ and\ \citenamefont {Sun}}]{PhysRevA.107.042424}%
  \BibitemOpen
  \bibfield  {author} {\bibinfo {author} {\bibfnamefont {L.}~\bibnamefont {Li}}, \bibinfo {author} {\bibfnamefont {C.}~\bibnamefont {Guo}}, \bibinfo {author} {\bibfnamefont {Q.}~\bibnamefont {Li}},\ and\ \bibinfo {author} {\bibfnamefont {X.}~\bibnamefont {Sun}},\ }\bibfield  {title} {\bibinfo {title} {Fast exact synthesis of two-qubit unitaries using a near-minimum number of $t$ gates},\ }\href {https://doi.org/10.1103/PhysRevA.107.042424} {\bibfield  {journal} {\bibinfo  {journal} {Phys. Rev. A}\ }\textbf {\bibinfo {volume} {107}},\ \bibinfo {pages} {042424} (\bibinfo {year} {2023})}\BibitemShut {NoStop}%
\bibitem [{\citenamefont {Perdew}\ \emph {et~al.}(1996)\citenamefont {Perdew}, \citenamefont {Burke},\ and\ \citenamefont {Ernzerhof}}]{pbe}%
  \BibitemOpen
  \bibfield  {author} {\bibinfo {author} {\bibfnamefont {J.~P.}\ \bibnamefont {Perdew}}, \bibinfo {author} {\bibfnamefont {K.}~\bibnamefont {Burke}},\ and\ \bibinfo {author} {\bibfnamefont {M.}~\bibnamefont {Ernzerhof}},\ }\bibfield  {title} {\bibinfo {title} {{Generalized Gradient Approximation Made Simple}},\ }\href@noop {} {\bibfield  {journal} {\bibinfo  {journal} {Physical Review Letters}\ }\textbf {\bibinfo {volume} {77}},\ \bibinfo {pages} {3865} (\bibinfo {year} {1996})}\BibitemShut {NoStop}%
\bibitem [{\citenamefont {Chamaki}\ \emph {et~al.}(2022)\citenamefont {Chamaki}, \citenamefont {Hadfield}, \citenamefont {Klymko}, \citenamefont {O'Gorman},\ and\ \citenamefont {Tubman}}]{chamaki2022selfconsistentquantumiterativelysparsified}%
  \BibitemOpen
  \bibfield  {author} {\bibinfo {author} {\bibfnamefont {D.~B.}\ \bibnamefont {Chamaki}}, \bibinfo {author} {\bibfnamefont {S.}~\bibnamefont {Hadfield}}, \bibinfo {author} {\bibfnamefont {K.}~\bibnamefont {Klymko}}, \bibinfo {author} {\bibfnamefont {B.}~\bibnamefont {O'Gorman}},\ and\ \bibinfo {author} {\bibfnamefont {N.~M.}\ \bibnamefont {Tubman}},\ }\href {https://arxiv.org/abs/2211.16522} {\bibinfo {title} {Self-consistent quantum iteratively sparsified hamiltonian method (squish): A new algorithm for efficient hamiltonian simulation and compression}} (\bibinfo {year} {2022}),\ \Eprint {https://arxiv.org/abs/2211.16522} {arXiv:2211.16522 [quant-ph]} \BibitemShut {NoStop}%
\bibitem [{\citenamefont {Wu}\ \emph {et~al.}(2024)\citenamefont {Wu}, \citenamefont {Hou}, \citenamefont {Li}, \citenamefont {He},\ and\ \citenamefont {Qiu}}]{PhysRevB.110.075133}%
  \BibitemOpen
  \bibfield  {author} {\bibinfo {author} {\bibfnamefont {J.}~\bibnamefont {Wu}}, \bibinfo {author} {\bibfnamefont {B.}~\bibnamefont {Hou}}, \bibinfo {author} {\bibfnamefont {W.}~\bibnamefont {Li}}, \bibinfo {author} {\bibfnamefont {Y.}~\bibnamefont {He}},\ and\ \bibinfo {author} {\bibfnamefont {D.~Y.}\ \bibnamefont {Qiu}},\ }\bibfield  {title} {\bibinfo {title} {Quasiparticle and excitonic properties of monolayer $1{T}^{\ensuremath{'}}$ ${\mathrm{wte}}_{2}$ within many-body perturbation theory},\ }\href {https://doi.org/10.1103/PhysRevB.110.075133} {\bibfield  {journal} {\bibinfo  {journal} {Phys. Rev. B}\ }\textbf {\bibinfo {volume} {110}},\ \bibinfo {pages} {075133} (\bibinfo {year} {2024})}\BibitemShut {NoStop}%
\bibitem [{\citenamefont {Rohlfing}\ and\ \citenamefont {Louie}(1998)}]{Rohlfing1998}%
  \BibitemOpen
  \bibfield  {author} {\bibinfo {author} {\bibfnamefont {M.}~\bibnamefont {Rohlfing}}\ and\ \bibinfo {author} {\bibfnamefont {S.~G.}\ \bibnamefont {Louie}},\ }\bibfield  {title} {\bibinfo {title} {{Electron-hole excitations in semiconductors and insulators}},\ }\href {https://doi.org/10.1103/PhysRevLett.81.2312} {\bibfield  {journal} {\bibinfo  {journal} {Physical Review Letters}\ }\textbf {\bibinfo {volume} {81}},\ \bibinfo {pages} {2312} (\bibinfo {year} {1998})}\BibitemShut {NoStop}%
\bibitem [{\citenamefont {Rohlfing}\ and\ \citenamefont {Louie}(2000)}]{Rohlfing2000}%
  \BibitemOpen
  \bibfield  {author} {\bibinfo {author} {\bibfnamefont {M.}~\bibnamefont {Rohlfing}}\ and\ \bibinfo {author} {\bibfnamefont {S.~G.}\ \bibnamefont {Louie}},\ }\bibfield  {title} {\bibinfo {title} {{Electron-hole excitations and optical spectra from first principles}},\ }\href {https://doi.org/10.1103/PhysRevB.62.4927} {\bibfield  {journal} {\bibinfo  {journal} {Physical Review B - Condensed Matter and Materials Physics}\ }\textbf {\bibinfo {volume} {62}},\ \bibinfo {pages} {4927} (\bibinfo {year} {2000})},\ \Eprint {https://arxiv.org/abs/0406203v3} {0406203v3 [arXiv:cond-mat]} \BibitemShut {NoStop}%
\bibitem [{\citenamefont {Kaneko}\ \emph {et~al.}(2012)\citenamefont {Kaneko}, \citenamefont {Seki},\ and\ \citenamefont {Ohta}}]{Kaneko2012}%
  \BibitemOpen
  \bibfield  {author} {\bibinfo {author} {\bibfnamefont {T.}~\bibnamefont {Kaneko}}, \bibinfo {author} {\bibfnamefont {K.}~\bibnamefont {Seki}},\ and\ \bibinfo {author} {\bibfnamefont {Y.}~\bibnamefont {Ohta}},\ }\bibfield  {title} {\bibinfo {title} {{Excitonic insulator state in the two-orbital Hubbard model: Variational cluster approach}},\ }\href {https://doi.org/10.1103/PhysRevB.85.165135} {\bibfield  {journal} {\bibinfo  {journal} {Physical Review B - Condensed Matter and Materials Physics}\ }\textbf {\bibinfo {volume} {85}},\ \bibinfo {pages} {1} (\bibinfo {year} {2012})}\BibitemShut {NoStop}%
\bibitem [{\citenamefont {Anselme~Martin}\ \emph {et~al.}(2022)\citenamefont {Anselme~Martin}, \citenamefont {Simon},\ and\ \citenamefont {Ran{\v c}i{\'c}}}]{anselmemartinSimulatingStronglyInteracting2022}%
  \BibitemOpen
  \bibfield  {author} {\bibinfo {author} {\bibfnamefont {B.}~\bibnamefont {Anselme~Martin}}, \bibinfo {author} {\bibfnamefont {P.}~\bibnamefont {Simon}},\ and\ \bibinfo {author} {\bibfnamefont {M.~J.}\ \bibnamefont {Ran{\v c}i{\'c}}},\ }\bibfield  {title} {\bibinfo {title} {Simulating strongly interacting {{Hubbard}} chains with the variational {{Hamiltonian}} ansatz on a quantum computer},\ }\href {https://doi.org/10.1103/PhysRevResearch.4.023190} {\bibfield  {journal} {\bibinfo  {journal} {Phys. Rev. Res.}\ }\textbf {\bibinfo {volume} {4}},\ \bibinfo {pages} {023190} (\bibinfo {year} {2022})}\BibitemShut {NoStop}%
\bibitem [{\citenamefont {Gustafson}\ \emph {et~al.}(2024)\citenamefont {Gustafson}, \citenamefont {Tiihonen}, \citenamefont {Chamaki}, \citenamefont {Sorourifar}, \citenamefont {Mullinax}, \citenamefont {Li}, \citenamefont {Maciejewski}, \citenamefont {Sawaya}, \citenamefont {Krogel}, \citenamefont {Neira},\ and\ \citenamefont {Tubman}}]{gustafson2024surrogateoptimizationvariationalquantum}%
  \BibitemOpen
  \bibfield  {author} {\bibinfo {author} {\bibfnamefont {E.~J.}\ \bibnamefont {Gustafson}}, \bibinfo {author} {\bibfnamefont {J.}~\bibnamefont {Tiihonen}}, \bibinfo {author} {\bibfnamefont {D.}~\bibnamefont {Chamaki}}, \bibinfo {author} {\bibfnamefont {F.}~\bibnamefont {Sorourifar}}, \bibinfo {author} {\bibfnamefont {J.~W.}\ \bibnamefont {Mullinax}}, \bibinfo {author} {\bibfnamefont {A.~C.~Y.}\ \bibnamefont {Li}}, \bibinfo {author} {\bibfnamefont {F.~B.}\ \bibnamefont {Maciejewski}}, \bibinfo {author} {\bibfnamefont {N.~P.}\ \bibnamefont {Sawaya}}, \bibinfo {author} {\bibfnamefont {J.~T.}\ \bibnamefont {Krogel}}, \bibinfo {author} {\bibfnamefont {D.~E.~B.}\ \bibnamefont {Neira}},\ and\ \bibinfo {author} {\bibfnamefont {N.~M.}\ \bibnamefont {Tubman}},\ }\href {https://arxiv.org/abs/2404.02951} {\bibinfo {title} {Surrogate optimization of variational quantum circuits}} (\bibinfo {year} {2024}),\ \Eprint {https://arxiv.org/abs/2404.02951} {arXiv:2404.02951 [quant-ph]} \BibitemShut {NoStop}%
\bibitem [{\citenamefont {Grimsley}\ \emph {et~al.}(2019)\citenamefont {Grimsley}, \citenamefont {Economou}, \citenamefont {Barnes},\ and\ \citenamefont {Mayhall}}]{Grimsley2019}%
  \BibitemOpen
  \bibfield  {author} {\bibinfo {author} {\bibfnamefont {H.~R.}\ \bibnamefont {Grimsley}}, \bibinfo {author} {\bibfnamefont {S.~E.}\ \bibnamefont {Economou}}, \bibinfo {author} {\bibfnamefont {E.}~\bibnamefont {Barnes}},\ and\ \bibinfo {author} {\bibfnamefont {N.~J.}\ \bibnamefont {Mayhall}},\ }\bibfield  {title} {\bibinfo {title} {{An adaptive variational algorithm for exact molecular simulations on a quantum computer}},\ }\href {https://doi.org/10.1038/s41467-019-10988-2} {\bibfield  {journal} {\bibinfo  {journal} {Nature Communications}\ }\textbf {\bibinfo {volume} {10}},\ \bibinfo {pages} {3007} (\bibinfo {year} {2019})}\BibitemShut {NoStop}%
\bibitem [{\citenamefont {Mullinax}\ \emph {et~al.}(2024)\citenamefont {Mullinax}, \citenamefont {Anastasiou}, \citenamefont {Larson}, \citenamefont {Economou},\ and\ \citenamefont {Tubman}}]{mullinax2024classical}%
  \BibitemOpen
  \bibfield  {author} {\bibinfo {author} {\bibfnamefont {J.~W.}\ \bibnamefont {Mullinax}}, \bibinfo {author} {\bibfnamefont {P.~G.}\ \bibnamefont {Anastasiou}}, \bibinfo {author} {\bibfnamefont {J.}~\bibnamefont {Larson}}, \bibinfo {author} {\bibfnamefont {S.~E.}\ \bibnamefont {Economou}},\ and\ \bibinfo {author} {\bibfnamefont {N.~M.}\ \bibnamefont {Tubman}},\ }\bibfield  {title} {\bibinfo {title} {Classical pre-optimization approach for adapt-vqe: Maximizing the potential of high-performance computing resources to improve quantum simulation of chemical applications},\ }\href@noop {} {\bibfield  {journal} {\bibinfo  {journal} {arXiv preprint arXiv:2411.07920}\ } (\bibinfo {year} {2024})}\BibitemShut {NoStop}%
\bibitem [{\citenamefont {Wing}\ \emph {et~al.}(2021)\citenamefont {Wing}, \citenamefont {Ohad}, \citenamefont {Haber}, \citenamefont {Filip}, \citenamefont {Gant}, \citenamefont {Neaton},\ and\ \citenamefont {Kronik}}]{Wing2021}%
  \BibitemOpen
  \bibfield  {author} {\bibinfo {author} {\bibfnamefont {D.}~\bibnamefont {Wing}}, \bibinfo {author} {\bibfnamefont {G.}~\bibnamefont {Ohad}}, \bibinfo {author} {\bibfnamefont {J.~B.}\ \bibnamefont {Haber}}, \bibinfo {author} {\bibfnamefont {M.~R.}\ \bibnamefont {Filip}}, \bibinfo {author} {\bibfnamefont {S.~E.}\ \bibnamefont {Gant}}, \bibinfo {author} {\bibfnamefont {J.~B.}\ \bibnamefont {Neaton}},\ and\ \bibinfo {author} {\bibfnamefont {L.}~\bibnamefont {Kronik}},\ }\bibfield  {title} {\bibinfo {title} {{Band gaps of crystalline solids from Wannier-localization–based optimal tuning of a screened range-separated hybrid functional}},\ }\href {https://doi.org/10.1073/pnas.2104556118} {\bibfield  {journal} {\bibinfo  {journal} {Proceedings of the National Academy of Sciences of the United States of America}\ }\textbf {\bibinfo {volume} {118}},\ \bibinfo {pages} {1} (\bibinfo {year} {2021})},\ \Eprint {https://arxiv.org/abs/2012.03278} {arXiv:2012.03278} \BibitemShut {NoStop}%
\bibitem [{\citenamefont {Ohad}\ \emph {et~al.}(2023)\citenamefont {Ohad}, \citenamefont {Gant}, \citenamefont {Wing}, \citenamefont {Haber}, \citenamefont {Camarasa-G{\'{o}}mez}, \citenamefont {Sagredo}, \citenamefont {Filip}, \citenamefont {Neaton},\ and\ \citenamefont {Kronik}}]{Ohad2023}%
  \BibitemOpen
  \bibfield  {author} {\bibinfo {author} {\bibfnamefont {G.}~\bibnamefont {Ohad}}, \bibinfo {author} {\bibfnamefont {S.~E.}\ \bibnamefont {Gant}}, \bibinfo {author} {\bibfnamefont {D.}~\bibnamefont {Wing}}, \bibinfo {author} {\bibfnamefont {J.~B.}\ \bibnamefont {Haber}}, \bibinfo {author} {\bibfnamefont {M.}~\bibnamefont {Camarasa-G{\'{o}}mez}}, \bibinfo {author} {\bibfnamefont {F.}~\bibnamefont {Sagredo}}, \bibinfo {author} {\bibfnamefont {M.~R.}\ \bibnamefont {Filip}}, \bibinfo {author} {\bibfnamefont {J.~B.}\ \bibnamefont {Neaton}},\ and\ \bibinfo {author} {\bibfnamefont {L.}~\bibnamefont {Kronik}},\ }\bibfield  {title} {\bibinfo {title} {{Optical absorption spectra of metal oxides from time-dependent density functional theory and many-body perturbation theory based on optimally-tuned hybrid functionals}},\ }\href {https://doi.org/10.1103/PhysRevMaterials.7.123803} {\bibfield  {journal} {\bibinfo  {journal} {Physical Review Materials}\ }\textbf {\bibinfo {volume} {7}},\ \bibinfo {pages} {123803} (\bibinfo
  {year} {2023})}\BibitemShut {NoStop}%
\bibitem [{\citenamefont {Filip}\ \emph {et~al.}(2021)\citenamefont {Filip}, \citenamefont {Haber},\ and\ \citenamefont {Neaton}}]{Filip2021}%
  \BibitemOpen
  \bibfield  {author} {\bibinfo {author} {\bibfnamefont {M.~R.}\ \bibnamefont {Filip}}, \bibinfo {author} {\bibfnamefont {J.~B.}\ \bibnamefont {Haber}},\ and\ \bibinfo {author} {\bibfnamefont {J.~B.}\ \bibnamefont {Neaton}},\ }\bibfield  {title} {\bibinfo {title} {{Phonon Screening of Excitons in Semiconductors: Halide Perovskites and beyond}},\ }\href {https://doi.org/10.1103/PhysRevLett.127.067401} {\bibfield  {journal} {\bibinfo  {journal} {Physical Review Letters}\ }\textbf {\bibinfo {volume} {127}},\ \bibinfo {pages} {67401} (\bibinfo {year} {2021})},\ \Eprint {https://arxiv.org/abs/2106.08697} {2106.08697} \BibitemShut {NoStop}%
\bibitem [{\citenamefont {Alvertis}\ \emph {et~al.}(2024{\natexlab{b}})\citenamefont {Alvertis}, \citenamefont {Haber}, \citenamefont {Li}, \citenamefont {Coveney}, \citenamefont {Louie}, \citenamefont {Filip},\ and\ \citenamefont {Neaton}}]{alvertis2023phonon}%
  \BibitemOpen
  \bibfield  {author} {\bibinfo {author} {\bibfnamefont {A.~M.}\ \bibnamefont {Alvertis}}, \bibinfo {author} {\bibfnamefont {J.~B.}\ \bibnamefont {Haber}}, \bibinfo {author} {\bibfnamefont {Z.}~\bibnamefont {Li}}, \bibinfo {author} {\bibfnamefont {C.~J.~N.}\ \bibnamefont {Coveney}}, \bibinfo {author} {\bibfnamefont {S.~G.}\ \bibnamefont {Louie}}, \bibinfo {author} {\bibfnamefont {M.~R.}\ \bibnamefont {Filip}},\ and\ \bibinfo {author} {\bibfnamefont {J.~B.}\ \bibnamefont {Neaton}},\ }\bibfield  {title} {\bibinfo {title} {{Phonon screening and dissociation of excitons at finite temperatures from first principles}},\ }\href {https://doi.org/10.1073/pnas.2403434121} {\bibfield  {journal} {\bibinfo  {journal} {Proceedings of the National Academy of Sciences}\ }\textbf {\bibinfo {volume} {121}},\ \bibinfo {pages} {e2403434121} (\bibinfo {year} {2024}{\natexlab{b}})}\BibitemShut {NoStop}%
\bibitem [{\citenamefont {Nomura}\ and\ \citenamefont {Arita}(2015)}]{Nomura2015}%
  \BibitemOpen
  \bibfield  {author} {\bibinfo {author} {\bibfnamefont {Y.}~\bibnamefont {Nomura}}\ and\ \bibinfo {author} {\bibfnamefont {R.}~\bibnamefont {Arita}},\ }\bibfield  {title} {\bibinfo {title} {{Ab initio downfolding for electron-phonon-coupled systems: Constrained density-functional perturbation theory}},\ }\href {https://doi.org/10.1103/PhysRevB.92.245108} {\bibfield  {journal} {\bibinfo  {journal} {Physical Review B - Condensed Matter and Materials Physics}\ }\textbf {\bibinfo {volume} {92}},\ \bibinfo {pages} {1} (\bibinfo {year} {2015})},\ \Eprint {https://arxiv.org/abs/1509.01138} {arXiv:1509.01138} \BibitemShut {NoStop}%
\bibitem [{\citenamefont {{Van Loon}}\ \emph {et~al.}(2021)\citenamefont {{Van Loon}}, \citenamefont {Berges},\ and\ \citenamefont {Wehling}}]{VanLoon2021}%
  \BibitemOpen
  \bibfield  {author} {\bibinfo {author} {\bibfnamefont {E.~G.}\ \bibnamefont {{Van Loon}}}, \bibinfo {author} {\bibfnamefont {J.}~\bibnamefont {Berges}},\ and\ \bibinfo {author} {\bibfnamefont {T.~O.}\ \bibnamefont {Wehling}},\ }\bibfield  {title} {\bibinfo {title} {{Downfolding approaches to electron-ion coupling: Constrained density-functional perturbation theory for molecules}},\ }\href {https://doi.org/10.1103/PhysRevB.103.205103} {\bibfield  {journal} {\bibinfo  {journal} {Physical Review B}\ }\textbf {\bibinfo {volume} {103}},\ \bibinfo {pages} {1} (\bibinfo {year} {2021})},\ \Eprint {https://arxiv.org/abs/2102.10072} {2102.10072} \BibitemShut {NoStop}%
\bibitem [{\citenamefont {Tubman}\ \emph {et~al.}(2025{\natexlab{a}})\citenamefont {Tubman}, \citenamefont {Coveney}, \citenamefont {Hsu}, \citenamefont {Montoya-Castillo}, \citenamefont {Filip}, \citenamefont {Neaton}, \citenamefont {Li}, \citenamefont {Vlcek},\ and\ \citenamefont {Alvertis}}]{tubman2025theoryabinitiodownfolding}%
  \BibitemOpen
  \bibfield  {author} {\bibinfo {author} {\bibfnamefont {N.~M.}\ \bibnamefont {Tubman}}, \bibinfo {author} {\bibfnamefont {C.~J.~N.}\ \bibnamefont {Coveney}}, \bibinfo {author} {\bibfnamefont {C.-E.}\ \bibnamefont {Hsu}}, \bibinfo {author} {\bibfnamefont {A.}~\bibnamefont {Montoya-Castillo}}, \bibinfo {author} {\bibfnamefont {M.~R.}\ \bibnamefont {Filip}}, \bibinfo {author} {\bibfnamefont {J.~B.}\ \bibnamefont {Neaton}}, \bibinfo {author} {\bibfnamefont {Z.}~\bibnamefont {Li}}, \bibinfo {author} {\bibfnamefont {V.}~\bibnamefont {Vlcek}},\ and\ \bibinfo {author} {\bibfnamefont {A.~M.}\ \bibnamefont {Alvertis}},\ }\href {https://arxiv.org/abs/2502.00103} {\bibinfo {title} {Theory of ab initio downfolding with arbitrary range electron-phonon coupling}} (\bibinfo {year} {2025}{\natexlab{a}}),\ \Eprint {https://arxiv.org/abs/2502.00103} {arXiv:2502.00103 [cond-mat.mtrl-sci]} \BibitemShut {NoStop}%
\bibitem [{\citenamefont {Tubman}\ \emph {et~al.}(2025{\natexlab{b}})\citenamefont {Tubman}, \citenamefont {Coveney}, \citenamefont {Hsu}, \citenamefont {Montoya-Castillo}, \citenamefont {Filip}, \citenamefont {Neaton}, \citenamefont {Li}, \citenamefont {Vlcek},\ and\ \citenamefont {Alvertis}}]{tubman2025phononmediatedelectronattractionsrtio3}%
  \BibitemOpen
  \bibfield  {author} {\bibinfo {author} {\bibfnamefont {N.~M.}\ \bibnamefont {Tubman}}, \bibinfo {author} {\bibfnamefont {C.~J.~N.}\ \bibnamefont {Coveney}}, \bibinfo {author} {\bibfnamefont {C.-E.}\ \bibnamefont {Hsu}}, \bibinfo {author} {\bibfnamefont {A.}~\bibnamefont {Montoya-Castillo}}, \bibinfo {author} {\bibfnamefont {M.~R.}\ \bibnamefont {Filip}}, \bibinfo {author} {\bibfnamefont {J.~B.}\ \bibnamefont {Neaton}}, \bibinfo {author} {\bibfnamefont {Z.}~\bibnamefont {Li}}, \bibinfo {author} {\bibfnamefont {V.}~\bibnamefont {Vlcek}},\ and\ \bibinfo {author} {\bibfnamefont {A.~M.}\ \bibnamefont {Alvertis}},\ }\href {https://arxiv.org/abs/2501.17230} {\bibinfo {title} {{Phonon-mediated electron attraction in SrTiO$_3$ via the generalized Fr\"ohlich and deformation potential mechanisms}}} (\bibinfo {year} {2025}{\natexlab{b}}),\ \Eprint {https://arxiv.org/abs/2501.17230} {arXiv:2501.17230 [cond-mat.supr-con]} \BibitemShut {NoStop}%
\bibitem [{\citenamefont {Nocedal}\ and\ \citenamefont {Wright}(1999)}]{nocedal1999numerical}%
  \BibitemOpen
  \bibfield  {author} {\bibinfo {author} {\bibfnamefont {J.}~\bibnamefont {Nocedal}}\ and\ \bibinfo {author} {\bibfnamefont {S.~J.}\ \bibnamefont {Wright}},\ }\href@noop {} {\emph {\bibinfo {title} {Numerical optimization}}}\ (\bibinfo  {publisher} {Springer},\ \bibinfo {year} {1999})\BibitemShut {NoStop}%
\bibitem [{\citenamefont {Hamann}(2013)}]{Hamann2013}%
  \BibitemOpen
  \bibfield  {author} {\bibinfo {author} {\bibfnamefont {D.~R.}\ \bibnamefont {Hamann}},\ }\bibfield  {title} {\bibinfo {title} {{Optimized norm-conserving Vanderbilt pseudopotentials}},\ }\href {https://doi.org/10.1103/PhysRevB.88.085117} {\bibfield  {journal} {\bibinfo  {journal} {Physical Review B - Condensed Matter and Materials Physics}\ }\textbf {\bibinfo {volume} {88}},\ \bibinfo {pages} {1} (\bibinfo {year} {2013})},\ \Eprint {https://arxiv.org/abs/1306.4707} {1306.4707} \BibitemShut {NoStop}%
\bibitem [{\citenamefont {van Setten}\ \emph {et~al.}(2018)\citenamefont {van Setten}, \citenamefont {Giantomassi}, \citenamefont {Bousquet}, \citenamefont {Verstraete}, \citenamefont {Hamann}, \citenamefont {Gonze},\ and\ \citenamefont {Rignanese}}]{VanSetten2018}%
  \BibitemOpen
  \bibfield  {author} {\bibinfo {author} {\bibfnamefont {M.~J.}\ \bibnamefont {van Setten}}, \bibinfo {author} {\bibfnamefont {M.}~\bibnamefont {Giantomassi}}, \bibinfo {author} {\bibfnamefont {E.}~\bibnamefont {Bousquet}}, \bibinfo {author} {\bibfnamefont {M.~J.}\ \bibnamefont {Verstraete}}, \bibinfo {author} {\bibfnamefont {D.~R.}\ \bibnamefont {Hamann}}, \bibinfo {author} {\bibfnamefont {X.}~\bibnamefont {Gonze}},\ and\ \bibinfo {author} {\bibfnamefont {G.~M.}\ \bibnamefont {Rignanese}},\ }\bibfield  {title} {\bibinfo {title} {{The PSEUDODOJO: Training and grading a 85 element optimized norm-conserving pseudopotential table}},\ }\href {https://doi.org/10.1016/j.cpc.2018.01.012} {\bibfield  {journal} {\bibinfo  {journal} {Computer Physics Communications}\ }\textbf {\bibinfo {volume} {226}},\ \bibinfo {pages} {39} (\bibinfo {year} {2018})},\ \Eprint {https://arxiv.org/abs/1710.10138} {arXiv:1710.10138} \BibitemShut {NoStop}%
\end{thebibliography}%

\end{document}